\documentclass{aa}
\usepackage{natbib}
\usepackage{ulem}
\usepackage{appendix}
\usepackage[varg]{txfonts}
\bibpunct{(}{)}{;}{a}{}{,} 
\usepackage{xcolor}
\usepackage{longtable}
\begin{document}

\title{IBIS-A: The IBIS data Archive
  \thanks{Research supported by the H2020 SOLARNET  grant 
    no. 824135}\fnmsep 
 } 

\subtitle{High resolution observations of \\
the solar photosphere and chromosphere with contextual data} 

\author{Ilaria Ermolli\inst{1}
  \and Fabrizio Giorgi\inst{1}
  \and Mariarita Murabito\inst{1}
  \and Marco Stangalini\inst{2,1}
    \and Vincenzo Guido\inst{1}
  \and Marco Molinaro\inst{3}
  \and Paolo Romano\inst{4}
  \and Salvatore L. Guglielmino\inst{4}
  \and Giorgio Viavattene\inst{1}
  \and Gianna Cauzzi \inst{5,6}
  \and Serena Criscuoli \inst{6}
    \and Kevin P. Reardon  \inst{6}
  \and Alexandra Tritschler \inst{6}
} 

\offprints{I. Ermolli, \email{ilaria.ermolli@inaf.it}}

\institute{INAF--Osservatorio Astronomico di Roma, via Frascati 33, 00078 Monte Porzio Catone, Italy 
  \and ASI Agenzia Spaziale Italiana, via della Ricerca Scientifica, Roma, Italy
  \and INAF---Osservatorio Astronomico di Trieste, Via Giambattista Tiepolo, 11, 34131 Trieste, Italy
  \and INAF---Osservatorio Astrofisico di Catania, Via S. Sofia 78, 95123 Catania, Italy
   \and INAF---Osservatorio Astrofisico di Arcetri, Largo Enrico Fermi 5, 50125 Firenze, Italy
   \and National Solar Observatory, 3665 Discovery Drive, Boulder CO 80303, USA
  } 

\date{Received x y 2021/ Accepted z w j}

\abstract {The IBIS data Archive (IBIS-A) stores data acquired with the Interferometric BIdimensional Spectropolarimeter (IBIS), which was  operated at the Dunn Solar Telescope of the US National Solar Observatory from June 2003 to June 2019. The instrument provided series of high-resolution narrow-band spectro-polarimetric imaging observations of the photosphere and chromosphere in the range 5800$-$8600 \AA~ and co-temporal broad-band observations in the same spectral range and with the same field-of-view of the polarimetric data. }
{We present the data currently stored in IBIS-A, as well as the interface utilized to explore such data and facilitate its scientific exploitation. 
To this  purpose we also describe the use of IBIS-A data in recent and undergoing studies relevant to Solar Physics and Space Weather research.}
{IBIS-A includes raw and calibrated observations, as well as science-ready data.  
The latter comprise  maps of the circular, linear, and net circular  polarization, and of the magnetic and velocity fields derived for a significant fraction of the series available in the archive. IBIS-A furthermore contains links to observations complementary to the IBIS data, such as co-temporal high-resolution observations of the solar atmosphere available from  
the instruments onboard the Hinode and IRIS satellites, and 
full-disc multiband images from 
INAF solar telescopes.} 
{IBIS-A currently consists of 30 TB of data taken with IBIS during 28 observing campaigns performed in 2008 and from 2012 to 2019 on 159 days.
29\% of the observations are released as Level 1 data calibrated for instrumental response and compensated for  residual seeing degradation, while 10\% of these data is also  available as Level 1.5 format as multi-dimensional arrays of circular, linear, and net circular polarization maps, and line-of-sight velocity patterns. 81\% of the photospheric calibrated series present Level 2 data with the view of the magnetic and velocity fields of the targets, as derived from data inversion with the Very Fast Inversion of the Stokes Vector code (VFISV). 
Metadata and movies of each calibrated and science-ready series are also available to help users evaluating observing conditions. }
{IBIS-A 
represents a unique resource for investigating the plasma processes in the solar atmosphere and the solar origin of Space Weather events.  
The archive presently contains 454 different series of observations.  A recently undertaken effort to preserve IBIS observations is expected to lead in the future to an increase of the raw measurements and fraction of processed data 
available in IBIS-A.
} 

\maketitle

\section{Introduction}
The solar atmosphere is a highly structured, complex, and strongly dynamic environment produced by the continuous interplay between un-magnetized and magnetized plasmas that permeate it \citep[e.g.][]{Rempel2011LRSP,Stein2012LRSP,Cheung2014LRSP,Vandriel2015LRSP,Bellot2019LRSP,Wiegelmann2021LRSP}. Such interplay 
drives the Sun's activity and radiative variability on time scales from seconds to centuries. Besides, it imposes  electromagnetic forces in the whole heliosphere, by modulating the solar particulate and magnetic flux   \citep[e.g.][]{Kilpua2017LRSP,Temmer2021LRSP}. The same interplay also holds a fundamental role in hugely diverse astrophysical systems. For example, in addition to being responsible for the evolution of magnetic features, coronal heating and wind acceleration in the solar and stellar atmospheres \citep[e.g.][]{Schmelz2003}, it is responsible for the acceleration of jets from active galactic nuclei and gamma ray bursts \citep[e.g.][]{Blandford2019}, and for the heating of the warm-ionized medium of galaxies \citep[e.g.][]{Padoan2011}. 
In all these systems, un-magnetized and magnetized plasmas interrelate at small spatial and temporal scales that are still beyond those achievable with current telescopes. However, these scales will be accessible soon in the Sun's atmosphere with the upcoming  4-m class solar telescopes, e.g.  the National
Science Foundation's Daniel K. Inouye Solar Telescope 
\citep[DKIST,][]{2020SoPh..295..172R,Rast2021} and European Solar Telescope   \citep[EST,][]{est,Schlichenmaier2019}.

Nevertheless, analysis of high-resolution multi-wavelength  spectro-polarimetric observations of the Sun's atmosphere performed over the past two decades have already yielded a wealth of information on the  interaction between un-magnetized and magnetized plasmas  
\citep[e.g.][]{Vandriel2015LRSP,Jess2015,Borrero2017,Bellot2019LRSP}. These observations include data   acquired 
with the Interferometric BIdimensional Spectrometer (IBIS, \citealp{cavallini2006}), which combined dual, tunable Fabry-Perot interferometers with a high-order adaptive optics system and short exposures to perform high-resolution solar studies.

IBIS was developed at the INAF Arcetri Astrophysical Observatory with contributions from the Universities of Florence and Rome “Tor Vergata”.   
It was installed  at the 
Dunn Solar Telescope  (DST) of the National Solar Observatory (NSO) in New Mexico (USA) in June 2003, on an optical bench fed by a high-order adaptive optics system. It was operated there by INAF and NSO from 2005 to 2019 as a facility instrument.  
In 2019 the instrument was dismantled  for refurbishing in light of its re-installation at a different telescope \citep{ermolli_ibis20}.

From 2003 to 2019 IBIS has provided data of the solar photosphere and chromosphere at high spectral, spatial and temporal resolution over a relatively large field-of-view (FOV, hereafter). Find more details in Sect. 2.2. These data  have often been acquired simultaneously to those from  other instruments installed at the DST, e.g. the Rapid Oscillations in the Solar Atmosphere   \citep[ROSA,][]{ROSA2010} and the Facility Infrared Spectrometer   \citep[FIRS,][]{jaeggli_2012}.
Besides,  
IBIS  
has often  
observed co-temporally with other instruments 
in the framework of coordinated observing campaigns, with  space-borne telescopes, e.g., Solar and Heliospheric Observatory/Michelson Doppler Imager \citep[SOHO/MDI,][]{mdi}, Hinode/Solar Optical Telescope \citep[Hinode/SOT,][]{Hinodesot}, Solar Dynamics Observatory/Helioseismic and Magnetic Imager  \citep[SDO/HMI,][]{hmi},  
SDO/Atmospheric Imaging Assembly  \citep[SDO/AIA,][]{aia},   
Interface Region Imaging Spectrograph \citep[IRIS,][]{iris}, and ground-based telescopes, e.g., New Solar Telescope \citep[NST,][]{cao2010} and 
Atacama Large Millimeter/submillimeter Array \citep[ALMA,][]{alma}. With the DST located in the mountains above the White Sands Missile Range, IBIS also co-observed with many solar sounding rocket payloads launched from the range, including  Extreme Ultraviolet Normal Incidence Spectrograph \citep[EUNIS,][]{eunis}, 
High Resolution Coronal Imager  \citep[Hi-C,][]{hic},  Chromospheric Lyman-Alpha Spectro-Polarimeter \citep[CLASP,][]{clasp}, and Very high Angular resolution Ultraviolet Telescope  \citep[VAULT,][]{vault}.
Instead, due to  different time zones and telescope operations, there are  few if any co-temporary observations between IBIS and other spectro-polarimetric instruments installed at the solar telescopes on Canary Islands,  e.g.  the CRisp Imaging Spectro-Polarimeter   
\citep[CRISP,][]{scharmer2008}  
and the CHROMospheric Imaging Spectrometer   \citep[CHROMIS,][]{Scharmer2019} operating at the Swedish 1-m Solar Telescope   \citep[SST,][]{scharmer2003}. 

Given the versatility of the instrument, IBIS data 
have been used to study a variety of topics, including: temporal evolution and 3D nature of plasma flow in quiet regions \citep{delmoro2007} and large-scale magnetic features \citep{giordano2008,sobotka2012,sobotka2013}; turbulent solar convection  \citep{viavattene2020,viavattene2021}; structure and brightness of small- \citep{viticchie2009,viticchie2010,romano2012} and large-scale magnetic features \citep{criscuoli2012}; dissipation of magnetic energy by current sheets above sunspot umbrae \citep{tritschler2008}; identification, formation and decay of active regions \citep{ermolli2017,zuccarello2009,murabito2016,murabito2017,schilliro2021}; formation, atmospheric stratification and brightening in penumbral regions \citep{romano2013,romano2014,murabito2019,Murabito2020ApJ,romano2020APJ}; acoustic oscillations and MHD wave propagation in the solar atmosphere  \citep{vecchio2007,vecchio2009,stangalini2011,stangalini2012,stangalini_2013,sobotka2016,abbasvand2020,houston2020} and in sunspots \citep{stangalini2018,stangalini2021RSPTA,stangalini2021Nat,stangalini2021}; First Ionization Potential (FIP) effect \citep{baker2021}; magnetic flux emergence from the photosphere to the chromosphere \citep{Murabito2020ApJ}; chromospheric vortices \citep{murabito2020AA}; flares \citep{kowalski2015,capparelli2017,romano2017}; sensitivity of spectral line diagnostics to  temperature variations and small-scale magnetic fields \citep{criscuoli2013} and new spectroscopic diagnostics of the solar chromosphere \citep{cauzzi2008,straus2008,cauzzi2009,reardon2009,judge2010,lipartito2014,molnar2019}.

Recently an effort has been  undertaken to collect to best effort the data acquired by the IBIS instrument and to preserve them in the permanent IBIS-Archive (IBIS-A). 
In this paper, we describe the public release of the data available in  IBIS-A.  
We introduce the IBIS observations and their processing in Sect. 2. We present the data available in IBIS-A and their quality in Sect. 3. We give an outline of the data access and 
examples of unexplored data available in 
IBIS-A in Sects. 4 and 5, respectively. Finally, we summarize this work and draw our conclusions in Sect. 6.

\section{Observations and data processing}

\subsection{IBIS}

The IBIS instrument \citep{cavallini2006}
basically consisted of two tunable Fabry-P\'erot interferometers (hereafter referred to as FPs) and narrow band interference 
filters (hereafter referred to as pre-filters) operating in a classical mount
over the spectral range 5800$-$8600 \AA~
(later extended down to 5400 \AA).  
During operations, 
the combination of the FPs, pre-filters and instrument control allowed tuning the instrument transmission in wavelength sequentially through spectral lines isolated by the pre-filters. 
The number of wavelength points sampled per line, as well as  the number of lines used in sequence, were defined by the observer 
depending on science target  and available pre-filters. 

It is worth noting that the layout and optical components of the IBIS instrument underwent improvements in time. For example, the addition of a polarimetric unit in 2006 allowed to expand measurements at the spectral positions sampled by the instrument to all the Stokes polarization states (I, Q, U, V). After this upgrade, the instrument also underwent the implementation of new detectors and control software in 2010. However, these changes did not affect the  main characteristics of the instrument described above. Moreover, from 2011 to 2019 the instrument had a rather stable condition with only minor modifications that did not significantly change the main  characteristics of the data acquired by  the instrument. 

 A thorough evaluation of the optical performance of IBIS is provided by \citet{reardon_cavallini2008} and \citet{righini2010}. In brief, the instrument had a
transmission profile with full width at half maximum (FWHM)  of 24.0  m\AA~ at 6328 \AA~ \citep{reardon_cavallini2008} and a pixel scale of 
0.082 arcsec per pixel (until 2010) or 0.098 arcsec per pixel  (after 2010). These values allowed observations at the diffraction limit of the DST ($\approx$ 0.16 arcsec at 5800 \AA).   
Find further details concerning the IBIS data in  Sect. 2.2.

\subsection{Observing campaigns}

A typical IBIS data set consists of measurements taken in sequence over multiple spectral lines (two or three lines, e.g. Fe I at 6302 \AA, Fe I at  6173 \AA, and Ca II at 8542 \AA), with each line sampled at several spectral positions (often between 10 and 30 positions), each position at six polarimetric modulation states (I$\pm \mathrm{[Q,U,V]}$) 
and each state with either single exposure or multiple exposures.
The data are characterized by a high spectral (R $>$ 200000), spatial ($\approx$ 0.16 -- 0.24 arcsec), and temporal (8 -- 15 frames per second) resolution, over a relatively large FOV (minimum linear dimension $>$ 40 arcsec, find more details in the following).  
IBIS  ensured a high wavelength stability in time of the instrumental profile (with a  drift of the order of 10 m/s over 10 hours), short exposure times to partially freeze the Earth’s atmospheric seeing (tens of ms), and nearly diffraction limited resolution over the whole FOV.

The instrument could be operated in both the spectroscopic
and spectro-polarimetric modes. 
It also allowed for a hybrid mode where full-Stokes  measurements were taken in some of the selected spectral lines while only spectral Stokes-I measurements were acquired for other lines. 
In standard spectroscopic mode the FOV was often circular with a diameter of $\approx$ 95 arcsec, 
 as defined by a circular mask  inserted in the entrance focal plane of the instrument. 
 In standard spectro-polarimetric mode the FOV was rectangular with a dimension of $\approx$ 40$\times$95 arcsec$^2$, as a result of the rectangular mask inserted in the focal entrance plane of the instrument so that the two 
 orthogonally linear polarized beams could be imaged onto the same detector. 

Both the spectroscopic and spectro-polarimetric observational modes were operated with acquisition of broad-band (hereafter referred to as BB) images simultaneous to the narrow-band (hereafter referred to as NB) data produced by the two FPs. The BB images allow for post-facto compensation of residual seeing degradation with enhancement of the image quality \citep{Lofdahl2016}.

\begin{figure*}
{
\centering
\includegraphics[scale=0.27,trim=20 40 28 50,clip]{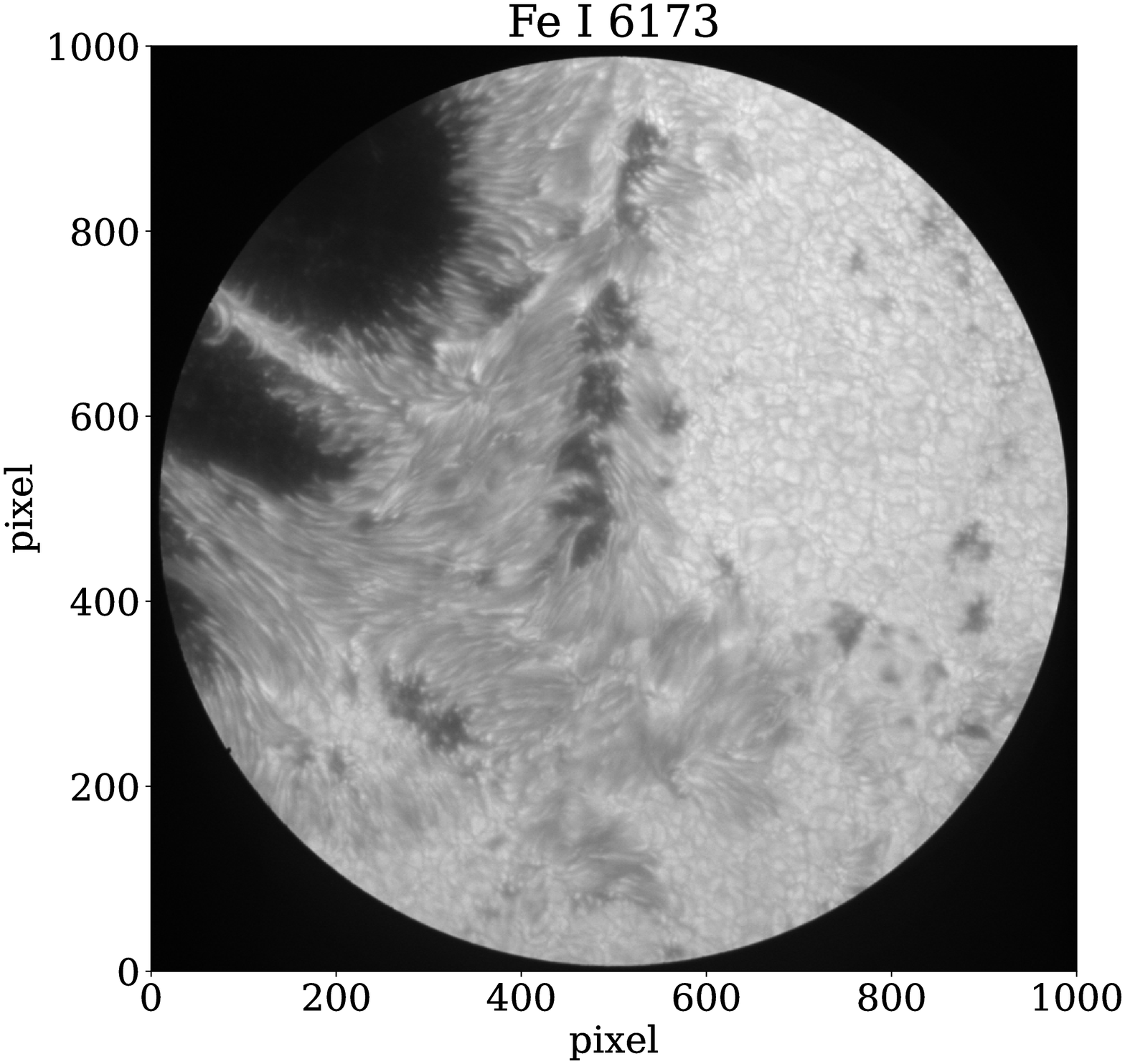}
\includegraphics[scale=0.25,trim=53 0 120 50,clip]{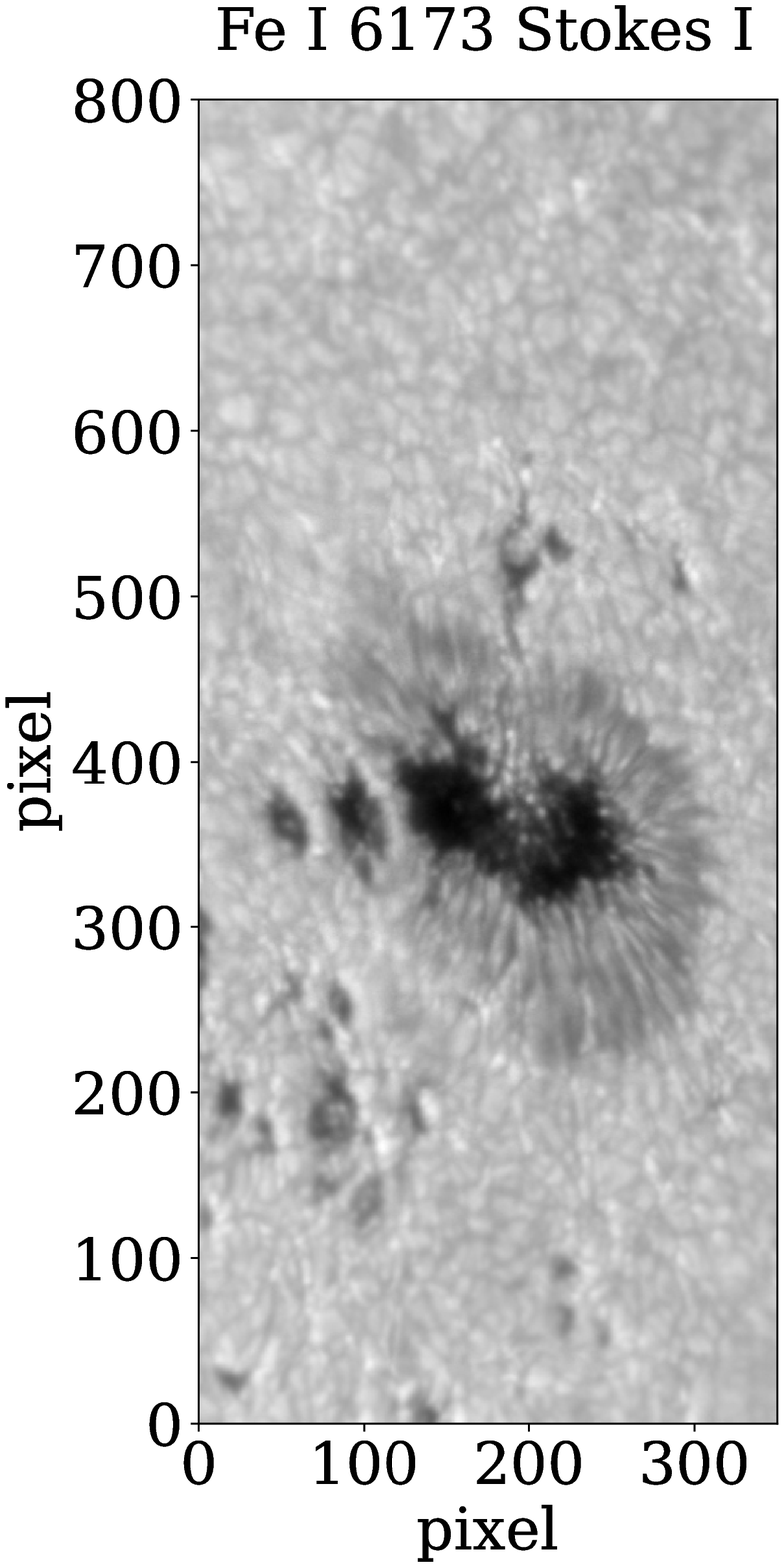}
\includegraphics[scale=0.25,trim=137 0 110 50,clip]{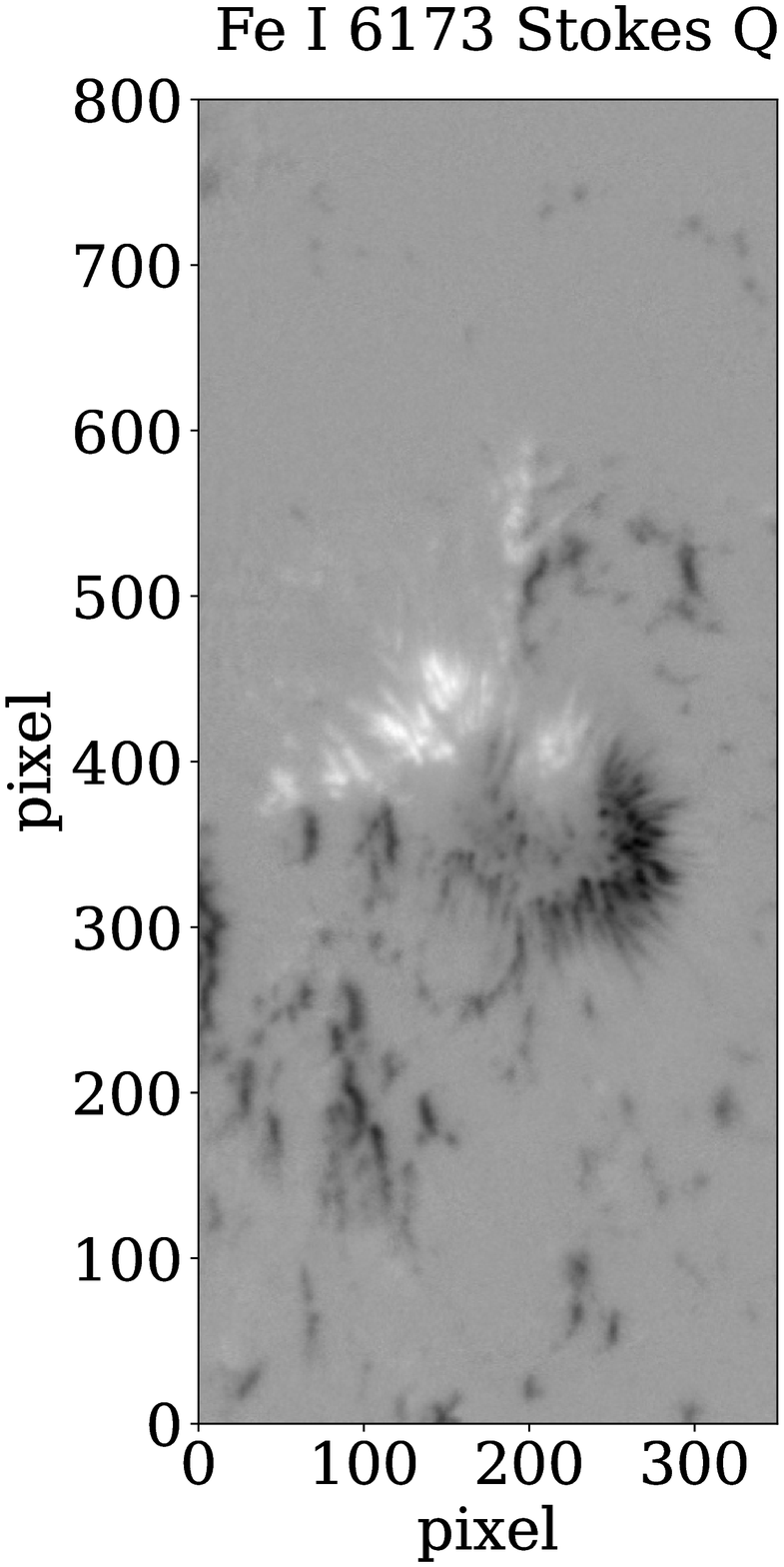}
\includegraphics[scale=0.25,trim=137 0 110 50,clip]{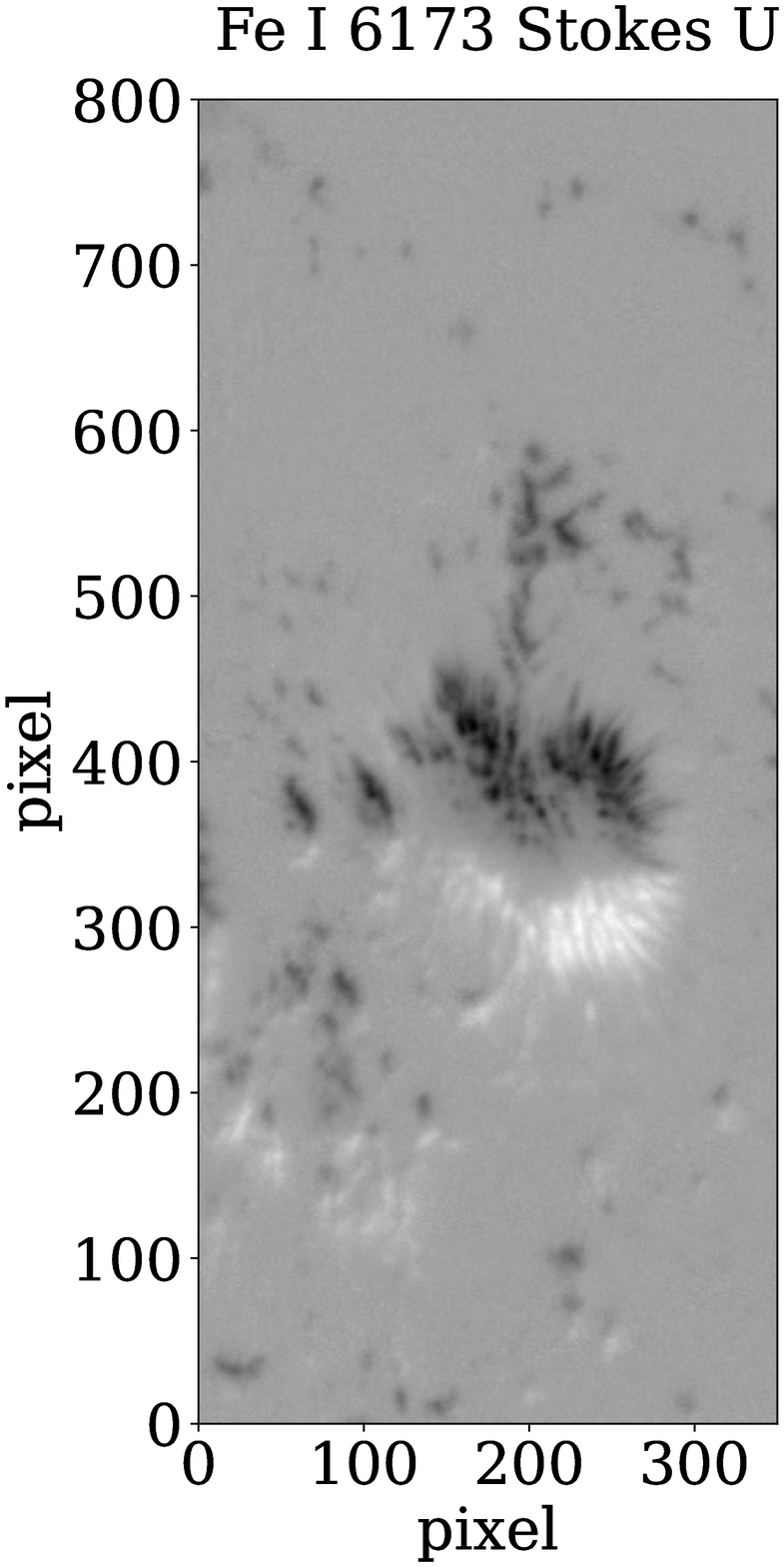}
\includegraphics[scale=0.25,trim=137 0 110 50,clip]{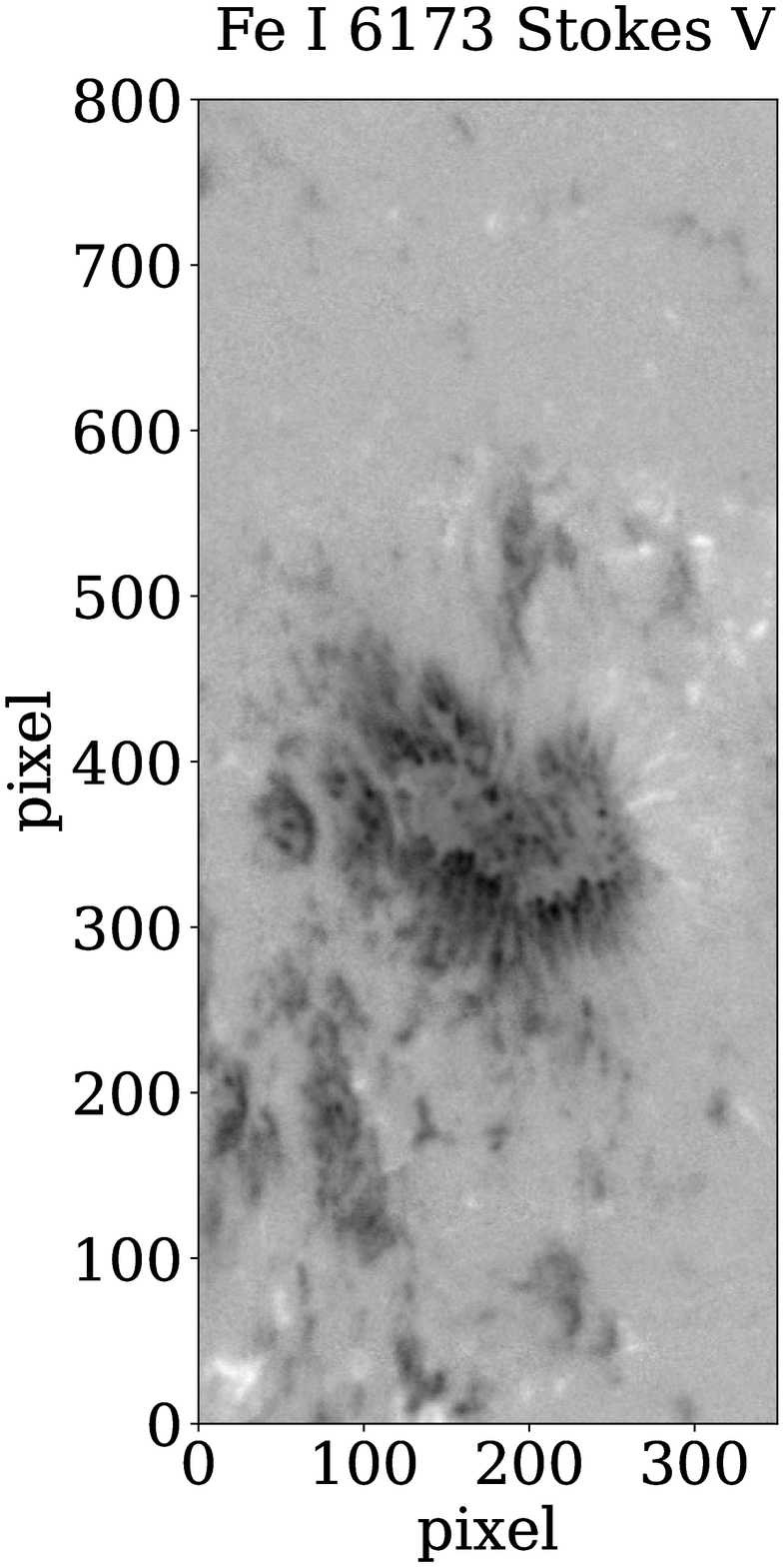}
}
\caption{From left to right: Examples of IBIS data acquired in standard spectroscopic  and spectro-polarimetric modes at the Fe I 6173 \AA~ line with circular (1000$\times$1000 pixels$^2$) and rectangular (350$\times$800 pixels$^2$) FOV on 25 October 2014, 15:32 UT and 13 May 2016, 13:38 UT at disc positions $\mathrm{\mu}$=0.85 and $\mathrm{\mu}$=0.68,  respectively. The spectroscopic and spectro-polarimetric Stokes I observations were taken in the line continuum, while the spectro-polarimetric Stokes Q and Stokes U and V data were acquired in the line core and line blue  wing, respectively. We show here the Level 1 data derived from the instrumental calibrations and MOMFBD restoration. See Sects. 2.2 and 3.1 for more details.}
\label{f1}
\end{figure*}

Figure \ref{f1} shows examples of IBIS data acquired in the standard  spectroscopic and spectro-polarimetric observational modes after instrumental calibrations and further processing described in Sect. 2.3. In particular, we show the spectroscopic observation at the photospheric Fe I 6173 \AA~ line of a large sunspot region with umbral and penumbral areas  imaged near the disc center, and the Stokes-I, -Q, -U, and -V measurements of a smaller sunspot region observed  closer to the limb. 

\begin{figure*}
{
\includegraphics[scale=0.45,trim=350 50 170 0,clip]{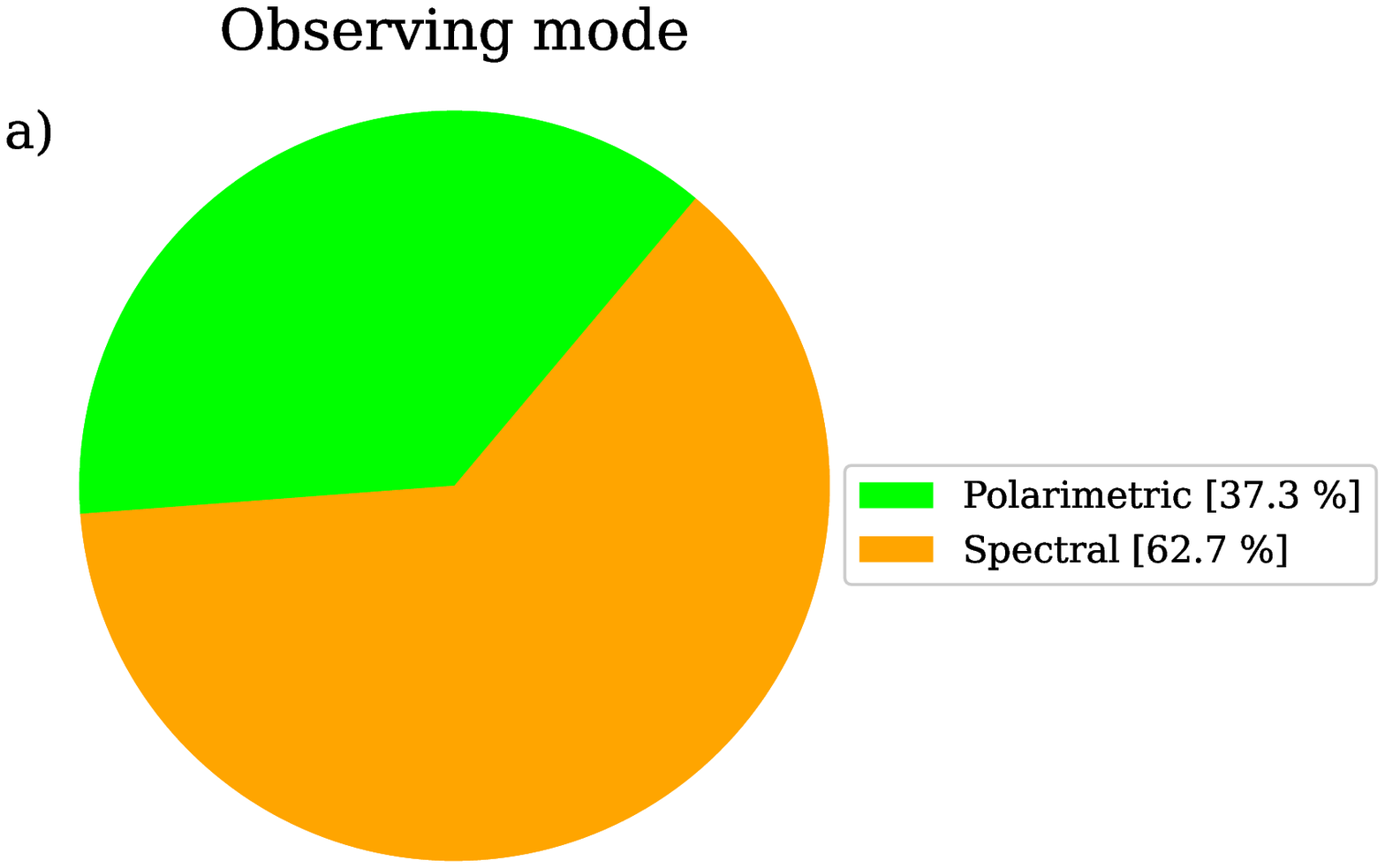}
\includegraphics[scale=0.45,trim=350 50 130 0,clip]{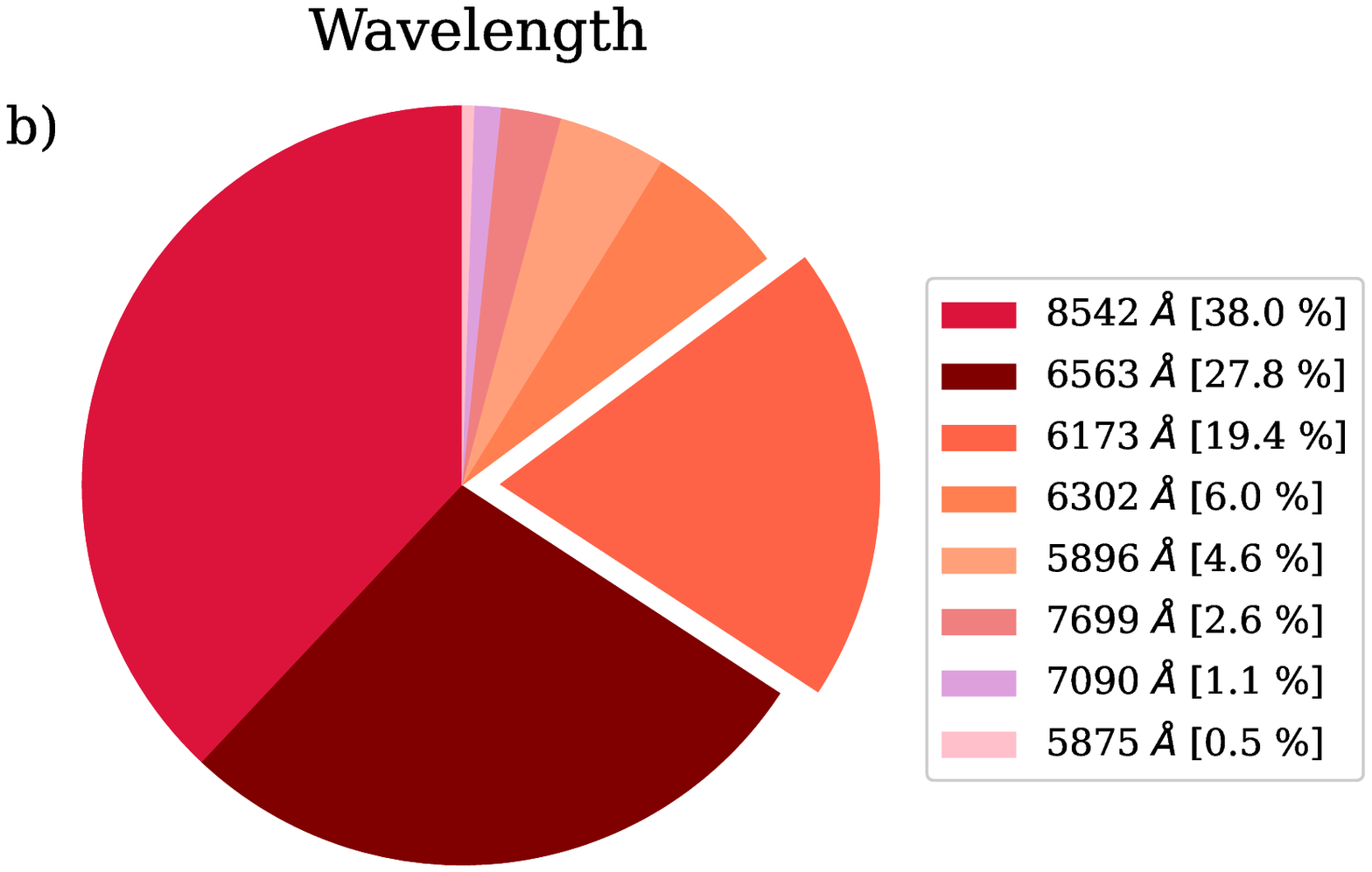}\\
\includegraphics[scale=0.45,trim=350 50 170 0,clip]{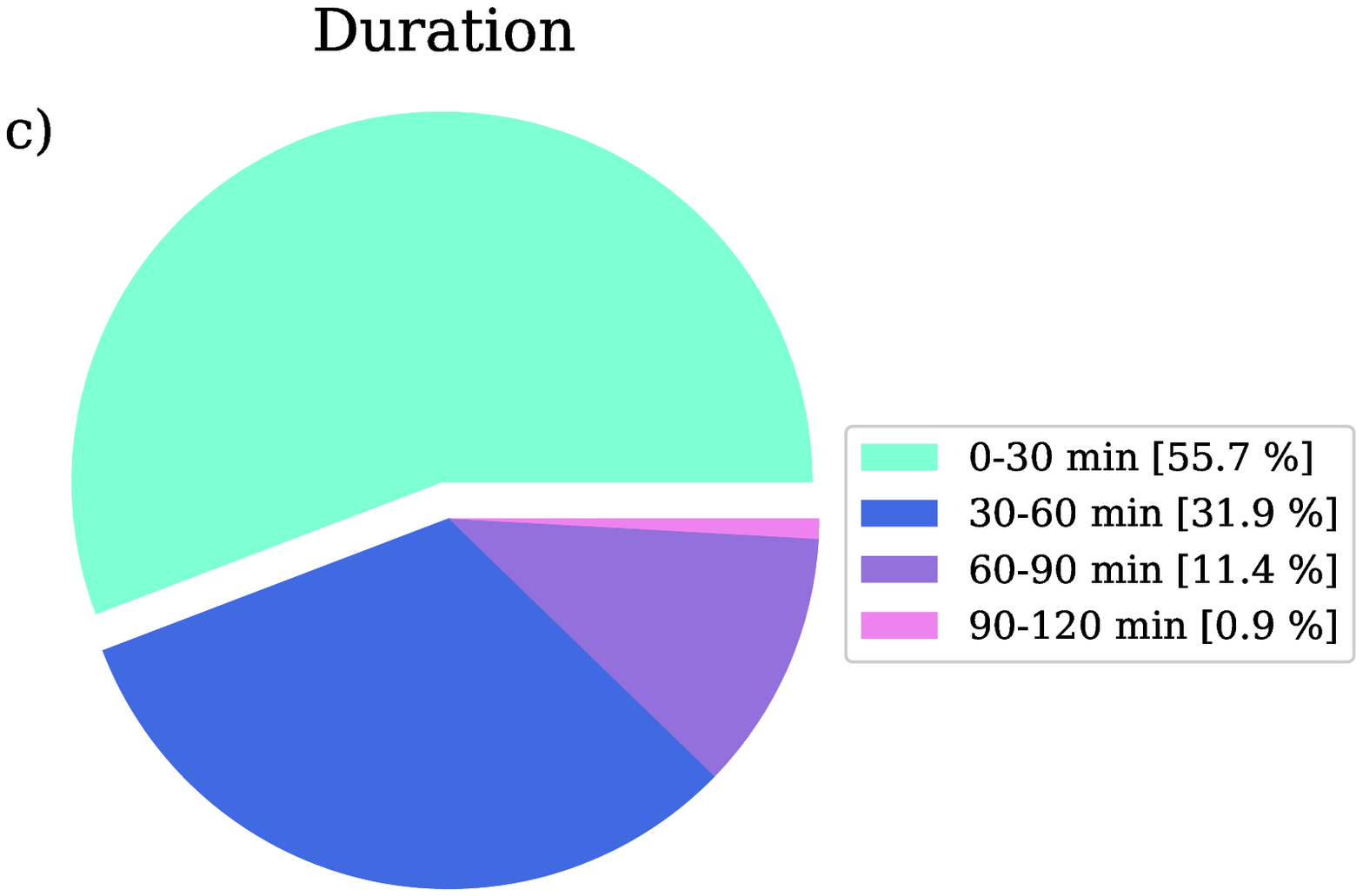}
\includegraphics[scale=0.45,trim=350 50 13 0,clip]{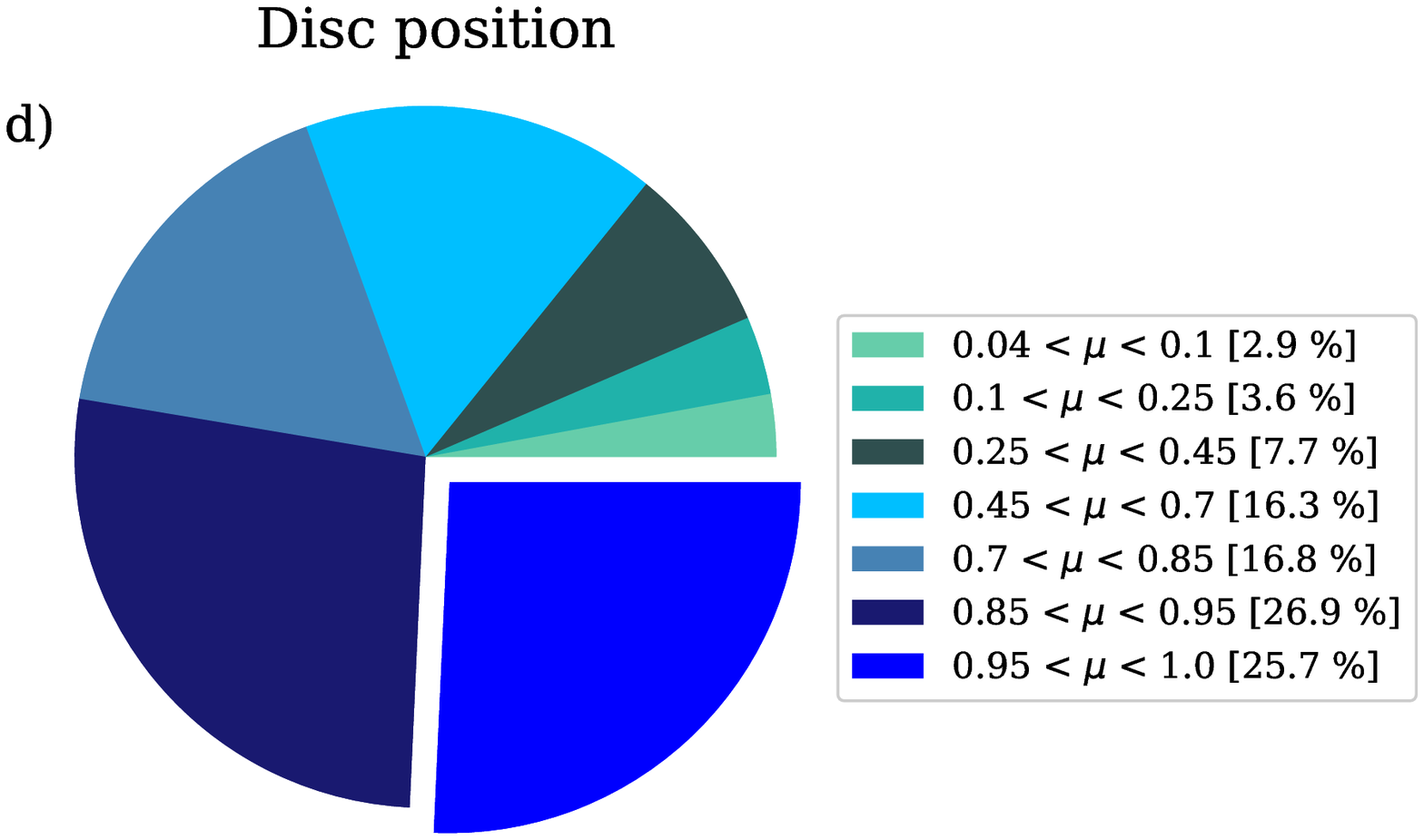}
}
\caption{Pie charts of the fraction (in \%) of : panel a) spectroscopic  and  spectro-polarimetric data available in IBIS-A, panel b) spectral lines (in \AA)  acquired during the observations, panel c) duration (in minutes) of the observational sequences, and panel d) $\mathrm{\mu}$ disc position of the FOV, for all the Level 0 data stored in the archive. $\mathrm{\mu=cos\theta}$ where $\mathrm{\theta}$ is the heliocentric angle.}
\label{f2}
\end{figure*}

\begin{table*}
\small
\center
\caption{\label{tab:generale} Overview of the Level 0 data available in IBIS-A.}
\begin{tabular}{lccccccccccccc}
\hline
\hline
Year    & N Series       & \multicolumn{4}{c}{ Target$\rm ^{(a)}$ [\%]}   & \multicolumn{8}{c}{N Data sets for each spectral line$\rm ^{(b)}$}  \\
        &                  & PO       & SP      & QS  & L            &   K I  & Fe I & Ca II &  H$\alpha$   & Fe I &  Fe I &   Na I D1 & He I                   \\
                &                  &        &       &   &             &   7699  & 6173  & 8542  &  6563  &    630.2  &   7090  & 5896  & 5875           \\
\hline
\hline  
2012    &     33            &  51.5 & 30.3 & 15.2 & 0       & 0 & 33 & 33 & 0 & 33 & 0 & 0 & 0    \\
2013    &     81            &  8.7  & 48.1 & 28.4 & 14.8  & 0 &	25 & 52 & 31& 0	 & 4 & 5 & 0 \\ 
2014    &     64            &  14.1 & 54.2 & 10.3 & 20.4  & 0 &	45 & 27	& 52& 0	 & 7 & 0 & 5   \\
2015    &     45            &  28.8 & 46.7 & 24.4 & 0       & 0 &	37 & 40	& 21& 16 & 0 & 0 & 0 \\
2016    &     26            &  19.2 & 46.2 & 34.6 & 0       & 0 &	17 & 26	& 17& 0	 & 0 & 1 & 0   \\
2017    &     20            &  40   & 35   & 25   & 0       & 0 &	16 & 20 & 10& 10 & 0 & 4 & 0  \\
2018    &     136           &  8.8  & 0.7  & 69.8 & 20.6  &25 &	14 &127	&104& 0  & 0 &27 & 0   \\
2019    &     47            &  32   & 0    & 63.4 & 4.3   & 0 &	1  & 46 & 37& 0  & 0 & 8 & 0   \\
\hline
\hline
\end{tabular}	
\begin{quotation}
\textbf{Notes.} 
$\rm ^{(a)}$ PO: pore, SP: sunspot, QS: quiet sun, L: Limb region. $\rm ^{(b)}$ Line core is given in \AA. 
\end{quotation}
\end{table*}

\begin{table*}
\small
\center
\caption{\label{tab:calibarted_data} Overview of the Level 1.5 and Level 2 data available in IBIS-A. }
\begin{tabular}{lclcccllcr}
\hline
\hline
Date            & Time       & Target$\rm ^{(a)}$ & Scans   & $\mu$ $\rm ^{(b)}$ & Cadence & Data$\rm ^{(c)}$ & Other data$\rm ^{(c)}$&Ref & Level$\rm ^{(d)}$ \\
YY-MM-DD        &    [UT]    &                    &           &                         & [s]     &                   processed & available &   &      \\
\hline
\hline  
2012-04-17      & 13:58:43 & PO            & 56          &   0.94      & 67     & Fe I 6173 &Fe I 6302, Ca II 8542 (I) & 1 &  2.0  \\
2012-04-17      & 15:09:06 & PO            &  68        &  0.94  &  67      &      Fe I 6173                                & Fe I 6302,  Ca II 8542 (I) &  &  2.0                     \\
2012-04-17      & 18:43:40 & PO         &  100       & 0.94              &    67    &      Fe I 6173                            & Fe I 6302, Ca II 8542 (I)             &    & 2.0                      \\
2013-10-07      & 17:39:08 &  SP        &  130       &     0.96   &   33       &   Fe I 6173 & Na I D1 5896            &    & 2.0   \\
2014-04-21      & 14:15:33 &  PO               & 15         &  0.98       &  116   & Fe I 6173 &  Ca II 8542           &    & 2.0\\
2014-10-14      & 15:15:57 &  PO               &  238       &  0.44       &  50    & Fe I 6173 & Ca II 8542, H$\alpha$ 6563 (I)     &    & 2.0\\
2015-04-10      & 14:34:25 &  SP               &  237       &  0.88       &  23    & Fe I 6173 & Ca II 8542              &    & 2.0\\
2015-04-11      & 14:14:40 &  PO               &  19        &  0.69       &  131   & Fe I 6173 & Ca II 8542    &    & 2.0\\
 2015-04-11                & 15:01:55 &           PO        & 18         &  0.88       & 131    &      Fe I 6173 & Ca II 8542                                  &  & 2.0\\
2015-04-30      & 13:52:44 &  QS               & 290        &  1.00       &    27  & Fe I 6173 & Ca II 8542 (I)            &    & 2.0\\
2015-05-01      & 14:17:52 &  QS               & 58         &  1.00       &   48   & Fe I 6173, Ca II 8542 &               & 2  & 1.5, 2.0 \\
2015-05-02      & 14:25:34 &  QS               & 50         &  1.00       &   44   & Fe I 6173, Ca II 8542        &        &    & 1.5, 2.0\\
2015-05-02                 & 15:08:33 &    QS               & 20         &  0.99       &    44    &    Fe I 6173, Ca II 8542   &                               &    & 1.5, 2.0                      \\
2015-05-02                 & 15:27:31 &   QS                & 79         &  0.98       &   44     &    Fe I 6173, Ca II 8542     &                             &    & 1.5, 2.0                     \\ 
2015-05-03      & 14:28:42 &  SP               &  85        &  0.74       &   44   & Fe I 6173, Ca II 8542           &     &    & 1.5, 2.0 \\
 2015-05-03                & 17:15:05 &  SP                 &  25        &  0.75       &   44    &        Fe I 6173, Ca II 8542    &                          &    & 1.5, 2.0                     \\ 
2016-05-13      & 13:38:48 &  SP               &  50        &  0.67       &   42   & Fe I 6173, Ca II 8542        &        &    & 1.5, 2.0\\
2016-05-13                 & 14:15:05 &  SP               &   58       &  0.67       &    42    &        Fe I 6173, Ca II 8542     &                         &    & 1.5, 2.0                      \\
2016-05-13                 & 15:00:27 &  PO               &   163      &  0.88       &   42     &       Fe I 6173, Ca II 8542     &                          &    & 1.5, 2.0                      \\
2016-05-19      & 13:37:43 &  SP               &  100       &  0.98       &    42  & Fe I 6173, Ca II 8542            &    &    & 1.5, 2.0\\
2016-05-20      & 13:53:06 &  SP               &   319      &  0.99       &    48  & Fe I 6173, Ca II 8542          &      & 3, 4, 5 & 1.5, 2.0 \\
2016-10-11      & 16:04:17 &  QS               & 29         &     1.00        &    49    &Fe I 6173& Ca II 8542, H$\alpha$ 6563 (I)   &    & 2.0\\
2016-10-11                 & 16:33:07 &  QS               & 31         &  1.00           &   49       &Fe I 6173 & Ca II 8542, H$\alpha$ 6563 (I)     &    & 2.0\\
2016-10-12      & 14:53:30 &  QS               & 100         &    1.00          &  49       &Fe I 6173& Ca II 8542, H$\alpha$ 6563 (I)    &    & 2.0\\
2017-05-05      & 14:06:00 &  PO               &   85       &  0.82       &    40  & Fe I 6173, Ca II 8542     &           &    & 1.5, 2.0\\
2017-05-05                 & 15:37:49 & QS                &  107       &  0.81       &   40     &    Fe I 6173, Ca II 8542                               &   &    & 1.5, 2.0  \\
2017-05-12      & 13:34:16 & QS                &   40       &  1.00       &    40  & Fe I 6173, Ca II 8542        &        &    & 1.5, 2.0\\

\hline\end{tabular}	
\begin{quotation}
\textbf{Notes.} 
 $\rm ^{(a)}$ Target: PO: pore, SP: sunspot, QS: quiet sun. $\rm ^{(b)}$ $\mu$ : $\mu=cos\theta$, where $\theta$ is the heliocentric  angle. $\rm ^{(c)}$ Spectral lines observed with IBIS. Line core is given in \AA. (I) specifies  spectroscopic data. $\rm ^{(d)}$ Level: 1.5 and 2.0 science-ready data described in the main text.  1: \citet{ermolli2017}, 2: \citet{murabito2020AA}, 3: \citet{stangalini2018}, 4: \citet{murabito2019}, 5: \citet{houston2020}. 
\end{quotation}
\end{table*}

\subsection{Data processing}
IBIS-A  includes the camera data   
as originally generated by instrument as well as processed data. 
The latter were obtained with a reduction package  built on the IBIS calibration pipeline \citep{criscuoli2014}
released by NSO\footnote{https://nso.edu/telescopes/dunn-solar-telescope/dst-pipelines/} and with other codes. The NSO pipeline includes calibration for the dark, flat-field response, and gain of the detectors, computation of the detector image scale and orientation, compensation for the artificial polarization introduced
by the instrumentation and by the telescope, calibrations for the systematic wavelength shift across the FOV due to the collimated mounting of the FPs, and for residual effects of the seeing. 
Besides, the data available in IBIS-A were calibrated for pre-filters transmission curves and residual polarization cross-talk between Stokes-I and Stokes-Q, -U, -V data.    
Moreover, the BB data were processed for image restoration by using the outcomes  
of the Multi-Object Multi-Frame Blind Deconvolution \citep[MOMFBD,][]{Loefdahl2012,vanNoort2005},  
in order to enhance the image quality over the full FOV. The package employed to process the data is written in IDL\footnote{Interactive Data Language, Harris Geospatial Solutions, Inc.}, and it is run under a Linux
Ubuntu distribution. The codes are run in a semi-automatic mode, which requires  inputs from the user at two steps, in order to select the FOV for alignment of NB and BB images and to compute the residual polarization cross-talk.  
At present the MOMFBD restoration, which is written in C++, is not integrated in
the data reduction package.

The IBIS measurements processed for the above steps form the Level 1 data in IBIS-A. At present, there are 133 calibrated  series out of the 454 series  of raw IBIS observations stored in the archive.  

A subset of the Level 1 observations  was further processed 
to generate Level 1.5 data. This subset consists of the calibrated spectro-polarimetric observations  
acquired at the photospheric Fe I 6173 \AA~ and chromospheric Ca II 8542 \AA~ lines under homogeneous  instrumental setup (i.e. same observing mode and spectral sampling).

The Level 1.5 data  comprise maps of the mean circular polarization (CP), mean  linear polarization (LP), and net circular polarization (NCP), and of the line-of-sight ($\mathrm{los}$) velocity field ($\mathrm{V_{los}}$). The mean CP signal was estimated by using two methods: 1) by computing the maximum amplitude of the Stokes-V spectral profile as reported in \citet{stangalini2021RSPTA} with the formula:
\begin{equation}
\mathrm{CP1 = \frac{| V_{max} |}{I_{cont}} \cdot sign ~(V_{max}) }
\end{equation} 
where $\mathrm{V_{max}}$ is the maximum amplitude of the Stokes-V parameter in the measured profile, and $\mathrm{I_{cont}}$ the local
continuum intensity, and 2) by considering all the values of the Stokes-V profile as set out by \citet{Martinez2011}.  The latter method was adapted to 
account for the actual  number $\mathrm{n}$ of spectral points sampled during the observations.    In particular, we applied the formula:
\begin{equation}
\mathrm{CP2 = \frac{1}{n ~\langle I_c \rangle}\sum _{i=1}^{n}\epsilon_i ~| V_i | }
\end{equation}
where $\mathrm{\epsilon}$ = 1 for the spectral positions of the line sampling on the blue wing of the line, $\mathrm{\epsilon}$ = -1 for the spectral positions on the red wing, and $\mathrm{\epsilon}$ = 0 for the line center position, $\mathrm{n=21}$ is the number of sampled spectral positions, and $i$ runs from the $\mathrm{1st}$ to the $21st$ wavelength position. $\mathrm{\langle I_c \rangle}$ and $\mathrm{V_i}$ are the values of the average line continuum intensity in a quiet Sun region within the FoV and of the Stokes-$\mathrm{V}$ parameter at the wavelength position $\mathrm{i}$.

We also estimated the LP signal with the method set out by \citet{Martinez2011},  by using the formula: 
\begin{equation}
\mathrm{LP = \frac{1}{n ~ \langle I_c \rangle}\sum _{i=1}^{n} \sqrt{~Q^2_i ~+ ~U^2_i}}
\end{equation}
where $\mathrm{Q_i}$ and $\mathrm{U_i}$ are the values of the Stokes-$\mathrm{Q}$ and Stokes-$\mathrm{U}$ parameters at the wavelength position $\mathrm{i}$.

The NCP, which is a measure of the asymmetry of the Stokes-V polarization signal, was estimated from the integral of the measured Stokes-V profile  
following \citet{Solanki&Montavon1993}. In particular, we applied the formula:
\begin{equation}
\mathrm{ NCP = \int _{\lambda _1} ^{\lambda _{n}} d \lambda ~ V (\lambda)} 
\end{equation}
 where the $\mathrm{\lambda}$,  $\mathrm{\lambda _1}$, and $\mathrm{\lambda _n}$  represent the wavelength, and wavelengths at the $\mathrm{1st}$ and $\mathrm{21st}$ 
 spectral positions of the sampled line, respectively.   
We computed   $\mathrm{V_{los}}$ by using the phase cross-correlation of Fourier transform as set out in Sect. 2.4 of 
 \citet{Schlichenmaier2000A&A}.

The Level 1.5 data were computed by utilizing the Matplotlib \citep{Matplotlib}, Scipy  \citep{Scipy}, and Numpy \citep{Numpy} libraries written in Python language. At present, there are 14 different series of Level 1.5 data available in the archive.

Finally, the Level 1 series from spectro-polarimetric observations at the Fe I 6173 \AA~ line were further considered to generate Level 2  data. In particular, all the scans of the above series were processed with inversion techniques. The core of these techniques relies on inference of the dynamics, magnetic and thermodynamics properties of the imaged  atmosphere from the measured Stokes profiles of single or multiple spectral lines that form at different heights in the solar atmosphere. This is achieved by solving the radiative transfer equation through codes aimed at working the inverse problem \citep{deltoro2016}. The Level 2 data in IBIS-A were obtained with the Very Fast Inversion of the Stokes Vector code \citep[VFISV,][]{borrero2011}, version 4.0, which performs a Milne-Eddington inversion of  measured Stokes profiles. At present, 31 data series have been inverted and a total of 2902 scans processed, which let  more than 80\% of the calibrated Fe I 6173 \AA~ line observations stored in the archive to also be available as Level 2 data.

The FOV size of Level 1 data\footnote{Currently available only for spectro-polarimetric series.} is 500$\times$1000 pixels$^2$, while the size of both the science-ready Level 1.5 and 2.0 data cover 350$\times$800 pixels$^2$, or $\approx$ 34$\times$78 arcsec$^2$.

\section{IBIS-A data}

The IBIS observing campaigns varied  depending on instrumental setup, observer, target, and science goals, thus the acquired data are very heterogeneous.  
However, almost all the IBIS data sets include the sampling of a photospheric line and a chromospheric line, either Ca II 8542 \AA~or H${\alpha}$ 6563 \AA, and sometimes of both these latter lines. From 2013 to 2017   
the chromospheric Ca II 8542 \AA~ observations were frequently carried out in spectro-polarimetric mode.

Figure \ref{f2} summarizes the main characteristics of the IBIS data available in IBIS-A. Figure \ref{f2} panels a) and b) show the fraction of spectroscopic and spectro-polarimetric series, and the spectral lines sampled in the observations, respectively. The spectro-polarimetric data are
37\% of the available measurements (in terms of number of sets). Over 19\%, 6\%, 4\% of the data refer to the photosphere sampled at the Fe I 6173 \AA~and 6302 \AA, and Na I at 5896 \AA~lines, respectively; 38\% and 27\% of the data report observations of the chromosphere at the Ca II 8542  \AA~and H${\alpha}$ 6563 \AA~lines, respectively. Figure \ref{f2} panels c) and d) display the fraction of sequences in the archive depending on their duration and solar disc position of the target, respectively. Almost 56\% of the data consist of sequences of observations lasting less than 30 minutes, but more than 12\% of the available data report on long-duration observations exceeding 1 hour. Besides, it is worth noting that duration in this statistics refers to the length of individual series. However, sometimes multiple series were run with minimal delay between each other and could thus be considered as a ``continuous'' observing block. 
About 53\% of the observations were taken at disc center ($\mu$> 0.85, where $\mu=cos\theta$ and $\theta$ is the heliocentric angle), while over 40\% of the data image the solar atmosphere at intermediate disc locations (0.25<$\mu$<0.85) and about 4\% very close  to the limb (0.04<$\mu$<0.1) or above it.

\begin{figure}
\centering
{
\includegraphics[scale=0.45,trim=350 0 0 0,clip]{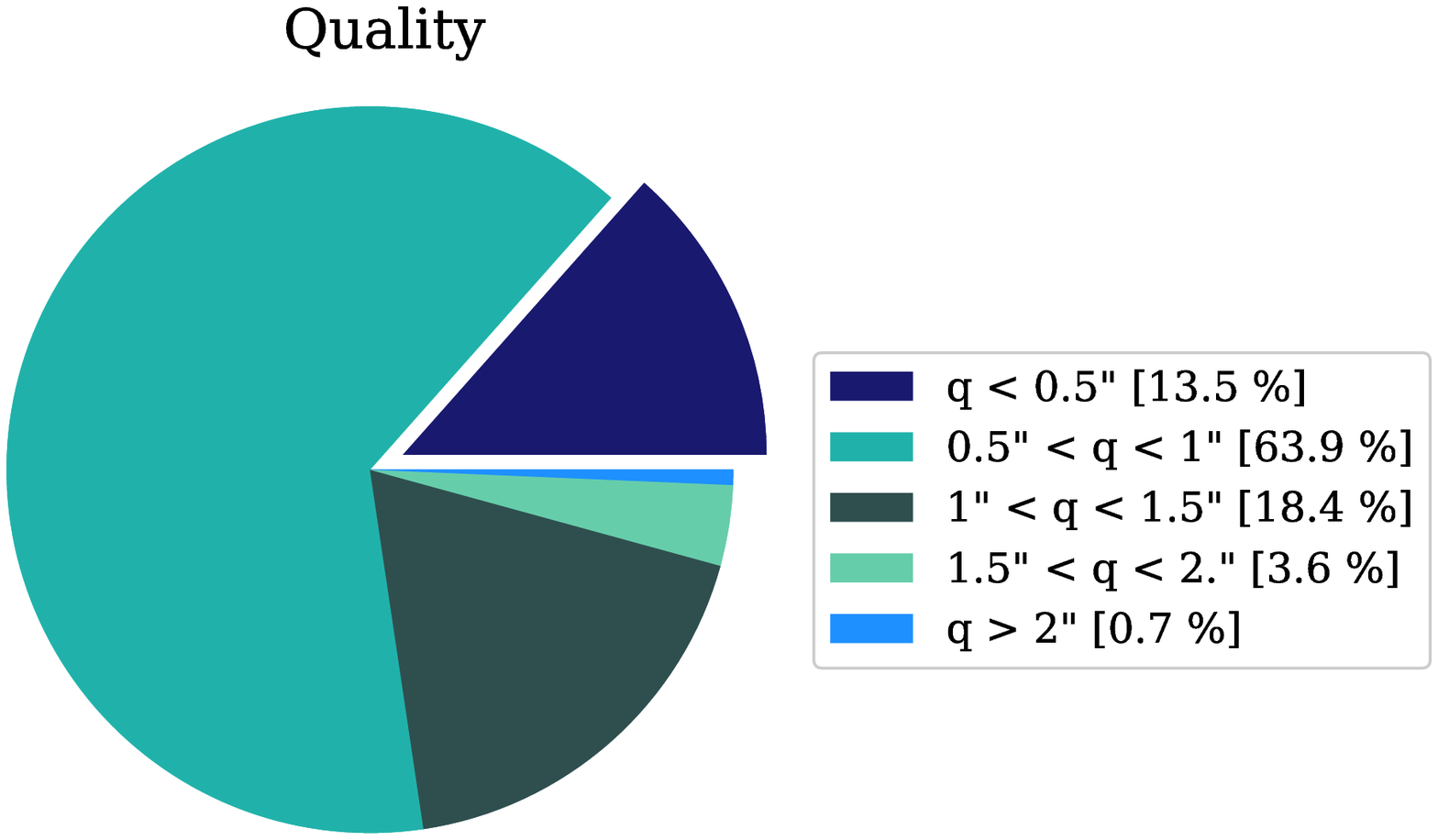}\\
\includegraphics[scale=0.45,trim=0 0 0 0,clip]{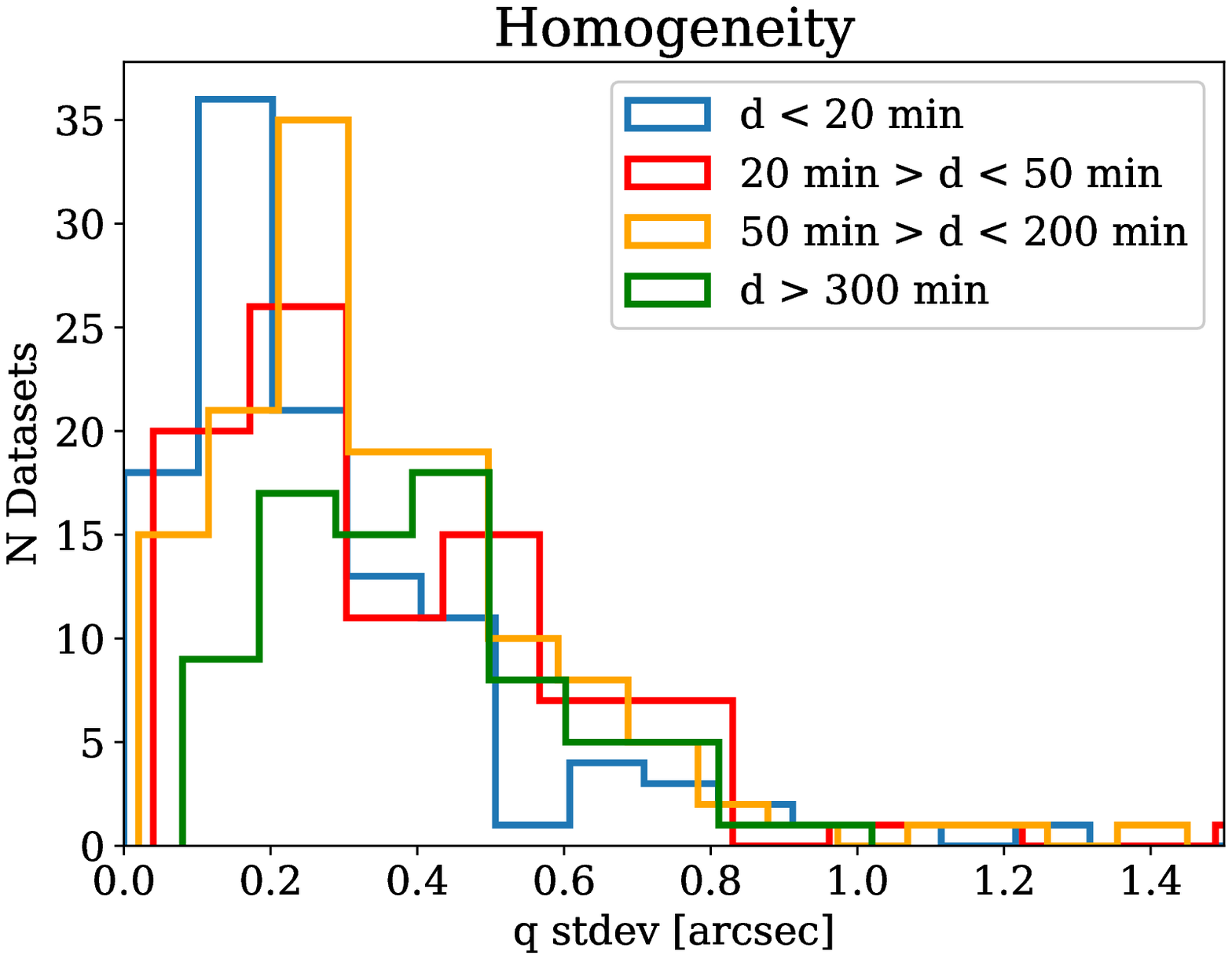}
}
\caption{Top panel: Pie chart of the IBIS-A data divided for the median value of the ``quality image'' parameter \textit{q} (in arcsec) during acquisition of Level 0 data. See Sect. 3.1 for more details. Bottom panel: Number of Level 0 data sets depending on their duration \textit{d} and homogeneity of the ``quality image'' parameter \textit{q} during observations. The homogeneity of the data set is described by the standard deviation of the  \textit{q}  values measured during the observing run.}
\label{f3}
\end{figure}

Figure \ref{f3} (top panel)  describes the IBIS-A data in terms of the  
median value of the ``quality image'' parameter \textit{q}. 
This quantity is a measure of the image resolution in arcsec due to the seeing. It is is generated by the DST control system from the monitoring of the variability of the Sun’s light intensity during the observing run; see  \citet{Seykora1993} for more details. Figure \ref{f3} shows that more than 77\% 
of the IBIS-A data were taken under very good and average seeing conditions, i.e. image resolution $<$ 1 arcsec, while about 4\% of the observations image the solar atmosphere at medium spatial resolution with image resolution $>$ 1.5 arcsec. About 18\% of the data are characterized by a resolution in the range 1-1.5 arcsec.

Figure \ref{f3} (bottom panel) provides information on the homogeneity of the data sets available in the archive. Here we show the number of data sets depending on the homogeneity of the ``image quality'' parameter \textit{q} during the observing run, for observations of given duration \textit{d}. The distribution of values in Fig. \ref{f3} (bottom panel) shows that a large fraction of the short duration data sets available in IBIS-A were taken under stable excellent and good seeing conditions, but this also holds for a significant number of observations exceeding one hour.

\subsection{Data format}

IBIS-A includes the data generated by the instrument, those processed for instrumental calibration and seeing restoration, and those further elaborated to derive physical quantities and science-ready data. These 
 diverse sets, which form the Level 0, Level 1, Level 1.5, and Level 2 data, respectively, are stored in the archive with different formats.

The Level 0 data ($\approx$ 25 TB of data) are archived as extended FITS format files in a main folder 
comprising two sub-directories. 
These include the NB and BB data, respectively, 
containing  data for both science and instrumental calibrations.
The
name of the files describes the nature of the data  (Science, Dark, Flat-Field, etc.). Each file comprises the data from a single  iteration of all pre-filters and wavelengths used during observations. General information about the data, which is common to the whole observing run, is stored in the primary FITS header, while information specific of a single exposure (acquisition time, wavelength position, exposure
time, telescope pointing, polarization state, etc) is stored in the extended FITS headers. Figures \ref{fhdr0} and \ref{fhdr02} in Appendix A give  examples of such information. Each FITS file consists of a four-dimensional array  with $\mathrm{(x, y, \lambda, Pol)}$-axes, where the spatial
$\mathrm{x}$ and $\mathrm{y}$ axes describe the instrument FOV, the $\mathrm{\lambda}$ axis covers the wavelength spectral points along the sampled line  and the  $\mathrm{Pol}$ axis represents the six polarimetric states (I+Q, I‐Q, I+V, I‐V, I+U, and I‐U)
measured by the instrument. 
The above structure and labelling of the Level 0 data follow the IBIS data definition and file arrangement given at the DST.

Table~\ref{tab:generale} gives an overview of the Level 0 data currently available in IBIS-A, with information on the number of data sets, the fraction of observations depending on the target and sampled spectral lines. The information is provided per year from 2012 to 2019. Table~\ref{tab:generale}  shows that there is a variety of observations at different lines, and targets available in IBIS-A, the latter including quiet Sun, sunspots,  pores, and limb regions. Some of the data also relate to plages and filaments.

The Level 1 data ($\approx$ 4 TB of data) are stored in the archive using the \textit{.sav} format generated by the IDL pipeline applied to the Level 0 observations.  The Level 1 files consist of four dimensional arrays with $\mathrm{(x, y, \lambda, Stokes)}$-axes, where the spatial $\mathrm{x}$ and $\mathrm{y}$ axes describe the instrument FOV, the $\mathrm{\lambda}$ axis renders the wavelength of the line sampling during the observation, and the $\mathrm{Stokes}$ axis represents the $\mathrm{(I, Q, U, V)}$ 
polarimetric states of the measurements.  Each   Level 1 file  also contains information about the data in the  \textit{info$\_$nb} variable, which is a structure with several tags and values. Figure \ref{fhdr1} in Appendix A lists the fields included in the \textit{info$\_$nb} variable. 
 The Level 1 data are stored with the naming convention \textit{llll-nbsss.pc.sav} and \textit{llll-nbsss.sav} for the spectro-polarimetric and purely spectroscopic data, respectively, where \textit{“llll”} is the wavelength in \AA~, and \textit{“sss”} is the scan number in the corresponding Level 0 data. Each \textit{.sav} file thus reports the data at the given time $t$ of the \textit{“sss”} scan during the sequence of observations.   
  The structure and naming of the Level 1 data follow the ones set in the NSO calibration pipeline.

Table~\ref{tablev1} in Appendix A summarizes key information on the Level 1 data currently available in IBIS-A, with details on targets, lines and number of spectral points sampled during observations, number of measurements at each spectral point and number of line scans. Table~\ref{tablev1}  makes it clear that most of these data currently include observations of the chromosphere at the Ca II 8542 \AA~ and H$\alpha$ 6563 \AA~ lines.
It is worth noting that the Level 1 data of IBIS-A can be accessed with the CRIsp SPectral EXplorer (CRISPEX) graphical user interface   \citep{vissers2012,lofdahl2021}, version 1.7.4, which was developed to manage the CRISP and CHROMIS data. To this purpose, the IBIS data format needs to be adapted to the FITS format accepted by the CRISPEX interface with a code available at the IBIS-A site specified in the following. The pipeline to reduce Level 0 data is also available  at the archive site. Other codes are available, e.g. to convert Level 1 data from \textit{.sav} to .FITS format.

Level 1.5 and Level 2 data ($\approx$ 15 GB and 20 GB of data, respectively) 
are stored in the archive in FITS format files\footnote{At present, FITS Standard Version 3 using CFITSIO Version 3.47.} as multi-dimensional arrays with $\mathrm{(x,y,var)}$-axes,  where $\mathrm{x}$ and $\mathrm{y}$ are the spatial coordinates in the FOV, and 
$\mathrm{var}$ indicates the physical quantities derived from the line measurements and VFISV data inversion, respectively. 
The physical quantities obtained from line measurements include estimates of the CP\footnote{From both the above-mentioned calculations applied.}, LP, NCP and $\mathrm{V_{los}}$, while those from the VFISV data inversion comprise  the magnetic field strength ($\mathrm{B}$), the inclination ($\mathrm{B_{inc}}$) and azimuth ($\mathrm{B_{azi}}$) angles of the magnetic field, the longitudinal component of the velocity field ($\mathrm{V_{los}}$), and the magnetic filling factor ($\mathrm{F_{fact}}$). Field azimuth ambiguity is unresolved. 

The Level 1.5 and Level 2 data comprehend
metadata information in the form of FITS header keywords that are stored in a primary header and two extended headers. The primary header contains
information derived from the Level 0 data to describe the observations with regard to
telescope pointing as Stonyhurst Heliographic solar coordinates (in degrees), solar $\mathrm{P0, L0, B0}$  ephemeris (in degrees), apparent solar photospheric disc  radius (in arcsec), Seykora scintillation monitor value (the ``quality image'' parameter given in arcsec), and duration of the line scan (in seconds). 
The information stored in the first and second extended headers concerns the physical quantities available in the data and spectral sampling of the observations employed to estimate them, respectively. Since Level 1.5 data are obtained from observations taken at the Fe I 6173 \AA~ and Ca II 8542 \AA~ lines, the latter extended header of Level 1.5 data includes information on both the sampled lines, while it only refers to Fe I 6173 \AA~ spectral measurements for Level 2 data.  It is worth noting that  
the  
metadata information delivered in the Level 1.5 and Level 2 data  
is compliant with the SOLARNET recommendations \citep{haugan2015,haugan2020}. 
In particular, a number of keywords employed in primary and extended headers  address the SOLARNET recommendations,  
e.g., the keywords  
specifying date and time of data series, wavelength, observing mode, exposure time, instrument and telescope employed, frame dimension, solar ephemeris, telescope pointing, data author,  data version, name of the procedures applied to the data, origin and level of the data.

Figures \ref{fhdr15}, \ref{fhdr151}, and  \ref{fhdr2} in Appendix A give examples of keywords in the FITS header of the Level 1.5 and Level 2  data. 
Level 1.5 and Level 2 files are stored in the archive with the naming  convention  $IBIS\_l\_YYYYMMDD\_HHMMSS$, where "$l$" shows the data level, $YYYYMMDD$ and $HHMMSS$ the date and time of the related observations. This file naming convention also addresses the SOLARNET recommendations.

Table \ref{tab:calibarted_data} summarizes the main characteristics of the Level 1.5 and Level 2 data available in IBIS-A. 
There is a variety of targets in science-ready data, including  pores, sunspots, and quiet Sun regions. Besides, a few series have further  co-temporal observations available than  the Fe I 6173 \AA~ and Ca II 8542 \AA~ ones considered to produce the Level 1.5 and Level 2 data, obtained at e.g. the photospheric Fe I 6302 \AA~ and chromospheric H$\alpha$ 6563 \AA~ lines.  Table~\ref{tablev2a} in Appendix A summarizes key information on the Level 2 data currently available in IBIS-A and further co-temporal spectral lines observed with IBIS.

\begin{figure*}
\centering
{
\includegraphics[scale=0.35,trim=0 130 30 100,clip]{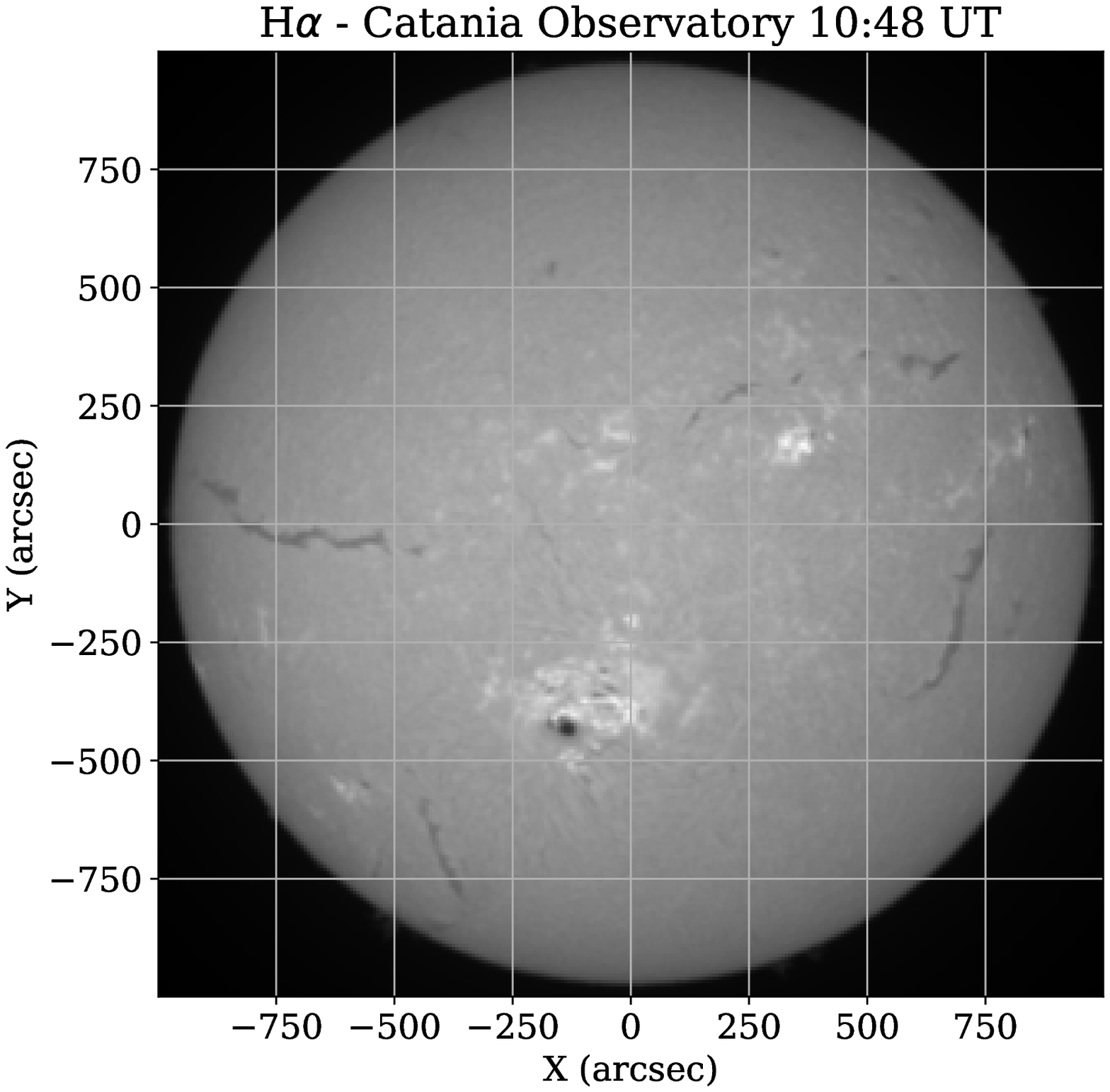} 
\includegraphics[scale=0.35,trim=0 130 30 100,clip]{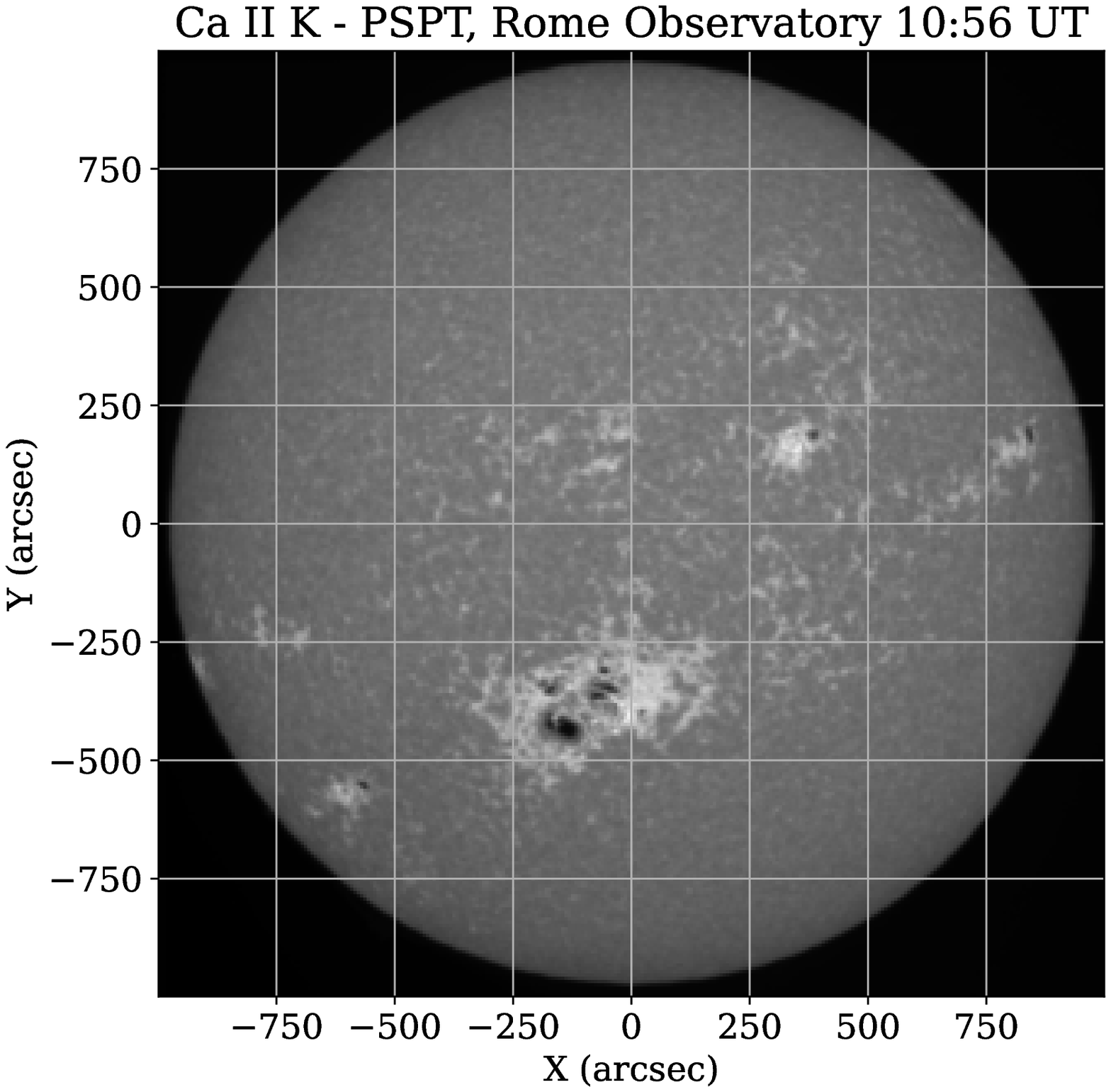} \\
\includegraphics[scale=0.4,trim=0 220 50 220,clip]{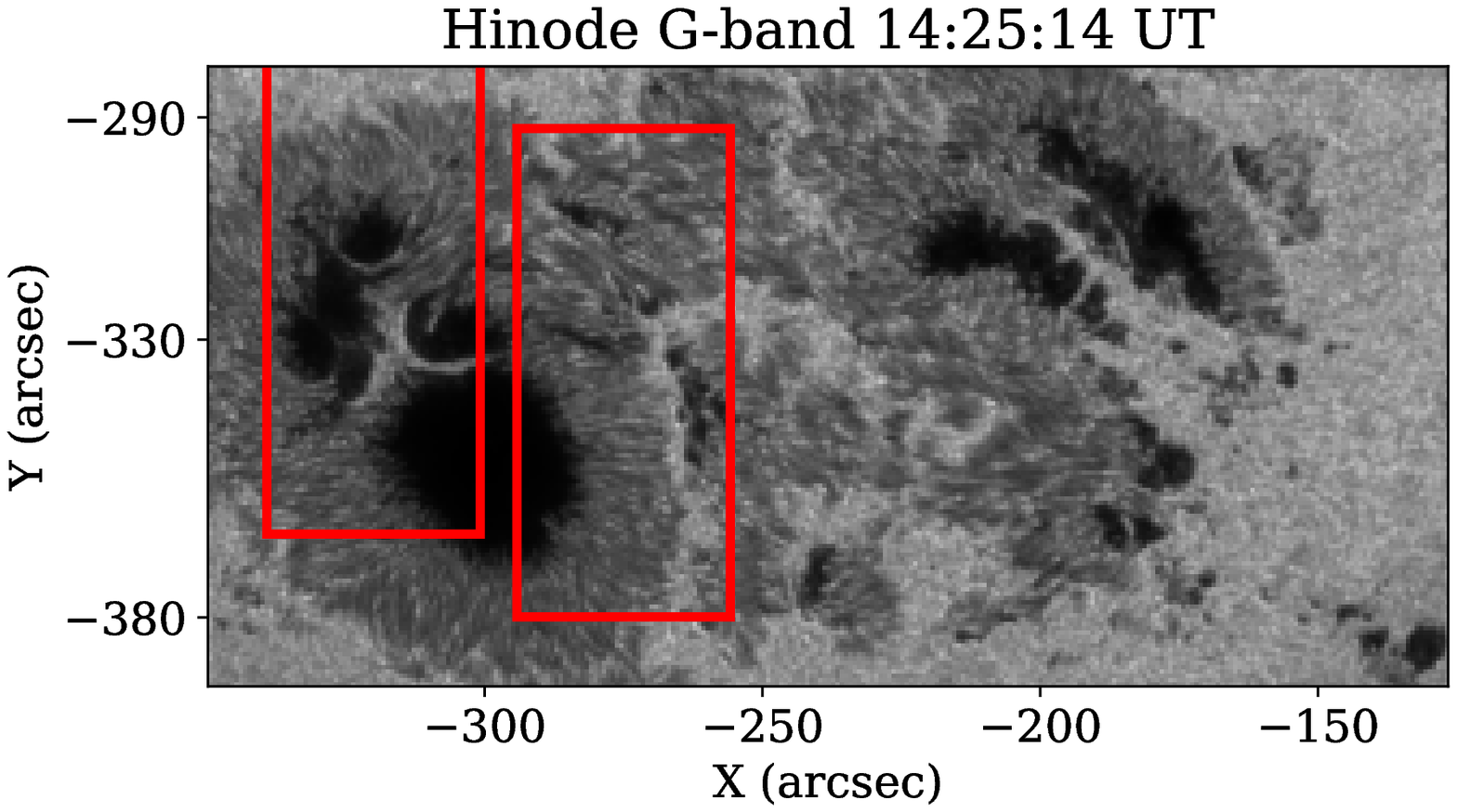} 
\includegraphics[scale=0.20,trim=0 0 50 0,clip]{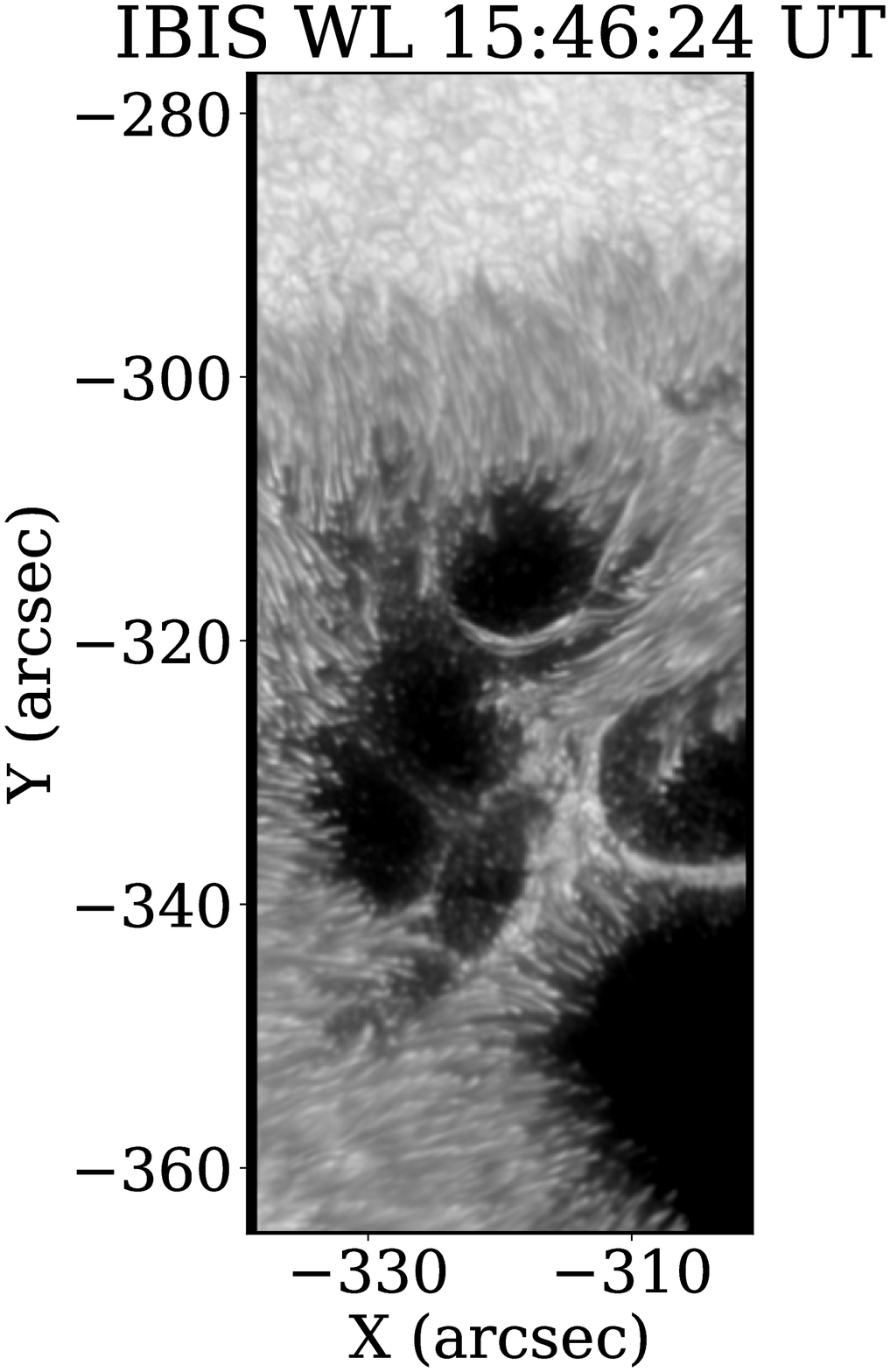}
\includegraphics[scale=0.20,trim=50 0 50 0,clip]{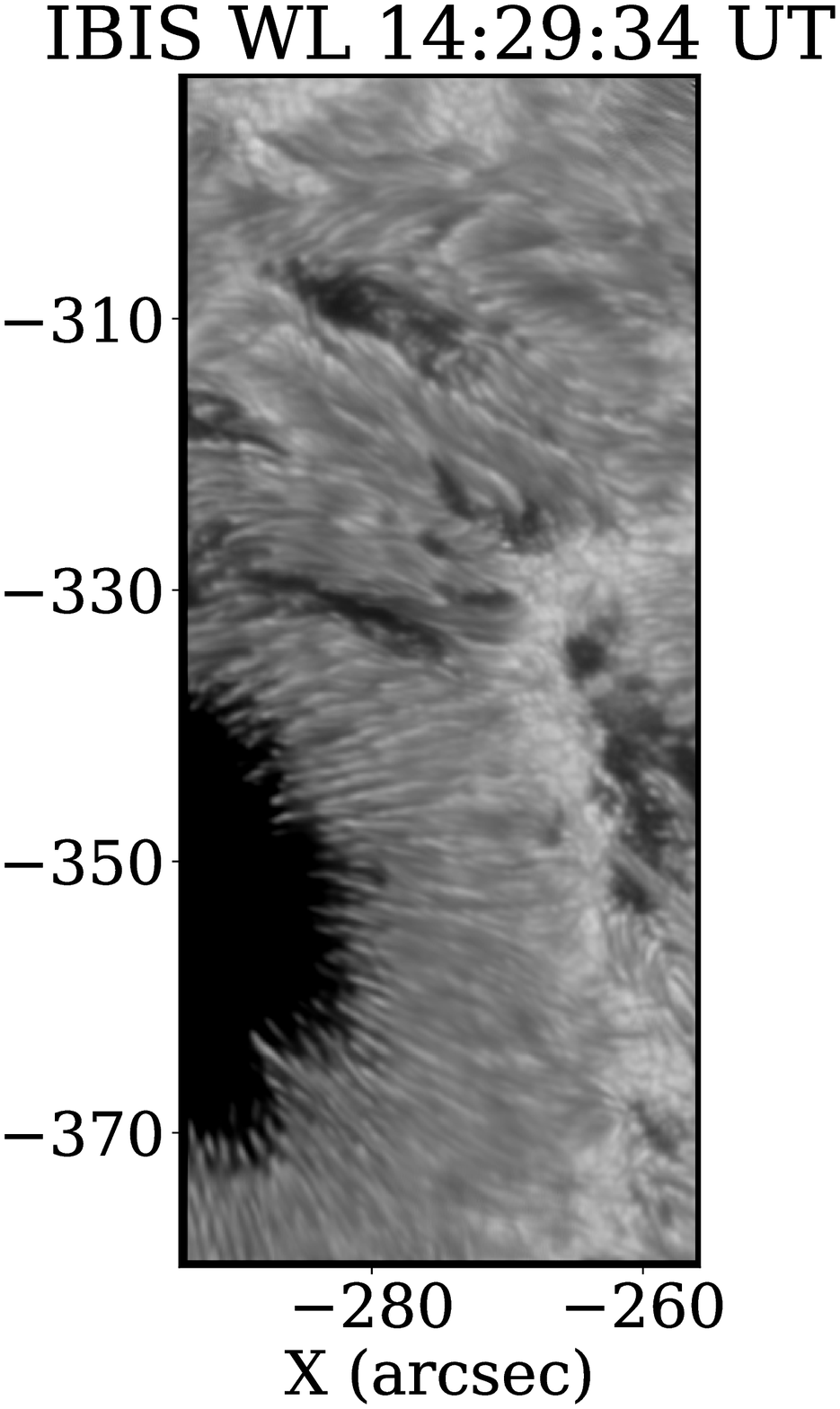}
}
\caption{Examples of contextual data available in IBIS-A. Full-disc observations of the solar chromosphere  acquired on 22 October 2014 at the  H$\alpha$ (top left panel) and Ca II K (top right panel) lines by the INAF Catania and Rome/PSPT telescopes, respectively, zoom of the Hinode/SOT observation (Level 1) 
of the central region of NOAA  AR 12192 (bottom left panel), also imaged by the IBIS data available in IBIS-A (bottom right panels).}
\label{f4}
\end{figure*}

\subsection{Contextual and coordinated observations}

The data stored at IBIS-A are characterized by the large volume of each target observation and the challenging  analysis that is typical of any spectro-polarimetric measurements.  
Nevertheless, the literature is rich in novel results derived from analysis of data either belonging to or similar to those in IBIS-A. Almost all these studies, as well as the ones listed in Sect. 1,  clearly show that combining spectro-polarimetric measurements such as those in IBIS-A with data from other instruments greatly enhances the return of the scientific analysis of the former observations, by expanding the FOV and range of atmospheric heights that can be investigated.  
In this light, the IBIS-A archive offers to its users links to observations complementary to the IBIS series available in the archive, such as those from co-temporal and near co-temporal high-resolution observations of the solar atmosphere available from the instruments onboard the Hinode/SOT \citep{Hinodesot} and IRIS \citep{iris} satellites, and full-disc filtergrams from INAF solar telescopes \citep{ermolli2011,romanobarra2021}.

Figure \ref{f4} shows examples of contextual data available in IBIS-A for the IBIS data set obtained on 22 October 2014 observing the active region (AR) NOAA 12192.  The contextual data include full-disc observations of the solar photosphere (not shown) and chromosphere at the  H$\alpha$ 6563 \AA~ and Ca II K 3933 \AA~ lines acquired by the INAF Catania and Rome/PSPT telescopes on the same day of the IBIS observations, and zoom of the central region of AR NOAA 12192 from Hinode/SOT observations almost co-temporal with the IBIS measurements stored in IBIS-A.
The latter pertain the inner part of the AR during the occurrence of a X1.6 flare.  
It is worth noting that at present  users are only provided with direct links to the  Hinode/SOT and IRIS data complementary to those available for each series in IBIS-A, without any processing to e.g. scale and target observations from the diverse instruments.


\section{Data access}

The IBIS-A data are stored on a RS24S3 SuperNAS server (96
TB, 64 GB Ram, 2 CPU Haswell 4C E5-2623V3 3G) at the
INAF Osservatorio Astronomico di Roma \footnote{The IBIS-A data can
be accessed at \url{http://ibis.oa-roma.inaf.it/IBISA/}.}.

The metadata of the Level 0 files were ingested into a
MySQL database (DB hereafter) by using a Python application.  
The Python Django framework has been used both to define the 
relational table schema using its object-relational internal 
mapping and to provide the web pages with simple query 
search interface.

The IBIS-A archive, so built, works on the server running 
Python 2.7.6, with Django version 1.9.1 on a MySQL 5.5 server.
It has recently been tested (to update the query service) 
on a test machine running Python 3.9.1, Django 3.2.9 and 
a MySQL 8.0 server.

The interface permits searching data according to
various criteria: solar target, disc position, observational mode,
data range; criteria can be combined in order to refine searches.
Based on the metadata stored on the IBIS-A DB, the data search
dynamically generates a HTML page with information about the
available data. 
The HTML tabular output is prepared using Django views for 
the search results and formatted in HTML using the internal
Django template tools.  
The products of the data search also include movies and plots
of the atmospheric seeing during observations, as basic information to allow users to assess the data quality. The movies for
quick look purposes were created by using the 2D plotting library Matplotlib (version 1.5.2).

Browsing IBIS-A does not require authentication, but users
have to register and login in order to be able to request and 
download IBIS-A data.

\begin{figure*}[t!]
{
\centering
\includegraphics[scale=0.65,trim=10 370 110 50,clip]{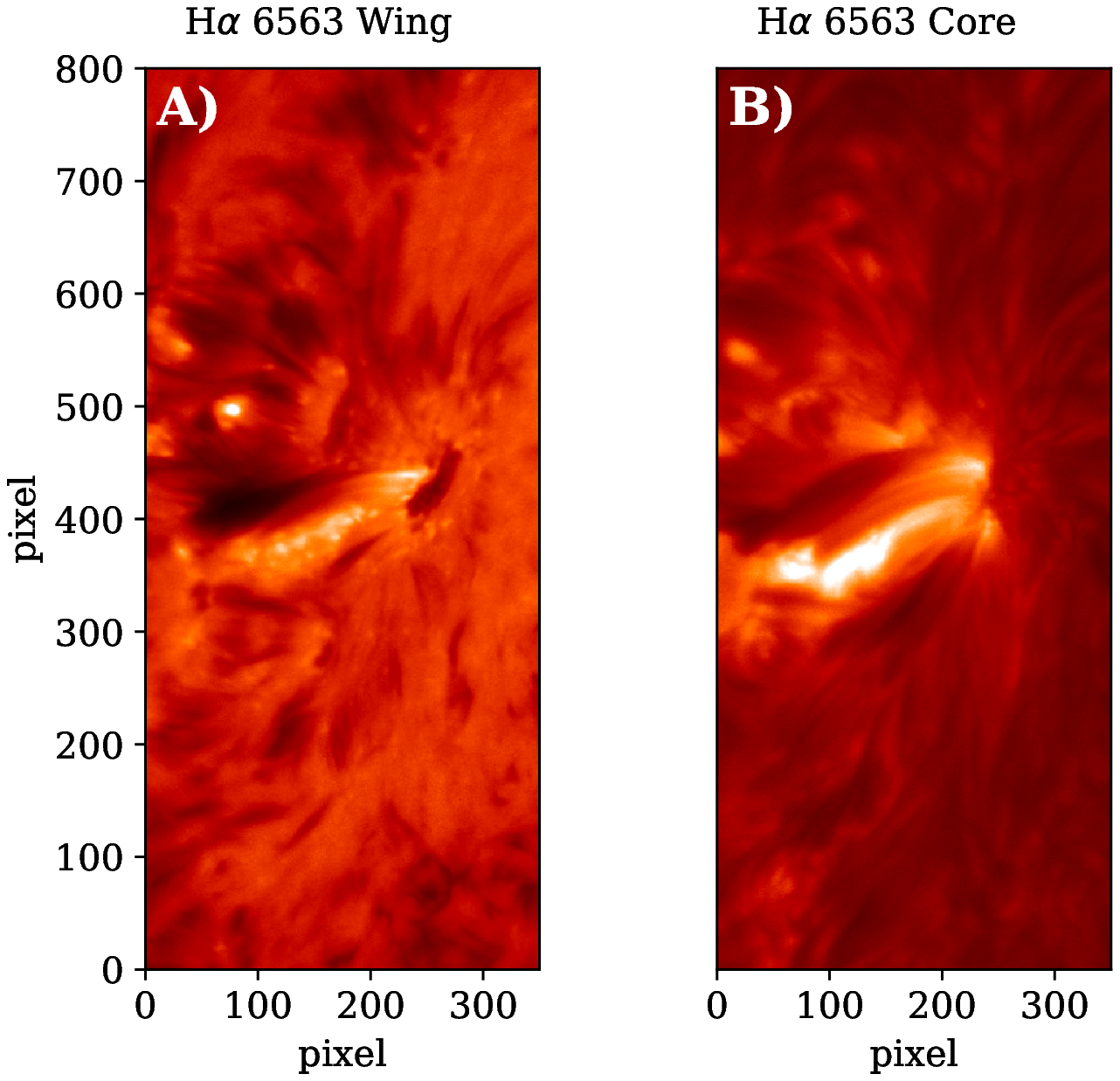}
\includegraphics[scale=0.65,trim=50 370 100 50,clip]{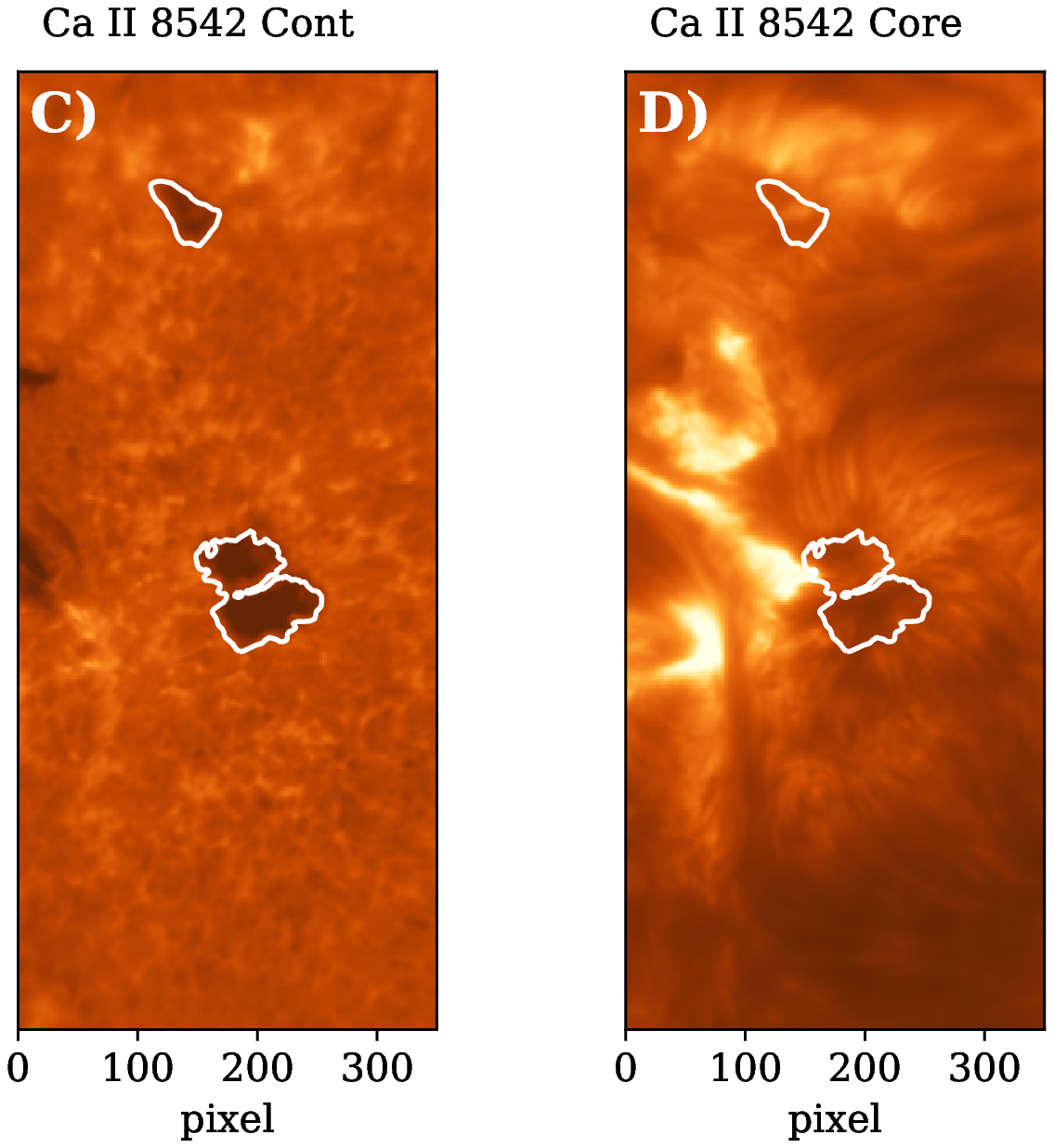}
}
\caption{Examples of Level 1 data of  chromospheric flaring regions available in  IBIS-A. The data were acquired  along the H$\alpha$ 6563 \AA 
~line, in the wing (panel A)  and core (panel B) of the line, on 10 October 2014, 14:20 UT  at disc position $\mu$=0.43, and along the  
Ca II 8542 \AA  ~line, in the far wing (panel C) and line core (panel D), on  13 May 2016, 15:00 UT, at disc position $\mu$=0.9. White lines  in right-most  panels show contours of magnetic pore regions corresponding to $I_{c}(x,y)=0.8\times I_{c}(quiet)$, where $I_{c}(x,y)$ and I$_{c} (quiet)$  are values of the line continuum intensity at the given position $(x,y)$ and in quiet region. 
}
\label{f5}
\end{figure*}

\begin{figure*}[t!]
{
\centering
\includegraphics[scale=0.56,trim=80 30 0 40,clip]{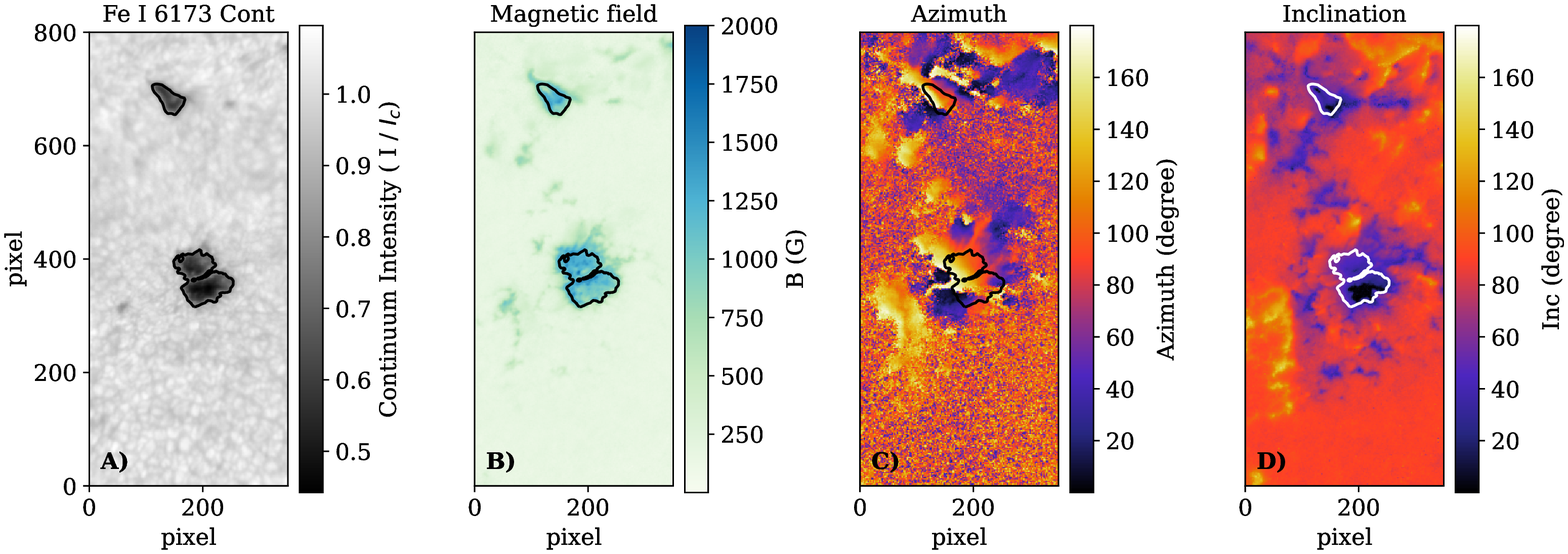}
\includegraphics[scale=0.56,trim=80 30 0 40,clip]{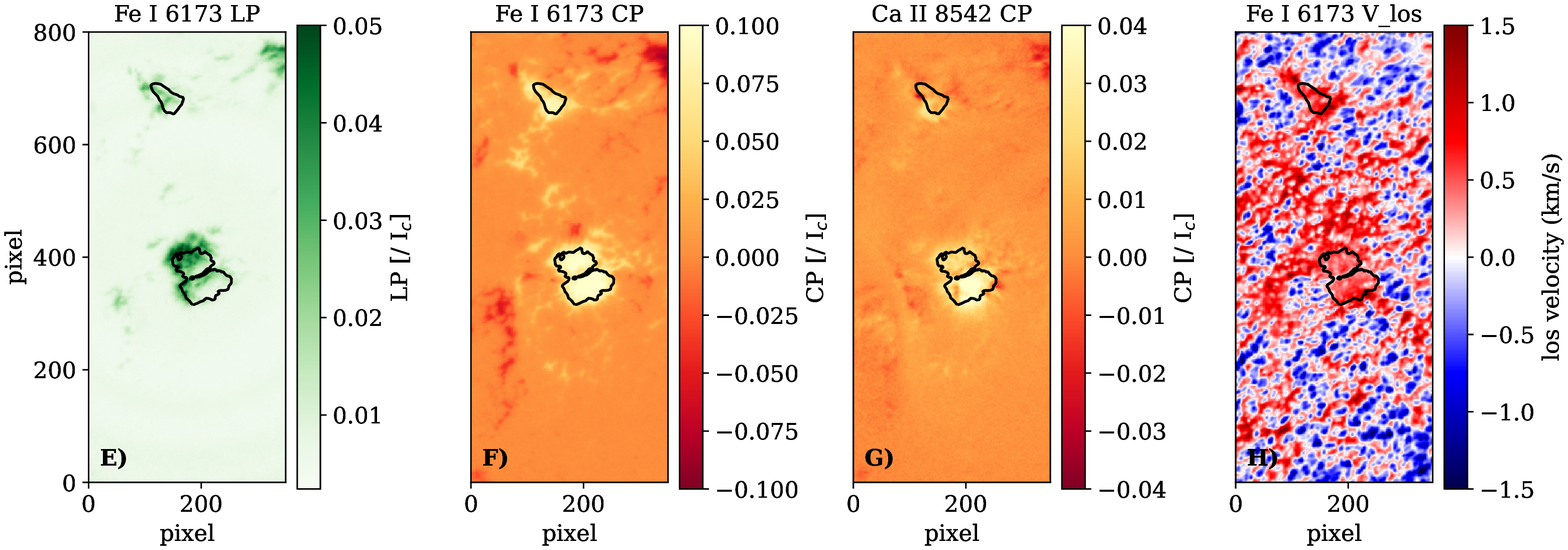}
}
\caption{Examples of Level 1, Level 1.5, and Level 2 data available in IBIS-A for the flaring region with magnetic pores shown in Fig. \ref{f5}. The observations were taken on 13 May 2016 at 15:00 UT.  
Top panels display Level 1 (panel A) and Level 2 (panels B-D) data, while bottom panels show Level 1.5 (panels E-H) data.
Top panels, from left to right: maps of the 
intensity measured on the FOV at the Fe I 6173 \AA~ line continuum and of the magnetic field strength,  azimuth and inclination angles derived from the VFISV inversion of the Fe I 6173 \AA~ line data. 
Bottom panels, from left to right:
maps of the linear and circular polarization estimated from the photospheric Fe I 6173 \AA~ line data, circular polarization from the chromospheric Ca II 8542 \AA~ observations, and $los$ velocity field at photopheric heights from Fe I 6173 \AA~ measurements. 
See caption of Fig. \ref{f5} for more details. }
\label{f6}
\end{figure*}

\begin{figure*}[t!]
{
\includegraphics[scale=0.56,trim=80 30 0 40,clip]{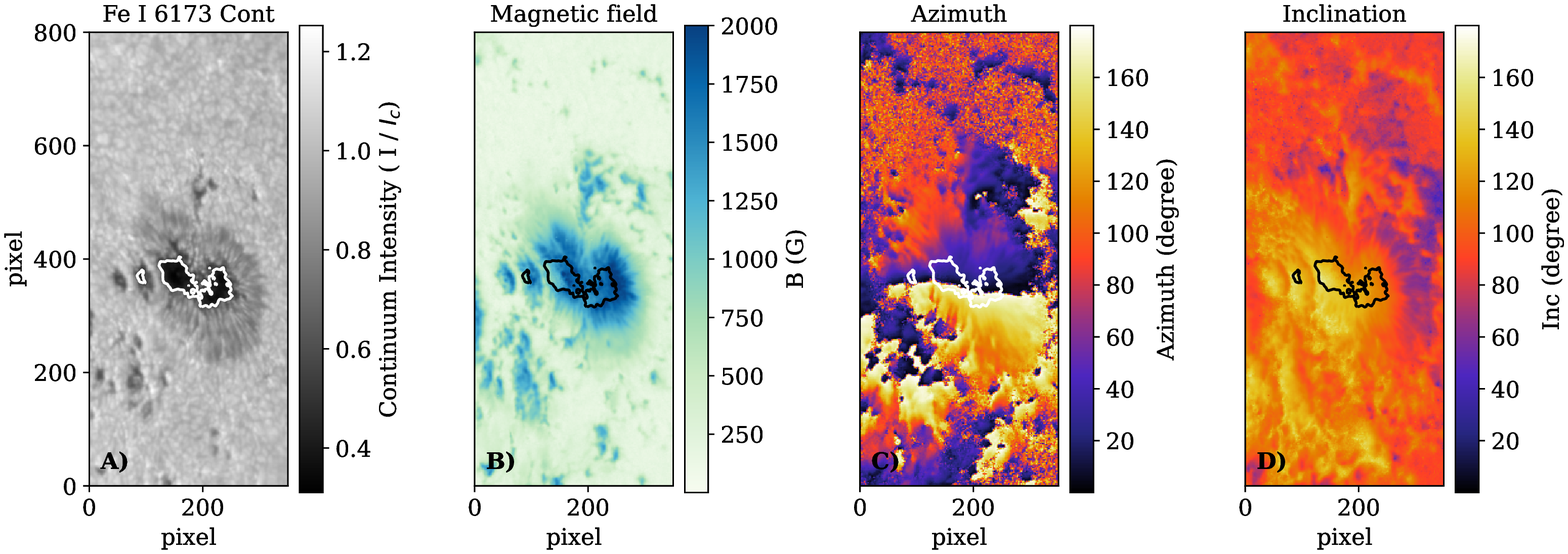} 
\includegraphics[scale=0.56,trim=80 30 0 40,clip]{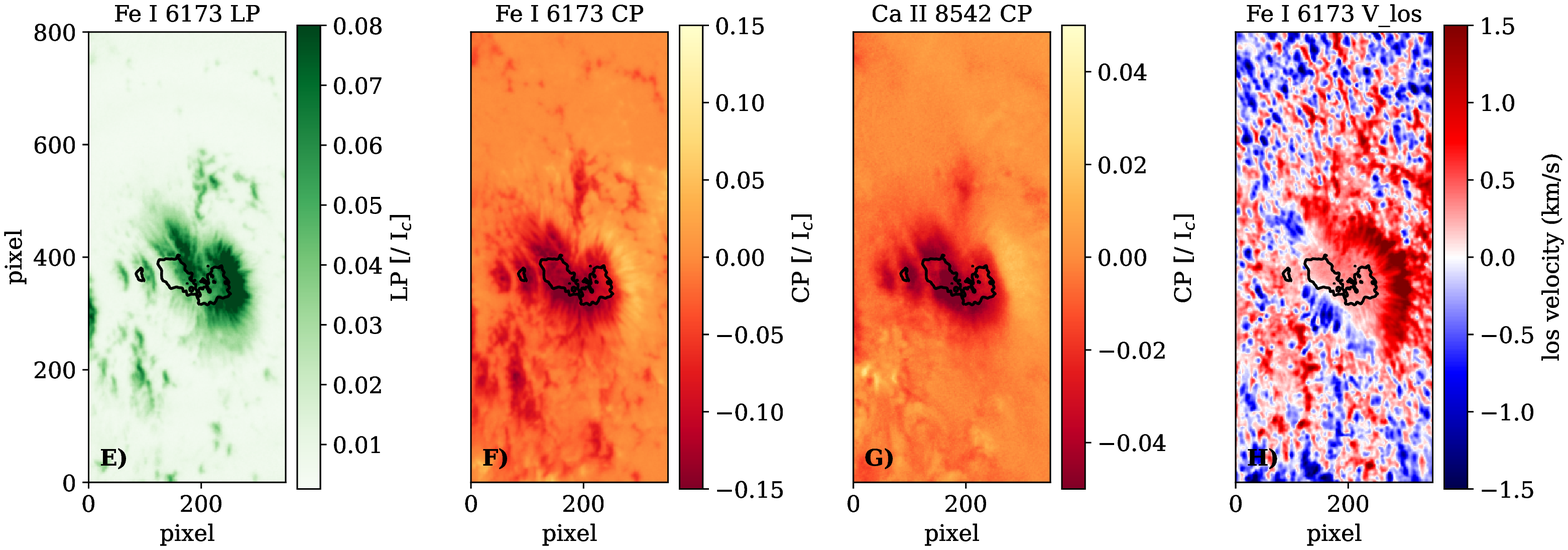}
}
\caption{Examples of Level 1, Level 1.5, and Level 2 data available in IBIS-A for the sunspot region observed on 13 May 2016, 13:38 UT at disc position $\mu$=0.68 shown in Fig. \ref{f1}. Top panels display Level 1 (panel A) and Level 2 (panels B-D) data, while bottom panels show Level 1.5 (panels E-H) data. See captions of Figs. \ref{f5} and \ref{f6} for more details. 
}
\label{f7}
\end{figure*}

\section{A glimpse at science with IBIS-A}

A significant fraction of the observations stored in IBIS-A are still scientifically unexplored. 
Figure \ref{f5} displays examples of these data. They concern calibrated observations of   flaring regions 
obtained at the photospheric Fe I 6173 \AA~ line and at the chromospheric  H$\alpha$ 6563 \AA~ (the latter is shown in Fig. \ref{f5} panels A and B at two different wavelengths) and 
Ca II 8542 \AA~ (Fig. \ref{f5} panels C and D) lines on 10 October 2014, 14:25 UT  and 13 May 2016, 15:00 UT, at disc positions $\mu$=0.47 and 
$\mu$=0.89, respectively.  
The observations of both these flaring regions  exceed 100 minutes.   

The atmosphere imaged in the wing of Ca II 8542 \AA~  (Fig. \ref{f5} panel C) is typical of the middle photosphere, with a clear pattern of reversed
granulation. Instead, both the wing of  H$\alpha$ 6563 \AA, and the core of both H$\alpha$ 6563 \AA~ and Ca II 8542 \AA~ (Fig. \ref{f5} panels A, B, D) show various
parts of the chromosphere, with the typical dazzling variety of features such as 
surges, bundles of arch filament systems, spicules, and localized and extended bright regions. 
Figure \ref{f6}  shows 
several science-ready data available in IBIS-A  for the flaring region with pores shown in panels C and D of Fig. \ref{f5}. In the left side of the FOV, the maps of CP derived from the photospheric Fe I 6173 \AA~ and chromospheric Ca II 8542 \AA~ data (Fig. \ref{f6} panels F and G, respectively), and the map of the field inclination obtained from the VFISV data inversion (Fig. \ref{f6} panel D)  
reveal magnetic field concentrations with opposite polarity with respect to that in the pores. At the same location, the chromospheric observations at the Ca II line core (Fig. \ref{f5} panel D) display a filament and a footpoint of the flaring region (see X=[0,100], Y=[300,400]). Magnetic signatures of these structures are also found in the chromospheric CP map from Ca II 8542 \AA~ data (Fig. \ref{f6} panel G). The maps of the magnetic field strength derived from inversion of the photospheric Fe I 6173 \AA~ data  (Fig. \ref{f6} panels B, C, and A,  respectively) show values  up to 1.4 kG in the pores, and field strength in the range of 0.4-0.6 kG in their surrounding area. There the maps of the photopsheric LP and V$_{los}$ from the Fe I 6173 \AA~ data (Fig. \ref{f6} panels E and H, respectively) show 
horizontal magnetic fields and velocity field with strong upflows.  

Figure \ref{f7} displays examples of the Level 1, Level 1.5, and Level 2 data available in IBIS-A for  the spectro-polarimetric observations of a sunspot region shown in Fig. \ref{f1}. That region was observed on 13 May 2016, 15:38 UT at disc position $\mathrm{\mu}$=0.67.  
It is worth noting that both Figs. \ref{f6} and \ref{f7} show values of the field inclination derived from data inversion in quiet Sun areas peaked around 90$\mathrm{^{\circ}}$. As  reported by  \citet{Borrero2011inc}, these values result from inversion of LP measurements characterized by a signal to noise ratio lower than 4.5. A more accurate estimate of the magnetic field inclination in quiet Sun regions is however beyond the scope of the present  work, which aims to provide quick-look science-ready data to IBIS-A users. 

\section{Summary and conclusions}

This paper presents the IBIS-A archive, which 
currently includes 30 TB of data taken with the Interferometric BIdimensional Spectropolarimeter (IBIS) during 28 observing campaigns carried across 2008 to 2019 on 159 days.
This presentation of the archive is meant to facilitate and foster the scientific usage of the IBIS-A data with examples of the data quality and their application in recent and undergoing studies relevant to solar processes and Space Weather science.

Indeed, since the implementation of IBIS-A in 2016, the raw
data available in the archive have increased, as well as the fraction
of calibrated measurements.
Also, links have been added pointing to
complementary data of the solar atmosphere available from
coordinated measurements performed with the instruments onboard
the Hinode and IRIS satellites, and the full-disc red
continuum, Ca II K, and H$\alpha$ observations of the photosphere and
chromosphere from the INAF synoptic solar telescopes. 

Furthermore, the archive has been populated with higher level products from inversion of the calibrated data. 
Multi-dimensional arrays with maps of circular,  linear, and net circular polarization, and $\mathrm{los}$ velocity patterns from  calibrated photospheric  Fe I 6173 \AA~  and chromospheric Ca II 8542 \AA~ series are also offered to registered users. Besides, 
more than 80\% of the calibrated Fe I 6173 \AA~ photospheric series have already been inverted with the VFISV code to offer a quick-look view of the magnetic and velocity fields in the archived targets. A further milestone will be reached from inversion of the remaining photospheric data and of the chromospheric series with the new NLTE DeSIRe code  
\citep{Quintero2021}. The IBIS-A team will also ingest more data in the archive when they will be  available to it.

IBIS-A data include metadata for proper interpretation and archiving of the observations available in the archive. At present, the keywords employed in the header of FITS Level  1.5 and Level 2 data are partly compliant\footnote{Following the definition in relevant documents.}  with the recently formulated SOLARNET recommendations.   
The complete standardization of the metadata information provided by the IBIS-A data planned in the near future will facilitate their usage by users with limited knowledge of the spectro-polarimetric observations. Besides, it 
will allow easy  inclusion of IBIS-A data in future Solar Virtual Observatories.

IBIS-A represents a unique resource for investigating the plasma processes in the solar atmosphere and the solar origin of Space Weather events. Besides, it represents the prototype of the archive that will host the data acquired with the updated version of the IBIS instrument, named IBIS 2.0, which is under development for installation at the Osservatorio del Roque de Los Muchachos in the Canary Islands.  

Browsing the IBIS-A does not require authentication, but users have to register and login in order to be able to request and download data. 
The IBIS-A team is open for technical support and help with the data analysis to foster the use of data stored in the archive in new science projects. In this respect, note that 
 IBIS-A   
 is currently maintained by the H2020 SOLARNET High-resolution Solar Physics Network project, which  aims at integrating the major European infrastructures in the field of high-resolution
solar physics.
Therefore, any publication that uses data contained in IBIS-A should acknowledge the above support by reporting 
``This study makes use of data collected in the IBIS-A archive, which has received funding from the European Union's Horizon 2020 research and innovation programme under Grant Agreements No 824135 (SOLARNET).''

\begin{acknowledgements}
The authors thank Doug Gilliam and Mike Bradford for acquisition of IBIS data throughout the years.
They also thank Juan Manuel Borrero and Han Uitenbroek for their work on IBIS data. The authors are grateful to the referee for carefully reading of the paper and for her/his comments.
	This research has received funding from the European Union's FP7 Capacities program under grant agreement No 312495 (SOLARNET) and from the European Union's Horizon 2020 Research and Innovation program under grant agreements No 824135 (SOLARNET) and No 739500 (PRE-EST).
	This work was also supported by the Italian MIUR-PRIN grant 2017 ''Circumterrestrial Environment: Impact of Sun--Earth Interaction'' and by the INAF Istituto Nazionale di Astrofisica. The Level 2 data available in IBIS-A were obtained from access to servers of the Italian CINECA and INAF. 
	IBIS has been designed and constructed by the INAF Osservatorio Astrofisico di Arcetri with contributions from the Universit\'a di Firenze, the Universit\'a di Roma Tor Vergata, and upgraded with further contributions from National Solar Observatory (NSO) and Queens University Belfast.
IBIS was operated with support of the NSO. The NSO is operated by the Association of Universities for Research in Astronomy, Inc., under cooperative agreement with the National Science Foundation. This research made use of NASA's Astrophysics Data System.
\end{acknowledgements}

\bibliographystyle{aa} 
\bibliography{ibisa_biblio} 

\begin{thebibliography}{104}
\expandafter\ifx\csname natexlab\endcsname\relax\def\natexlab#1{#1}\fi

\bibitem[{{Abbasvand} {et~al.}(2020){Abbasvand}, {Sobotka}, {Heinzel},
  {{\v{S}}vanda}, {Jur{\v{c}}{\'a}k}, {del Moro}, \&
  {Berrilli}}]{abbasvand2020}
{Abbasvand}, V., {Sobotka}, M., {Heinzel}, P., {et~al.} 2020, \apj, 890, 22

\bibitem[{{Baker} {et~al.}(2021){Baker}, {Stangalini}, {Valori}, {Brooks},
  {To}, {van Driel-Gesztelyi}, {D{\'e}moulin}, {Stansby}, {Jess}, \&
  {Jafarzadeh}}]{baker2021}
{Baker}, D., {Stangalini}, M., {Valori}, G., {et~al.} 2021, \apj, 907, 16

\bibitem[{{Bellot Rubio} \& {Orozco Su{\'a}rez}(2019)}]{Bellot2019LRSP}
{Bellot Rubio}, L. \& {Orozco Su{\'a}rez}, D. 2019, Living Reviews in Solar
  Physics, 16, 1

\bibitem[{{Blandford} {et~al.}(2019){Blandford}, {Meier}, \&
  {Readhead}}]{Blandford2019}
{Blandford}, R., {Meier}, D., \& {Readhead}, A. 2019, \araa, 57, 467

\bibitem[{{Borrero} {et~al.}(2017){Borrero}, {Jafarzadeh}, {Sch{\"u}ssler}, \&
  {Solanki}}]{Borrero2017}
{Borrero}, J.~M., {Jafarzadeh}, S., {Sch{\"u}ssler}, M., \& {Solanki}, S.~K.
  2017, \ssr, 210, 275

\bibitem[{{Borrero} \& {Kobel}(2011)}]{Borrero2011inc}
{Borrero}, J.~M. \& {Kobel}, P. 2011, \aap, 527, A29

\bibitem[{{Borrero} {et~al.}(2011){Borrero}, {Tomczyk}, {Kubo},
  {Socas-Navarro}, {Schou}, {Couvidat}, \& {Bogart}}]{borrero2011}
{Borrero}, J.~M., {Tomczyk}, S., {Kubo}, M., {et~al.} 2011, \solphys, 273, 267

\bibitem[{{Brosius} {et~al.}(2014){Brosius}, {Daw}, \& {Rabin}}]{eunis}
{Brosius}, J.~W., {Daw}, A.~N., \& {Rabin}, D.~M. 2014, \apj, 790, 112

\bibitem[{{Cao} {et~al.}(2010){Cao}, {Gorceix}, {Coulter}, {W{\"o}ger}, {Ahn},
  {Shumko}, {Varsik}, {Coulter}, \& {Goode}}]{cao2010}
{Cao}, W., {Gorceix}, N., {Coulter}, R., {et~al.} 2010, in Society of
  Photo-Optical Instrumentation Engineers (SPIE) Conference Series, Vol. 7735,
  Ground-based and Airborne Instrumentation for Astronomy III, ed. I.~S.
  {McLean}, S.~K. {Ramsay}, \& H.~{Takami}, 77355V

\bibitem[{{Capparelli} {et~al.}(2017){Capparelli}, {Zuccarello}, {Romano},
  {Sim{\~o}es}, {Fletcher}, {Kuridze}, {Mathioudakis}, {Keys}, {Cauzzi}, \&
  {Carlsson}}]{capparelli2017}
{Capparelli}, V., {Zuccarello}, F., {Romano}, P., {et~al.} 2017, \apj, 850, 36

\bibitem[{{Cauzzi} {et~al.}(2009){Cauzzi}, {Reardon}, {Rutten}, {Tritschler},
  \& {Uitenbroek}}]{cauzzi2009}
{Cauzzi}, G., {Reardon}, K., {Rutten}, R.~J., {Tritschler}, A., \&
  {Uitenbroek}, H. 2009, \aap, 503, 577

\bibitem[{{Cauzzi} {et~al.}(2008){Cauzzi}, {Reardon}, {Uitenbroek},
  {Cavallini}, {Falchi}, {Falciani}, {Janssen}, {Rimmele}, {Vecchio}, \&
  {W{\"o}ger}}]{cauzzi2008}
{Cauzzi}, G., {Reardon}, K.~P., {Uitenbroek}, H., {et~al.} 2008, \aap, 480, 515

\bibitem[{{Cavallini}(2006)}]{cavallini2006}
{Cavallini}, F. 2006, \solphys, 236, 415

\bibitem[{{Cheung} \& {Isobe}(2014)}]{Cheung2014LRSP}
{Cheung}, M. C.~M. \& {Isobe}, H. 2014, Living Reviews in Solar Physics, 11, 3

\bibitem[{{Collados} {et~al.}(2013){Collados}, {Bettonvil}, {Cavaller},
  {Ermolli}, {Gelly}, {P{\'e}rez}, {Socas-Navarro}, {Soltau}, {Volkmer}, \&
  {EST Team}}]{est}
{Collados}, M., {Bettonvil}, F., {Cavaller}, L., {et~al.} 2013, \memsai, 84,
  379

\bibitem[{{Criscuoli} {et~al.}(2012){Criscuoli}, {Del Moro}, {Giannattasio},
  {Viticchi{\'e}}, {Giorgi}, {Ermolli}, {Zuccarello}, \&
  {Berrilli}}]{criscuoli2012}
{Criscuoli}, S., {Del Moro}, D., {Giannattasio}, F., {et~al.} 2012, \aap, 546,
  A26

\bibitem[{{Criscuoli} {et~al.}(2013){Criscuoli}, {Ermolli}, {Uitenbroek}, \&
  {Giorgi}}]{criscuoli2013}
{Criscuoli}, S., {Ermolli}, I., {Uitenbroek}, H., \& {Giorgi}, F. 2013, \apj,
  763, 144

\bibitem[{{Criscuoli} \& {Tritschler}({2014})}]{criscuoli2014}
{Criscuoli}, S. \& {Tritschler}, A. {2014}, {IBIS Data Reduction Notes}, Tech.
  Rep. NS05, {National Solar Observatory, Sacramento Peak}

\bibitem[{{De Pontieu} {et~al.}(2014){De Pontieu}, {Title}, {Lemen}, {Kushner},
  {Akin}, {Allard}, {Berger}, {Boerner}, {Cheung}, {Chou}, {Drake}, {Duncan},
  {Freeland}, {Heyman}, {Hoffman}, {Hurlburt}, {Lindgren}, {Mathur}, {Rehse},
  {Sabolish}, {Seguin}, {Schrijver}, {Tarbell}, {W{\"u}lser}, {Wolfson},
  {Yanari}, {Mudge}, {Nguyen-Phuc}, {Timmons}, {van Bezooijen}, {Weingrod},
  {Brookner}, {Butcher}, {Dougherty}, {Eder}, {Knagenhjelm}, {Larsen},
  {Mansir}, {Phan}, {Boyle}, {Cheimets}, {DeLuca}, {Golub}, {Gates}, {Hertz},
  {McKillop}, {Park}, {Perry}, {Podgorski}, {Reeves}, {Saar}, {Testa}, {Tian},
  {Weber}, {Dunn}, {Eccles}, {Jaeggli}, {Kankelborg}, {Mashburn}, {Pust},
  {Springer}, {Carvalho}, {Kleint}, {Marmie}, {Mazmanian}, {Pereira}, {Sawyer},
  {Strong}, {Worden}, {Carlsson}, {Hansteen}, {Leenaarts}, {Wiesmann},
  {Aloise}, {Chu}, {Bush}, {Scherrer}, {Brekke}, {Martinez-Sykora}, {Lites},
  {McIntosh}, {Uitenbroek}, {Okamoto}, {Gummin}, {Auker}, {Jerram}, {Pool}, \&
  {Waltham}}]{iris}
{De Pontieu}, B., {Title}, A.~M., {Lemen}, J.~R., {et~al.} 2014, \solphys, 289,
  2733

\bibitem[{{Del Moro} {et~al.}(2007){Del Moro}, {Giordano}, \&
  {Berrilli}}]{delmoro2007}
{Del Moro}, D., {Giordano}, S., \& {Berrilli}, F. 2007, \aap, 472, 599

\bibitem[{{del Toro Iniesta} \& {Ruiz Cobo}(2016)}]{deltoro2016}
{del Toro Iniesta}, J.~C. \& {Ruiz Cobo}, B. 2016, Living Reviews in Solar
  Physics, 13, 4

\bibitem[{{Ermolli} {et~al.}(2020){Ermolli}, {Cirami}, {Calderone}, {Del Moro},
  {Romano}, {Viavattene}, {Coretti}, {Giorgi}, {Baldini}, {Di Marcantonio},
  {Giovannelli}, {Guglielmino}, {Murabito}, {Pedichini}, {Piazzesi},
  {Aliverti}, {Redaelli}, {Berrilli}, \& {Zuccarello}}]{ermolli_ibis20}
{Ermolli}, I., {Cirami}, R., {Calderone}, G., {et~al.} 2020, in Society of
  Photo-Optical Instrumentation Engineers (SPIE) Conference Series, Vol. 11447,
  Society of Photo-Optical Instrumentation Engineers (SPIE) Conference Series,
  114470Z

\bibitem[{{Ermolli} {et~al.}(2011){Ermolli}, {Criscuoli}, \&
  {Giorgi}}]{ermolli2011}
{Ermolli}, I., {Criscuoli}, S., \& {Giorgi}, F. 2011, Contributions of the
  Astronomical Observatory Skalnate Pleso, 41, 73

\bibitem[{{Ermolli} {et~al.}(2017){Ermolli}, {Cristaldi}, {Giorgi},
  {Giannattasio}, {Stangalini}, {Romano}, {Tritschler}, \&
  {Zuccarello}}]{ermolli2017}
{Ermolli}, I., {Cristaldi}, A., {Giorgi}, F., {et~al.} 2017, \aap, 600, A102

\bibitem[{{Giordano} {et~al.}(2008){Giordano}, {Berrilli}, {Del Moro}, \&
  {Penza}}]{giordano2008}
{Giordano}, S., {Berrilli}, F., {Del Moro}, D., \& {Penza}, V. 2008, \aap, 489,
  747

\bibitem[{Harris {et~al.}(2020)Harris, Millman, van~der Walt, Gommers,
  Virtanen, Cournapeau, Wieser, Taylor, Berg, Smith, Kern, Picus, Hoyer, van
  Kerkwijk, Brett, Haldane, del R{\'{i}}o, Wiebe, Peterson,
  G{\'{e}}rard-Marchant, Sheppard, Reddy, Weckesser, Abbasi, Gohlke, \&
  Oliphant}]{Numpy}
Harris, C.~R., Millman, K.~J., van~der Walt, S.~J., {et~al.} 2020, Nature, 585,
  357

\bibitem[{{Haugan} \& {Fredvik}(2015)}]{haugan2015}
{Haugan}, S. V.~H. \& {Fredvik}, T. 2015, {Document on Standards for Data
  Archiving and VO, Deliverable D20.4, SOLARNET (EC 7th FP grant 312495}, Tech.
  Rep. D20.4, {Institute of Theoretical Astrophysics, University of Oslo}

\bibitem[{{Haugan} \& {Fredvik}(2020)}]{haugan2020}
{Haugan}, S. V.~H. \& {Fredvik}, T. 2020, arXiv e-prints, arXiv:2011.12139

\bibitem[{{Houston} {et~al.}(2020){Houston}, {Jess}, {Keppens}, {Stangalini},
  {Keys}, {Grant}, {Jafarzadeh}, {McFetridge}, {Murabito}, {Ermolli}, \&
  {Giorgi}}]{houston2020}
{Houston}, S.~J., {Jess}, D.~B., {Keppens}, R., {et~al.} 2020, \apj, 892, 49

\bibitem[{Hunter(2007)}]{Matplotlib}
Hunter, J.~D. 2007, Computing in Science \& Engineering, 9, 90

\bibitem[{{Jaeggli} {et~al.}(2012){Jaeggli}, {Lin}, \&
  {Uitenbroek}}]{jaeggli_2012}
{Jaeggli}, S.~A., {Lin}, H., \& {Uitenbroek}, H. 2012, \apj, 745, 133

\bibitem[{{Jess} {et~al.}(2010){Jess}, {Mathioudakis}, {Christian}, {Keenan},
  {Ryans}, \& {Crockett}}]{ROSA2010}
{Jess}, D.~B., {Mathioudakis}, M., {Christian}, D.~J., {et~al.} 2010, \solphys,
  261, 363

\bibitem[{{Jess} {et~al.}(2015){Jess}, {Morton}, {Verth}, {Fedun}, {Grant}, \&
  {Giagkiozis}}]{Jess2015}
{Jess}, D.~B., {Morton}, R.~J., {Verth}, G., {et~al.} 2015, \ssr, 190, 103

\bibitem[{{Judge} {et~al.}(2010){Judge}, {Tritschler}, {Uitenbroek}, {Reardon},
  {Cauzzi}, \& {de Wijn}}]{judge2010}
{Judge}, P.~G., {Tritschler}, A., {Uitenbroek}, H., {et~al.} 2010, \apj, 710,
  1486

\bibitem[{{Kilpua} {et~al.}(2017){Kilpua}, {Koskinen}, \&
  {Pulkkinen}}]{Kilpua2017LRSP}
{Kilpua}, E., {Koskinen}, H. E.~J., \& {Pulkkinen}, T.~I. 2017, Living Reviews
  in Solar Physics, 14, 5

\bibitem[{{Kobayashi} {et~al.}(2014){Kobayashi}, {Cirtain}, {Winebarger},
  {Korreck}, {Golub}, {Walsh}, {De Pontieu}, {DeForest}, {Title}, {Kuzin},
  {Savage}, {Beabout}, {Beabout}, {Podgorski}, {Caldwell}, {McCracken},
  {Ordway}, {Bergner}, {Gates}, {McKillop}, {Cheimets}, {Platt}, {Mitchell}, \&
  {Windt}}]{hic}
{Kobayashi}, K., {Cirtain}, J., {Winebarger}, A.~R., {et~al.} 2014, \solphys,
  289, 4393

\bibitem[{{Kowalski} {et~al.}(2015){Kowalski}, {Cauzzi}, \&
  {Fletcher}}]{kowalski2015}
{Kowalski}, A.~F., {Cauzzi}, G., \& {Fletcher}, L. 2015, \apj, 798, 107

\bibitem[{{Lemen} {et~al.}(2012){Lemen}, {Title}, {Akin}, {Boerner}, {Chou},
  {Drake}, {Duncan}, {Edwards}, {Friedlaender}, {Heyman}, {Hurlburt}, {Katz},
  {Kushner}, {Levay}, {Lindgren}, {Mathur}, {McFeaters}, {Mitchell}, {Rehse},
  {Schrijver}, {Springer}, {Stern}, {Tarbell}, {Wuelser}, {Wolfson}, {Yanari},
  {Bookbinder}, {Cheimets}, {Caldwell}, {Deluca}, {Gates}, {Golub}, {Park},
  {Podgorski}, {Bush}, {Scherrer}, {Gummin}, {Smith}, {Auker}, {Jerram},
  {Pool}, {Soufli}, {Windt}, {Beardsley}, {Clapp}, {Lang}, \& {Waltham}}]{aia}
{Lemen}, J.~R., {Title}, A.~M., {Akin}, D.~J., {et~al.} 2012, \solphys, 275, 17

\bibitem[{{Lipartito} {et~al.}(2014){Lipartito}, {Judge}, {Reardon}, \&
  {Cauzzi}}]{lipartito2014}
{Lipartito}, I., {Judge}, P.~G., {Reardon}, K., \& {Cauzzi}, G. 2014, \apj,
  785, 109

\bibitem[{{L{\"o}fdahl}(2012)}]{Loefdahl2012}
{L{\"o}fdahl}, M. 2012, IAU Special Session, 6, E3.05

\bibitem[{{L{\"o}fdahl}(2016)}]{Lofdahl2016}
{L{\"o}fdahl}, M. 2016, in Astronomical Society of the Pacific Conference
  Series, Vol. 504, Coimbra Solar Physics Meeting: Ground-based Solar
  Observations in the Space Instrumentation Era, ed. I.~{Dorotovic}, C.~E.
  {Fischer}, \& M.~{Temmer}, 111

\bibitem[{{L{\"o}fdahl} {et~al.}(2021){L{\"o}fdahl}, {Hillberg}, {de la Cruz
  Rodr{\'\i}guez}, {Vissers}, {Andriienko}, {Scharmer}, {Haugan}, \&
  {Fredvik}}]{lofdahl2021}
{L{\"o}fdahl}, M.~G., {Hillberg}, T., {de la Cruz Rodr{\'\i}guez}, J., {et~al.}
  2021, \aap, 653, A68

\bibitem[{{Mart{\'\i}nez Pillet} {et~al.}(2011){Mart{\'\i}nez Pillet}, {Del
  Toro Iniesta}, {{\'A}lvarez-Herrero}, {Domingo}, {Bonet}, {Gonz{\'a}lez
  Fern{\'a}ndez}, {L{\'o}pez Jim{\'e}nez}, {Pastor}, {Gasent Blesa}, {Mellado},
  {Piqueras}, {Aparicio}, {Balaguer}, {Ballesteros}, {Belenguer}, {Bellot
  Rubio}, {Berkefeld}, {Collados}, {Deutsch}, {Feller}, {Girela}, {Grauf},
  {Heredero}, {Herranz}, {Jer{\'o}nimo}, {Laguna}, {Meller}, {Men{\'e}ndez},
  {Morales}, {Orozco Su{\'a}rez}, {Ramos}, {Reina}, {Ramos}, {Rodr{\'\i}guez},
  {S{\'a}nchez}, {Uribe-Patarroyo}, {Barthol}, {Gandorfer}, {Knoelker},
  {Schmidt}, {Solanki}, \& {Vargas Dom{\'\i}nguez}}]{Martinez2011}
{Mart{\'\i}nez Pillet}, V., {Del Toro Iniesta}, J.~C., {{\'A}lvarez-Herrero},
  A., {et~al.} 2011, \solphys, 268, 57

\bibitem[{{Molnar} {et~al.}(2019){Molnar}, {Reardon}, {Chai}, {Gary},
  {Uitenbroek}, {Cauzzi}, \& {Cranmer}}]{molnar2019}
{Molnar}, M.~E., {Reardon}, K.~P., {Chai}, Y., {et~al.} 2019, \apj, 881, 99

\bibitem[{{Murabito} {et~al.}(2019){Murabito}, {Ermolli}, {Giorgi},
  {Stangalini}, {Guglielmino}, {Jafarzadeh}, {Socas-Navarro}, {Romano}, \&
  {Zuccarello}}]{murabito2019}
{Murabito}, M., {Ermolli}, I., {Giorgi}, F., {et~al.} 2019, \apj, 873, 126

\bibitem[{{Murabito} {et~al.}(2020{\natexlab{a}}){Murabito}, {Guglielmino},
  {Ermolli}, {Stangalini}, \& {Giorgi}}]{Murabito2020ApJ}
{Murabito}, M., {Guglielmino}, S.~L., {Ermolli}, I., {Stangalini}, M., \&
  {Giorgi}, F. 2020{\natexlab{a}}, \apj, 890, 96

\bibitem[{{Murabito} {et~al.}(2017){Murabito}, {Romano}, {Guglielmino}, \&
  {Zuccarello}}]{murabito2017}
{Murabito}, M., {Romano}, P., {Guglielmino}, S.~L., \& {Zuccarello}, F. 2017,
  \apj, 834, 76

\bibitem[{{Murabito} {et~al.}(2016){Murabito}, {Romano}, {Guglielmino},
  {Zuccarello}, \& {Solanki}}]{murabito2016}
{Murabito}, M., {Romano}, P., {Guglielmino}, S.~L., {Zuccarello}, F., \&
  {Solanki}, S.~K. 2016, \apj, 825, 75

\bibitem[{{Murabito} {et~al.}(2020{\natexlab{b}}){Murabito}, {Shetye},
  {Stangalini}, {Verwichte}, {Arber}, {Ermolli}, {Giorgi}, \&
  {Goffrey}}]{murabito2020AA}
{Murabito}, M., {Shetye}, J., {Stangalini}, M., {et~al.} 2020{\natexlab{b}},
  \aap, 639, A59

\bibitem[{{Narukage} {et~al.}(2016){Narukage}, {McKenzie}, {Ishikawa},
  {Trujillo-Bueno}, {De Pontieu}, {Kubo}, {Ishikawa}, {Kano}, {Suematsu},
  {Yoshida}, {Rachmeler}, {Kobayashi}, {Cirtain}, {Winebarger}, {Asensio
  Ramos}, {del Pino Aleman}, {{\v{S}}t{\k{e}}p{\'a}n}, {Belluzzi},
  {Larruquert}, {Auch{\`e}re}, {Leenaarts}, \& {Carlsson}}]{clasp}
{Narukage}, N., {McKenzie}, D.~E., {Ishikawa}, R., {et~al.} 2016, in Society of
  Photo-Optical Instrumentation Engineers (SPIE) Conference Series, Vol. 9905,
  Space Telescopes and Instrumentation 2016: Ultraviolet to Gamma Ray, ed.
  J.-W.~A. {den Herder}, T.~{Takahashi}, \& M.~{Bautz}, 990508

\bibitem[{{Padoan} \& {Nordlund}(2011)}]{Padoan2011}
{Padoan}, P. \& {Nordlund}, {\r{A}}. 2011, \apj, 730, 40

\bibitem[{{Rast} {et~al.}(2021){Rast}, {Bello Gonz{\'a}lez}, {Bellot Rubio},
  {Cao}, {Cauzzi}, {Deluca}, {de Pontieu}, {Fletcher}, {Gibson}, {Judge},
  {Katsukawa}, {Kazachenko}, {Khomenko}, {Landi}, {Mart{\'\i}nez Pillet},
  {Petrie}, {Qiu}, {Rachmeler}, {Rempel}, {Schmidt}, {Scullion}, {Sun},
  {Welsch}, {Andretta}, {Antolin}, {Ayres}, {Balasubramaniam}, {Ballai},
  {Berger}, {Bradshaw}, {Campbell}, {Carlsson}, {Casini}, {Centeno}, {Cranmer},
  {Criscuoli}, {Deforest}, {Deng}, {Erd{\'e}lyi}, {Fedun}, {Fischer},
  {Gonz{\'a}lez Manrique}, {Hahn}, {Harra}, {Henriques}, {Hurlburt}, {Jaeggli},
  {Jafarzadeh}, {Jain}, {Jefferies}, {Keys}, {Kowalski}, {Kuckein}, {Kuhn},
  {Kuridze}, {Liu}, {Liu}, {Longcope}, {Mathioudakis}, {McAteer}, {McIntosh},
  {McKenzie}, {Miralles}, {Morton}, {Muglach}, {Nelson}, {Panesar}, {Parenti},
  {Parnell}, {Poduval}, {Reardon}, {Reep}, {Schad}, {Schmit}, {Sharma},
  {Socas-Navarro}, {Srivastava}, {Sterling}, {Suematsu}, {Tarr}, {Tiwari},
  {Tritschler}, {Verth}, {Vourlidas}, {Wang}, {Wang}, {NSO and DKIST Project},
  {DKIST Instrument Scientists}, {DKIST Science Working Group}, \& {DKIST
  Critical Science Plan Community}}]{Rast2021}
{Rast}, M.~P., {Bello Gonz{\'a}lez}, N., {Bellot Rubio}, L., {et~al.} 2021,
  \solphys, 296, 70

\bibitem[{{Reardon} \& {Cavallini}(2008)}]{reardon_cavallini2008}
{Reardon}, K.~P. \& {Cavallini}, F. 2008, \aap, 481, 897

\bibitem[{{Reardon} {et~al.}(2009){Reardon}, {Uitenbroek}, \&
  {Cauzzi}}]{reardon2009}
{Reardon}, K.~P., {Uitenbroek}, H., \& {Cauzzi}, G. 2009, \aap, 500, 1239

\bibitem[{{Rempel} \& {Schlichenmaier}(2011)}]{Rempel2011LRSP}
{Rempel}, M. \& {Schlichenmaier}, R. 2011, Living Reviews in Solar Physics, 8,
  3

\bibitem[{{Righini} {et~al.}(2010){Righini}, {Cavallini}, \&
  {Reardon}}]{righini2010}
{Righini}, A., {Cavallini}, F., \& {Reardon}, K.~P. 2010, \aap, 515, A85

\bibitem[{{Rimmele} {et~al.}(2020){Rimmele}, {Warner}, {Keil}, {Goode},
  {Kn{\"o}lker}, {Kuhn}, {Rosner}, {McMullin}, {Casini}, {Lin}, {W{\"o}ger},
  {von der L{\"u}he}, {Tritschler}, {Davey}, {de Wijn}, {Elmore}, {Fehlmann},
  {Harrington}, {Jaeggli}, {Rast}, {Schad}, {Schmidt}, {Mathioudakis},
  {Mickey}, {Anan}, {Beck}, {Marshall}, {Jeffers}, {Oschmann}, {Beard},
  {Berst}, {Cowan}, {Craig}, {Cross}, {Cummings}, {Donnelly}, {de Vanssay},
  {Eigenbrot}, {Ferayorni}, {Foster}, {Galapon}, {Gedrites}, {Gonzales},
  {Goodrich}, {Gregory}, {Guzman}, {Guzzo}, {Hegwer}, {Hubbard}, {Hubbard},
  {Johansson}, {Johnson}, {Liang}, {Liang}, {McQuillen}, {Mayer}, {Newman},
  {Onodera}, {Phelps}, {Puentes}, {Richards}, {Rimmele}, {Sekulic}, {Shimko},
  {Simison}, {Smith}, {Starman}, {Sueoka}, {Summers}, {Szabo}, {Szabo},
  {Wampler}, {Williams}, \& {White}}]{2020SoPh..295..172R}
{Rimmele}, T.~R., {Warner}, M., {Keil}, S.~L., {et~al.} 2020, \solphys, 295,
  172

\bibitem[{{Romano} {et~al.}(2012){Romano}, {Berrilli}, {Criscuoli}, {Del Moro},
  {Ermolli}, {Giorgi}, {Viticchi{\'e}}, \& {Zuccarello}}]{romano2012}
{Romano}, P., {Berrilli}, F., {Criscuoli}, S., {et~al.} 2012, \solphys, 280,
  407

\bibitem[{{Romano} {et~al.}(2017){Romano}, {Falco}, {Guglielmino}, \&
  {Murabito}}]{romano2017}
{Romano}, P., {Falco}, M., {Guglielmino}, S.~L., \& {Murabito}, M. 2017, \apj,
  837, 173

\bibitem[{{Romano} {et~al.}(2013){Romano}, {Frasca}, {Guglielmino}, {Ermolli},
  {Tritschler}, {Reardon}, \& {Zuccarello}}]{romano2013}
{Romano}, P., {Frasca}, D., {Guglielmino}, S.~L., {et~al.} 2013, \apjl, 771, L3

\bibitem[{{Romano} {et~al.}(2022){Romano}, {Guglielmino}, {Costa}, {Falco},
  {Buttaccio}, {Costa}, {Martinetti}, {Occhipinti}, {Spadaro}, {Ventura},
  {Capuano}, \& {Zuccarello}}]{romanobarra2021}
{Romano}, P., {Guglielmino}, S.~L., {Costa}, P., {et~al.} 2022, \solphys, 297,
  7

\bibitem[{{Romano} {et~al.}(2020){Romano}, {Murabito}, {Guglielmino},
  {Zuccarello}, \& {Falco}}]{romano2020APJ}
{Romano}, P., {Murabito}, M., {Guglielmino}, S.~L., {Zuccarello}, F., \&
  {Falco}, M. 2020, \apj, 899, 129

\bibitem[{{Romano} {et~al.}(2014){Romano}, {Zuccarello}, {Guglielmino}, \&
  {Zuccarello}}]{romano2014}
{Romano}, P., {Zuccarello}, F.~P., {Guglielmino}, S.~L., \& {Zuccarello}, F.
  2014, \apj, 794, 118

\bibitem[{{Ruiz Cobo} {et~al.}(2022){Ruiz Cobo}, {Quintero Noda}, {Gafeira},
  {Uitenbroek}, {Orozco Su{\'a}rez}, \& {P{\'a}ez Ma{\~n}{\'a}}}]{Quintero2021}
{Ruiz Cobo}, B., {Quintero Noda}, C., {Gafeira}, R., {et~al.} 2022, arXiv
  e-prints, arXiv:2202.02226

\bibitem[{{Scharmer} {et~al.}(2003){Scharmer}, {Bjelksjo}, {Korhonen},
  {Lindberg}, \& {Petterson}}]{scharmer2003}
{Scharmer}, G.~B., {Bjelksjo}, K., {Korhonen}, T.~K., {Lindberg}, B., \&
  {Petterson}, B. 2003, in Society of Photo-Optical Instrumentation Engineers
  (SPIE) Conference Series, Vol. 4853, Innovative Telescopes and
  Instrumentation for Solar Astrophysics, ed. S.~L. {Keil} \& S.~V. {Avakyan},
  341--350

\bibitem[{{Scharmer} {et~al.}(2019){Scharmer}, {L{\"o}fdahl}, {Sliepen}, \& {de
  la Cruz Rodr{\'\i}guez}}]{Scharmer2019}
{Scharmer}, G.~B., {L{\"o}fdahl}, M.~G., {Sliepen}, G., \& {de la Cruz
  Rodr{\'\i}guez}, J. 2019, \aap, 626, A55

\bibitem[{{Scharmer} {et~al.}(2008){Scharmer}, {Narayan}, {Hillberg}, {de la
  Cruz Rodriguez}, {L{\"o}fdahl}, {Kiselman}, {S{\"u}tterlin}, {van Noort}, \&
  {Lagg}}]{scharmer2008}
{Scharmer}, G.~B., {Narayan}, G., {Hillberg}, T., {et~al.} 2008, \apjl, 689,
  L69

\bibitem[{{Scherrer} {et~al.}(1995){Scherrer}, {Bogart}, {Bush}, {Hoeksema},
  {Kosovichev}, {Schou}, {Rosenberg}, {Springer}, {Tarbell}, {Title},
  {Wolfson}, {Zayer}, \& {MDI Engineering Team}}]{mdi}
{Scherrer}, P.~H., {Bogart}, R.~S., {Bush}, R.~I., {et~al.} 1995, \solphys,
  162, 129

\bibitem[{{Scherrer} {et~al.}(2012){Scherrer}, {Schou}, {Bush}, {Kosovichev},
  {Bogart}, {Hoeksema}, {Liu}, {Duvall}, {Zhao}, {Title}, {Schrijver},
  {Tarbell}, \& {Tomczyk}}]{hmi}
{Scherrer}, P.~H., {Schou}, J., {Bush}, R.~I., {et~al.} 2012, \solphys, 275,
  207

\bibitem[{{Schilliro} \& {Romano}(2021)}]{schilliro2021}
{Schilliro}, F. \& {Romano}, P. 2021, \mnras, 503, 2676

\bibitem[{{Schlichenmaier} {et~al.}(2019){Schlichenmaier}, {Bellot Rubio},
  {Collados}, {Erdelyi}, {Feller}, {Fletcher}, {Jurcak}, {Khomenko},
  {Leenaarts}, {Matthews}, {Belluzzi}, {Carlsson}, {Dalmasse}, {Danilovic},
  {G{\"o}m{\"o}ry}, {Kuckein}, {Manso Sainz}, {Martinez Gonzalez},
  {Mathioudakis}, {Ortiz}, {Riethm{\"u}ller}, {Rouppe van der Voort}, {Simoes},
  {Trujillo Bueno}, {Utz}, \& {Zuccarello}}]{Schlichenmaier2019}
{Schlichenmaier}, R., {Bellot Rubio}, L.~R., {Collados}, M., {et~al.} 2019,
  arXiv e-prints, arXiv:1912.08650

\bibitem[{{Schlichenmaier} \& {Schmidt}(2000)}]{Schlichenmaier2000A&A}
{Schlichenmaier}, R. \& {Schmidt}, W. 2000, \aap, 358, 1122

\bibitem[{{Schmelz}(2003)}]{Schmelz2003}
{Schmelz}, J.~T. 2003, Advances in Space Research, 32, 895

\bibitem[{{Seykora}(1993)}]{Seykora1993}
{Seykora}, E.~J. 1993, \solphys, 145, 389

\bibitem[{{Sobotka} {et~al.}(2012){Sobotka}, {Del Moro}, {Jur{\v{c}}{\'a}k}, \&
  {Berrilli}}]{sobotka2012}
{Sobotka}, M., {Del Moro}, D., {Jur{\v{c}}{\'a}k}, J., \& {Berrilli}, F. 2012,
  \aap, 537, A85

\bibitem[{{Sobotka} {et~al.}(2016){Sobotka}, {Heinzel}, {{\v{S}}vanda},
  {Jur{\v{c}}{\'a}k}, {del Moro}, \& {Berrilli}}]{sobotka2016}
{Sobotka}, M., {Heinzel}, P., {{\v{S}}vanda}, M., {et~al.} 2016, \apj, 826, 49

\bibitem[{{Sobotka} {et~al.}(2013){Sobotka}, {{\v{S}}vanda},
  {Jur{\v{c}}{\'a}k}, {Heinzel}, {Del Moro}, \& {Berrilli}}]{sobotka2013}
{Sobotka}, M., {{\v{S}}vanda}, M., {Jur{\v{c}}{\'a}k}, J., {et~al.} 2013, \aap,
  560, A84

\bibitem[{{Solanki} \& {Montavon}(1993)}]{Solanki&Montavon1993}
{Solanki}, S.~K. \& {Montavon}, C.~A.~P. 1993, \aap, 275, 283

\bibitem[{{Stangalini} {et~al.}(2021{\natexlab{a}}){Stangalini}, {Baker},
  {Valori}, {Jess}, {Jafarzadeh}, {Murabito}, {To}, {Brooks}, {Ermolli},
  {Giorgi}, \& {MacBride}}]{stangalini2021RSPTA}
{Stangalini}, M., {Baker}, D., {Valori}, G., {et~al.} 2021{\natexlab{a}},
  Philosophical Transactions of the Royal Society of London Series A, 379,
  20200216

\bibitem[{{Stangalini} {et~al.}(2013){Stangalini}, {Berrilli}, \&
  {Consolini}}]{stangalini_2013}
{Stangalini}, M., {Berrilli}, F., \& {Consolini}, G. 2013, \aap, 559, A88

\bibitem[{{Stangalini} {et~al.}(2011){Stangalini}, {Del Moro}, {Berrilli}, \&
  {Jefferies}}]{stangalini2011}
{Stangalini}, M., {Del Moro}, D., {Berrilli}, F., \& {Jefferies}, S.~M. 2011,
  \aap, 534, A65

\bibitem[{{Stangalini} {et~al.}(2021{\natexlab{b}}){Stangalini}, {Erd{\'e}lyi},
  {Boocock}, {Tsiklauri}, {Nelson}, {Del Moro}, {Berrilli}, \&
  {Kors{\'o}s}}]{stangalini2021Nat}
{Stangalini}, M., {Erd{\'e}lyi}, R., {Boocock}, C., {et~al.}
  2021{\natexlab{b}}, Nature Astronomy

\bibitem[{{Stangalini} {et~al.}(2012){Stangalini}, {Giannattasio}, {Del Moro},
  \& {Berrilli}}]{stangalini2012}
{Stangalini}, M., {Giannattasio}, F., {Del Moro}, D., \& {Berrilli}, F. 2012,
  \aap, 539, L4

\bibitem[{{Stangalini} {et~al.}(2018){Stangalini}, {Jafarzadeh}, {Ermolli},
  {Erd{\'e}lyi}, {Jess}, {Keys}, {Giorgi}, {Murabito}, {Berrilli}, \& {Del
  Moro}}]{stangalini2018}
{Stangalini}, M., {Jafarzadeh}, S., {Ermolli}, I., {et~al.} 2018, \apj, 869,
  110

\bibitem[{{Stangalini} {et~al.}(2021{\natexlab{c}}){Stangalini}, {Jess},
  {Verth}, {Fedun}, {Fleck}, {Jafarzadeh}, {Keys}, {Murabito}, {Calchetti},
  {Aldhafeeri}, {Berrilli}, {Del Moro}, {Jefferies}, {Terradas}, \&
  {Soler}}]{stangalini2021}
{Stangalini}, M., {Jess}, D.~B., {Verth}, G., {et~al.} 2021{\natexlab{c}},
  \aap, 649, A169

\bibitem[{{Stein}(2012)}]{Stein2012LRSP}
{Stein}, R.~F. 2012, Living Reviews in Solar Physics, 9, 4

\bibitem[{{Straus} {et~al.}(2008){Straus}, {Fleck}, {Jefferies}, {Cauzzi},
  {McIntosh}, {Reardon}, {Severino}, \& {Steffen}}]{straus2008}
{Straus}, T., {Fleck}, B., {Jefferies}, S.~M., {et~al.} 2008, \apjl, 681, L125

\bibitem[{{Temmer}(2021)}]{Temmer2021LRSP}
{Temmer}, M. 2021, Living Reviews in Solar Physics, 18, 4

\bibitem[{{Tritschler} {et~al.}(2008){Tritschler}, {Uitenbroek}, \&
  {Reardon}}]{tritschler2008}
{Tritschler}, A., {Uitenbroek}, H., \& {Reardon}, K. 2008, \apjl, 686, L45

\bibitem[{{Tsuneta} {et~al.}(2008){Tsuneta}, {Ichimoto}, {Katsukawa}, {Nagata},
  {Otsubo}, {Shimizu}, {Suematsu}, {Nakagiri}, {Noguchi}, {Tarbell}, {Title},
  {Shine}, {Rosenberg}, {Hoffmann}, {Jurcevich}, {Kushner}, {Levay}, {Lites},
  {Elmore}, {Matsushita}, {Kawaguchi}, {Saito}, {Mikami}, {Hill}, \&
  {Owens}}]{Hinodesot}
{Tsuneta}, S., {Ichimoto}, K., {Katsukawa}, Y., {et~al.} 2008, \solphys, 249,
  167

\bibitem[{{van Driel-Gesztelyi} \& {Green}(2015)}]{Vandriel2015LRSP}
{van Driel-Gesztelyi}, L. \& {Green}, L.~M. 2015, Living Reviews in Solar
  Physics, 12, 1

\bibitem[{{van Noort} {et~al.}(2005){van Noort}, {Rouppe van der Voort}, \&
  {L{\"o}fdahl}}]{vanNoort2005}
{van Noort}, M., {Rouppe van der Voort}, L., \& {L{\"o}fdahl}, M.~G. 2005,
  \solphys, 228, 191

\bibitem[{{Vecchio} {et~al.}(2009){Vecchio}, {Cauzzi}, \&
  {Reardon}}]{vecchio2009}
{Vecchio}, A., {Cauzzi}, G., \& {Reardon}, K.~P. 2009, \aap, 494, 269

\bibitem[{{Vecchio} {et~al.}(2007){Vecchio}, {Cauzzi}, {Reardon}, {Janssen}, \&
  {Rimmele}}]{vecchio2007}
{Vecchio}, A., {Cauzzi}, G., {Reardon}, K.~P., {Janssen}, K., \& {Rimmele}, T.
  2007, \aap, 461, L1

\bibitem[{{Viavattene} {et~al.}(2020){Viavattene}, {Consolini}, {Giovannelli},
  {Berrilli}, {Del Moro}, {Giannattasio}, {Penza}, \&
  {Calchetti}}]{viavattene2020}
{Viavattene}, G., {Consolini}, G., {Giovannelli}, L., {et~al.} 2020, Entropy,
  22, 716

\bibitem[{{Viavattene} {et~al.}(2021){Viavattene}, {Murabito}, {Guglielmino},
  {Ermolli}, {Consolini}, {Giorgi}, \& {Jafarzadeh}}]{viavattene2021}
{Viavattene}, G., {Murabito}, M., {Guglielmino}, S.~L., {et~al.} 2021, Entropy,
  23, 413

\bibitem[{Virtanen {et~al.}(2020)Virtanen, Gommers, Oliphant, Haberland, Reddy,
  Cournapeau, Burovski, Peterson, Weckesser, Bright, {van der Walt}, Brett,
  Wilson, Millman, Mayorov, Nelson, Jones, Kern, Larson, Carey, Polat, Feng,
  Moore, {VanderPlas}, Laxalde, Perktold, Cimrman, Henriksen, Quintero, Harris,
  Archibald, Ribeiro, Pedregosa, {van Mulbregt}, \& {SciPy 1.0
  Contributors}}]{Scipy}
Virtanen, P., Gommers, R., Oliphant, T.~E., {et~al.} 2020, Nature Methods, 17,
  261

\bibitem[{{Vissers} \& {Rouppe van der Voort}(2012)}]{vissers2012}
{Vissers}, G. \& {Rouppe van der Voort}, L. 2012, \apj, 750, 22

\bibitem[{{Viticchi{\'e}} {et~al.}(2009){Viticchi{\'e}}, {Del Moro},
  {Berrilli}, {Bellot Rubio}, \& {Tritschler}}]{viticchie2009}
{Viticchi{\'e}}, B., {Del Moro}, D., {Berrilli}, F., {Bellot Rubio}, L., \&
  {Tritschler}, A. 2009, \apjl, 700, L145

\bibitem[{{Viticchi{\'e}} {et~al.}(2010){Viticchi{\'e}}, {Del Moro},
  {Criscuoli}, \& {Berrilli}}]{viticchie2010}
{Viticchi{\'e}}, B., {Del Moro}, D., {Criscuoli}, S., \& {Berrilli}, F. 2010,
  \apj, 723, 787

\bibitem[{{Vourlidas} {et~al.}(2016){Vourlidas}, {Beltran}, {Chintzoglou},
  {Eisenhower}, {Korendyke}, {Feldman}, {Moser}, {Shea}, {Johnson-Rambert},
  {McMullin}, {Stenborg}, {Shepler}, \& {Roberts}}]{vault}
{Vourlidas}, A., {Beltran}, S.~T., {Chintzoglou}, G., {et~al.} 2016, Journal of
  Astronomical Instrumentation, 5, 1640003

\bibitem[{{Wedemeyer} {et~al.}(2020){Wedemeyer}, {Szydlarski}, {Jafarzadeh},
  {Eklund}, {Guevara Gomez}, {Bastian}, {Fleck}, {de la Cruz Rodriguez},
  {Rodger}, \& {Carlsson}}]{alma}
{Wedemeyer}, S., {Szydlarski}, M., {Jafarzadeh}, S., {et~al.} 2020, \aap, 635,
  A71

\bibitem[{{Wiegelmann} \& {Sakurai}(2021)}]{Wiegelmann2021LRSP}
{Wiegelmann}, T. \& {Sakurai}, T. 2021, Living Reviews in Solar Physics, 18, 1

\bibitem[{{Zuccarello} {et~al.}(2009){Zuccarello}, {Romano}, {Guglielmino},
  {Centrone}, {Criscuoli}, {Ermolli}, {Berrilli}, \& {Del
  Moro}}]{zuccarello2009}
{Zuccarello}, F., {Romano}, P., {Guglielmino}, S.~L., {et~al.} 2009, \aap, 500,
  L5

\end{thebibliography}

\appendix

\section{Data summary and Metadata information}

\begin{figure*}
\centering
{
\includegraphics[scale=0.9,trim=10 220 50 60,clip]{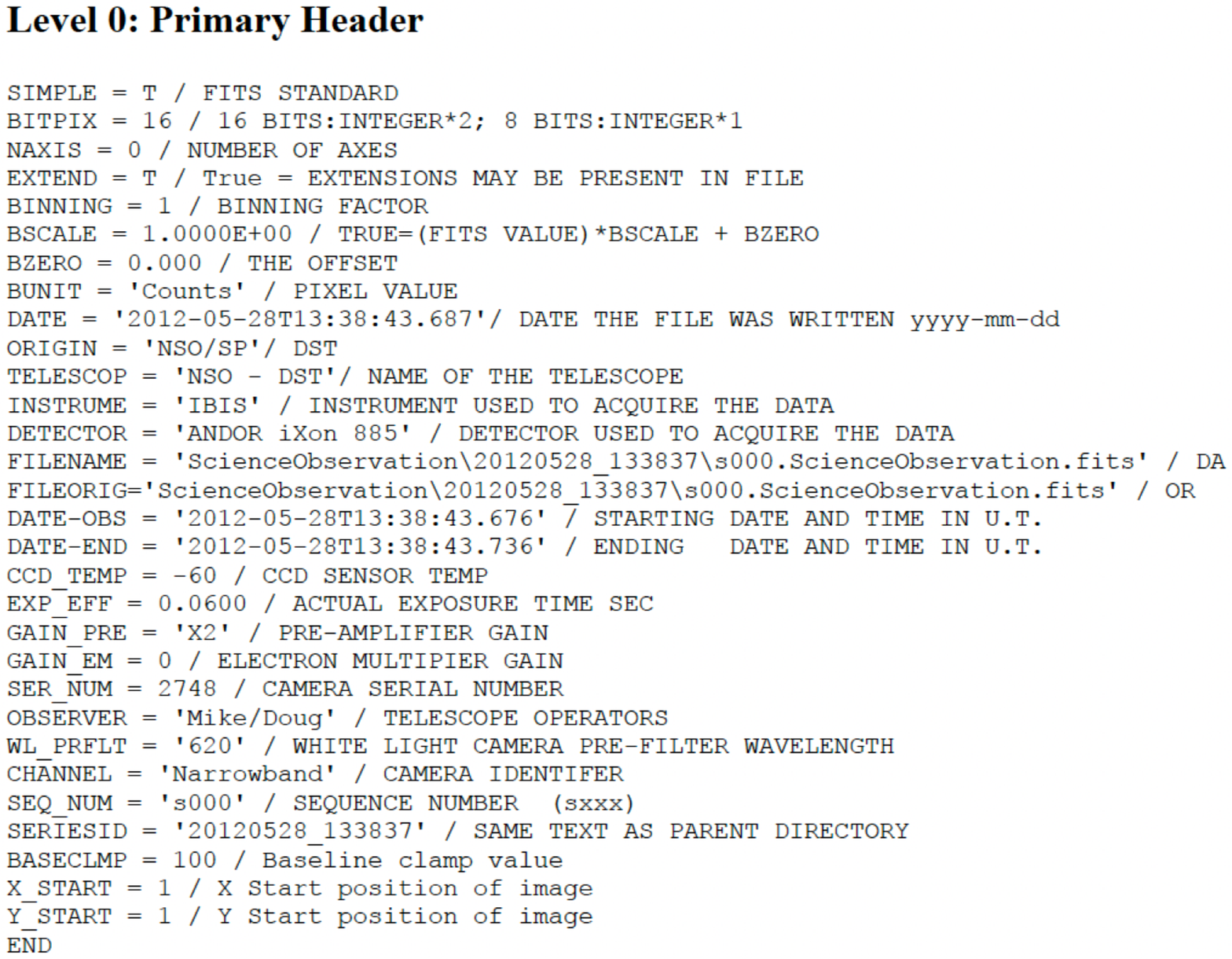}
}
\caption{Example of metadata information stored in the FITS primary header of the Level 0 data available in IBIS-A for the observations taken on 28 May 2012 at 13:38:43 UT.
}
\label{fhdr0}
\end{figure*}

\begin{figure*}
\centering
{
\includegraphics[scale=1,trim=50 100 10 100,clip]{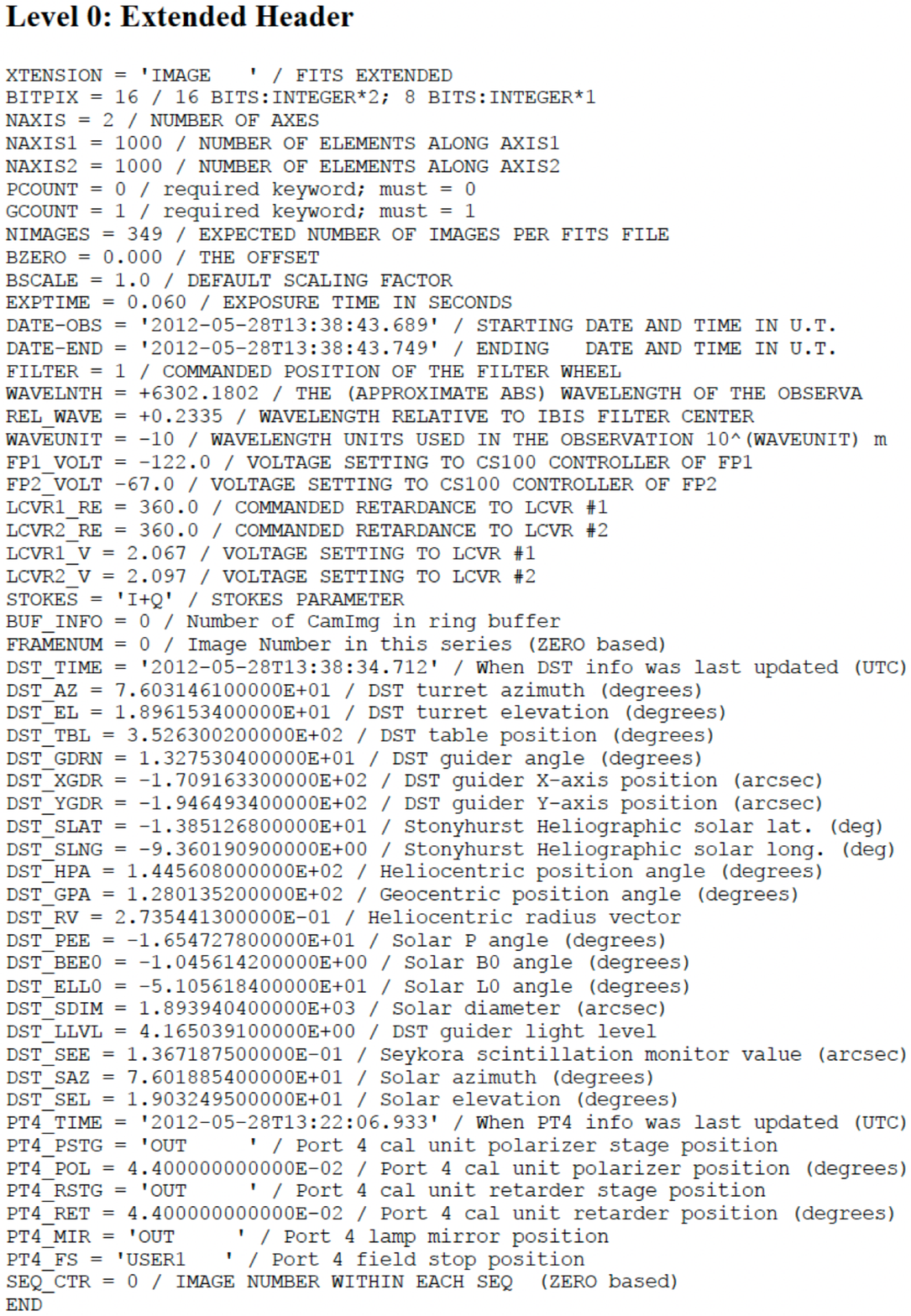}
}
\caption{Example of metadata information stored in the FITS extended header of the Level 0 data available in IBIS-A for the observations taken on 28 May 2012 at 13:38:43 UT.
}
\label{fhdr02}
\end{figure*}

\begin{figure*}
\centering
{
\includegraphics[scale=1,trim=50 200 10 100,clip]{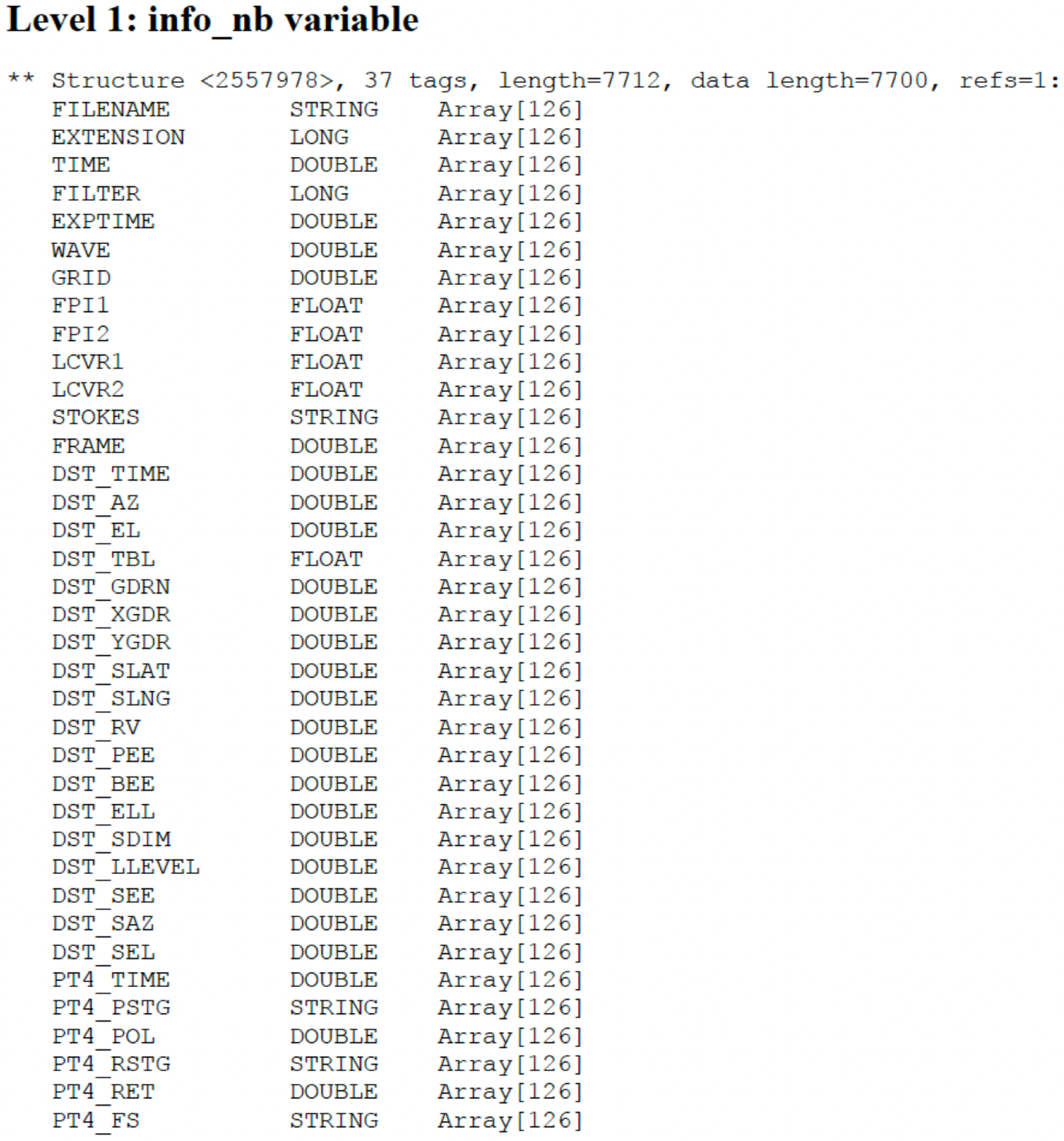}
}
\caption{Example of the several fields with information stored in the \textit{info$\_$nb} variable, which is relevant to the Level 1 data derived from the spectro-polarimetric observations taken at the Fe I 617.3 nm line on 13 May 2016 at  13:38:48 UT.
}
\label{fhdr1}
\end{figure*}


\begin{figure*}
\centering
{
\includegraphics[scale=1,trim=50 200 10 200,clip]{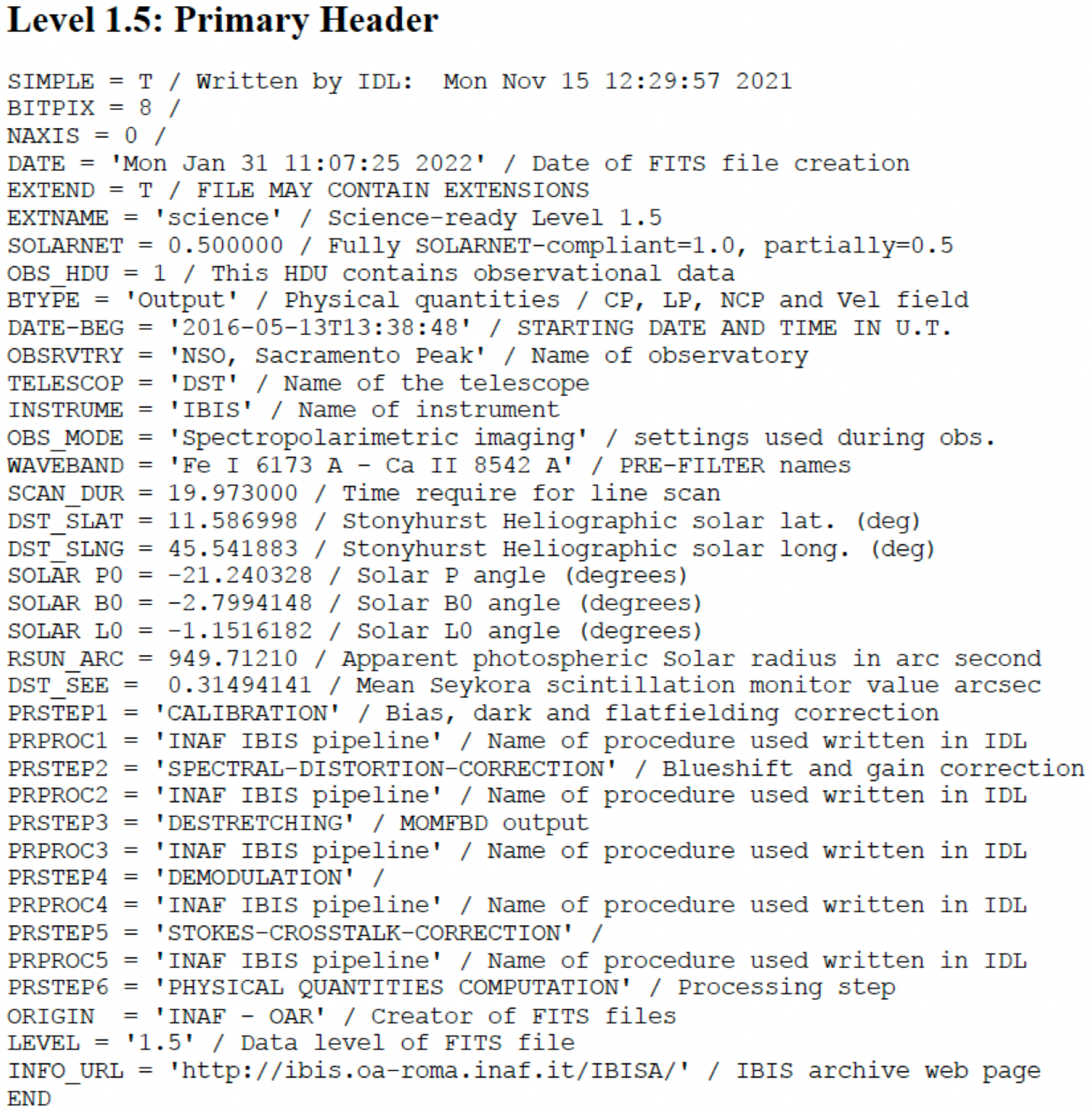}
}
\caption{Example of metadata information stored in the FITS  primary header of the Level 1.5 data available in IBIS-A for the observations taken on 13 May 2016 at  13:38:48 UT. 
}
\label{fhdr15}
\end{figure*}

\begin{figure*}
\centering
{
\includegraphics[scale=0.9,trim=50 150 10 150,clip]{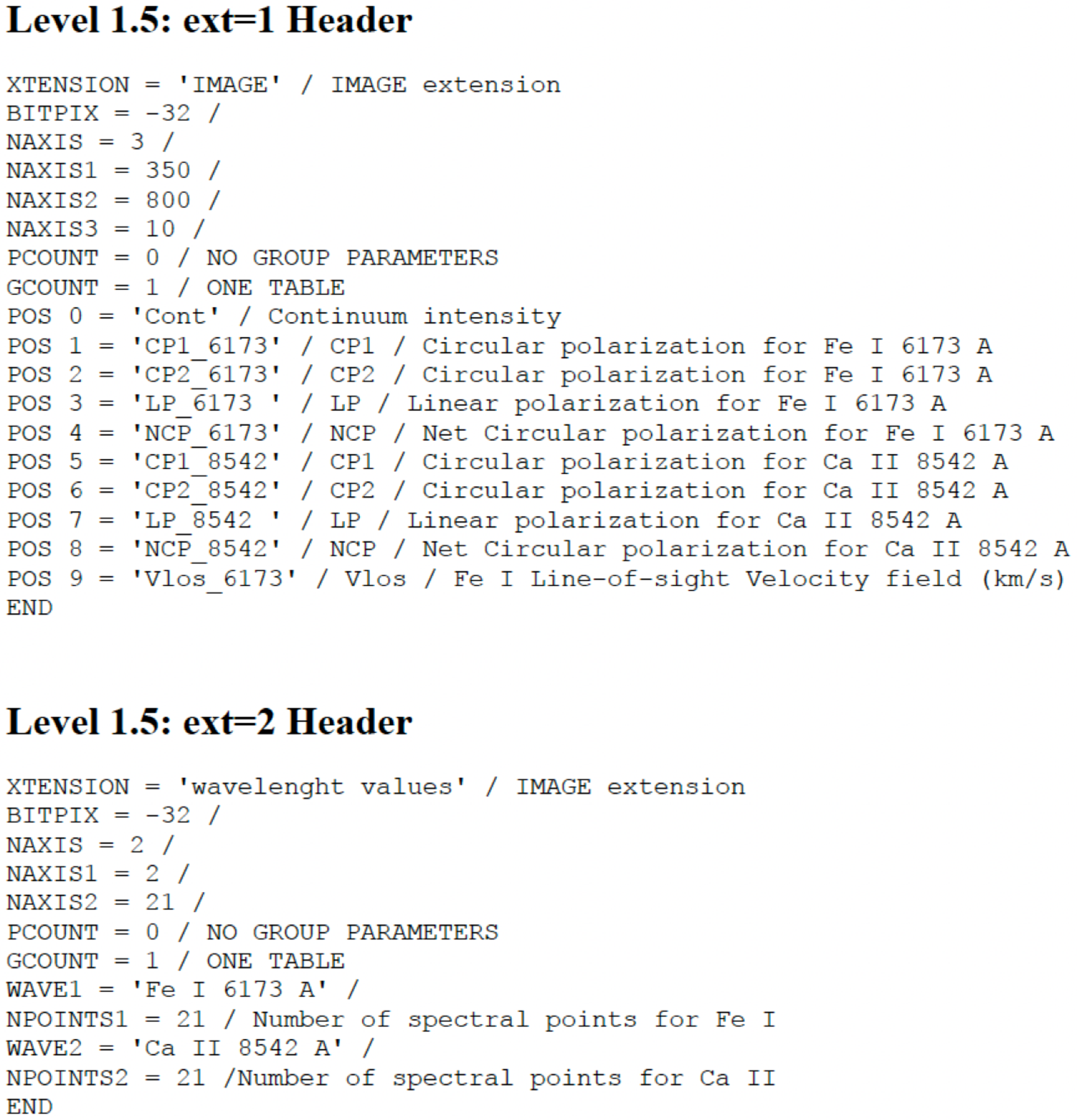}
}
\caption{Example of metadata information stored in the FITS  extended header of the Level 1.5 data available in IBIS-A for the observations taken on 13 May 2016 at  13:38:48 UT. 
}
\label{fhdr151}
\end{figure*}

\begin{figure*}
\centering
{
\includegraphics[scale=0.8,trim=50 160 10 170,clip]{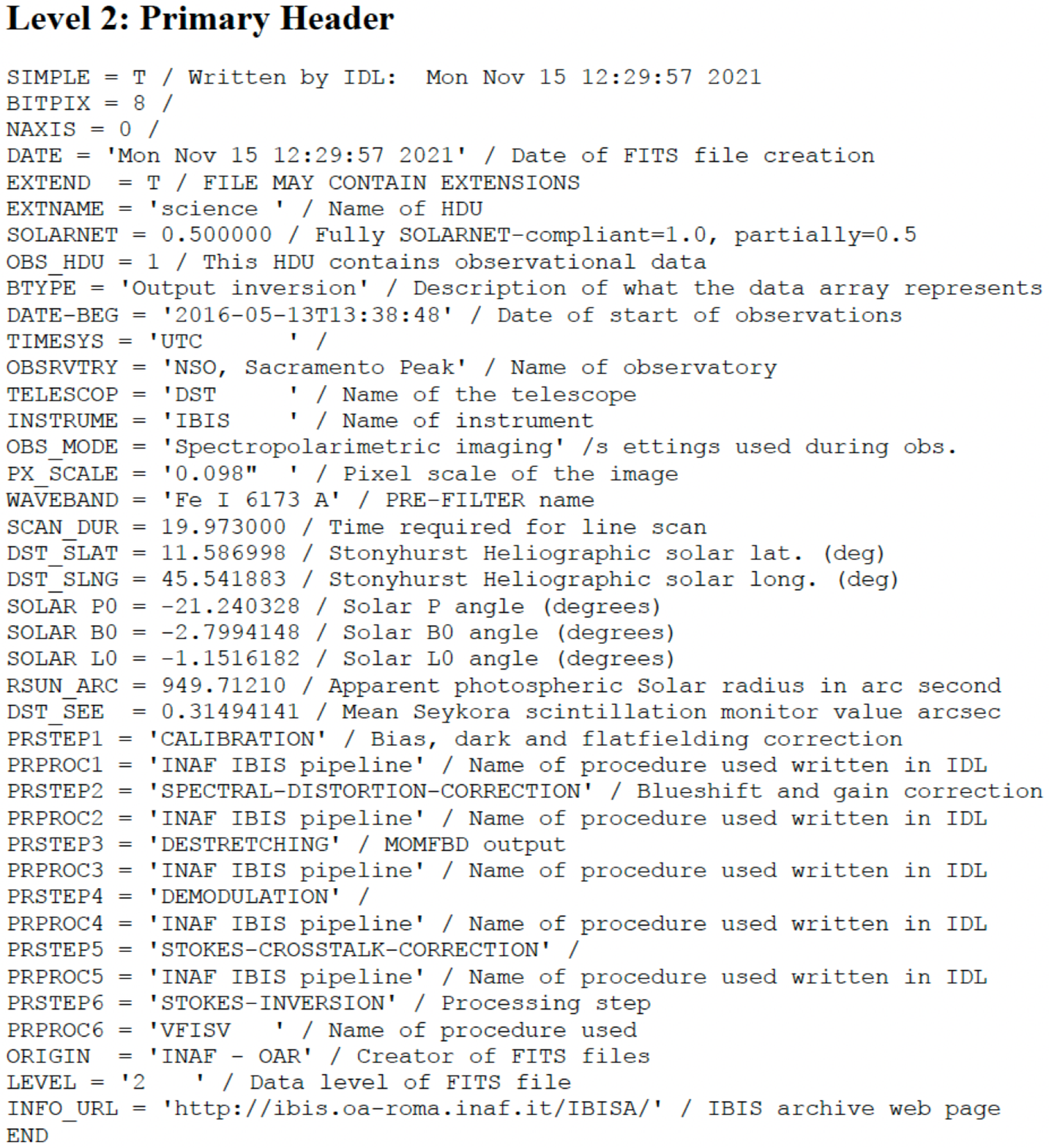}}
\includegraphics[scale=0.8,trim=110 240 10 240,clip]{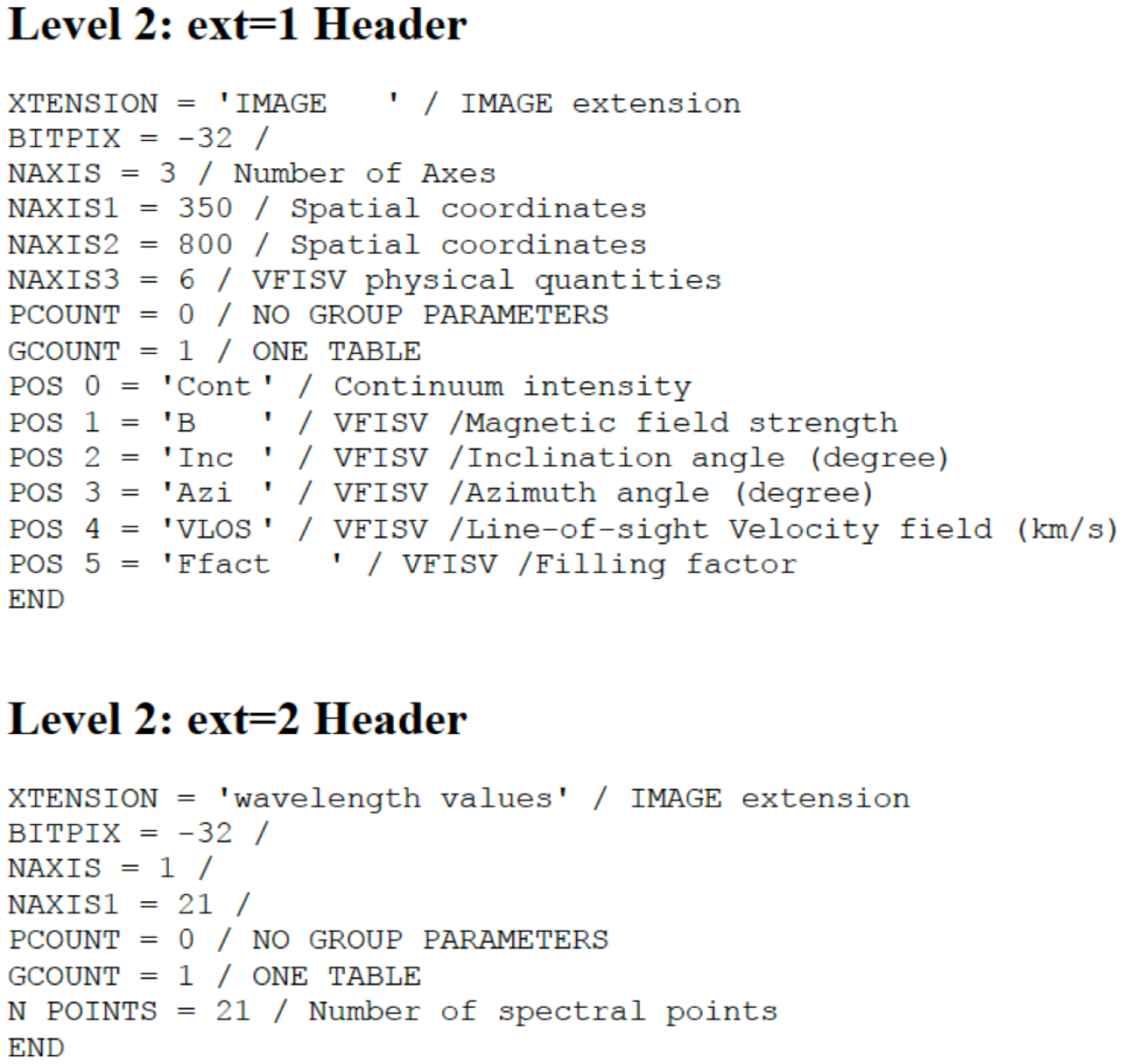}
\caption{Example of metadata information stored in the FITS  header of the Level 2 data available in IBIS-A for the observations taken on 13 May 2016 at  13:38 UT. 
}
\label{fhdr2}
\end{figure*}

Table \ref{tablev1} summarizes information on the Level 1 data available in IBIS-A. These data consist of the observations processed for instrumental calibrations and restored for residual image degradation by seeing using the MOMFBD method. 
Table  \ref{tablev2a}  summarizes information on the Level 2 data available in IBIS-A. These data consist of the observations processed for data inversion using the VFISV code. 
Find more details  concerning the Level 1 and Level 2 data  in Sect. 2.3.

Figures  \ref{fhdr0} and  \ref{fhdr02} show examples of metadata information provided in the keywords of the primary and extended headers of FITS files of the  Level 0 data available in IBIS-A. Figure \ref{fhdr1} gives an example of the fields with information stored in the variable  \textit{info$\_$nb} relevant to Level 1 data.  Figures \ref{fhdr15} and \ref{fhdr151} show examples of metadata information provided in the keywords of the primary and extended headers of FITS files of the  Level 1.5 data, respectively. Figure \ref{fhdr2} displays same information for Level 2 data.  
See Sect. 3.1 for more details.

\onecolumn
\begin{longtable}{lcllc}
\caption{\label{tablev1} Overview of the Level 1 data available in IBIS-A.}\\
\hline\hline
Date         & Time        & Target$\rm ^{(a)}$ & Lambda$\rm ^{(b)}$ Points$\rm ^{(c)}$ & Scans \\
YY-MM-DD     &    [UT]     &                    &                          &      \\
\hline
\endfirsthead
\caption{continued.}\\
\hline\hline
Date         & Time        & Target$\rm ^{(a)}$ & Lambda$\rm ^{(b)}$ Points$\rm ^{(c)}$ & Scans \\
YY-MM-DD     &    [UT]     &                    &                          &      \\
\hline
\endhead
\hline 
\endfoot
2008-10-15   & 16:30:00    & PO                 & 6173 IQUV (21), 8542 I (21)       & 80      \\
2012-04-17   & 13:58:43    & PO                 & 6302 IQUV (30), 6173 IQUV (24) 8542 I (25) & 56\\
             & 15:09:06    & PO                 & 6302 IQUV (30), 6173 IQUV (24) 8542 I (25) & 68\\
             & 18:13:40    & PO                 & 6302 IQUV (30), 6173 IQUV (24) 8542 I (25) & 100\\
2013-02-05   & 18:18:21    & SP                 & 6173 IQUV (5), 6563 I (14), 8542 IQUV (25) & 61\\
2013-10-07   & 17:39:08    & SP                 & 6173 IQUV (10), 5896 IQUV (23)             & 137\\
2013-10-20   & 15:14:22    & SP                 & 6173 IQUV (10), 7090 I (10), 8542 IQUV (11)  & 175\\
             & 17:37:20    & QS                 & 5896 IQUV (10)                               & 157\\
             & 18:11:51    & QS                 & 5896 IQUV (10)                               & 137\\
2014-04-15   & 14:41:21    & PO                 & 6173 IQUV (5), 8542 I (11)                   & 300\\
2014-04-21   & 14:15:33    & PO                 & 6173 IQUV (10x6), 8542 IQUV (11x6)         & 15\\
             & 15:17:03    & PO                 & 6173 IQUV (10x6), 8542 IQUV (11x6),        & 15\\
2014-10-14   & 15:15:57    & PO                 & 6173 IQUV (20), 6563 I (25), 8542 IQUV (25)& 238\\
2014-10-22   & 14:29:34    & SP                 & 6173 IQUV (20), 6563 I (25), 8542 IQUV (25)& 84\\
             & 15:46:24    & SP                 & 6173 IQUV (20), 6563 I (25), 8542 IQUV (25)& 17\\
             & 16:05:36    & SP                 & 6173 IQUV (20), 6563 I (25), 8542 IQUV (25)& 37\\
             & 17:43:40    & PO                 & 6173 IQUV (20), 6563 I (25), 8542 IQUV (25)& 31\\
2015-04-10   & 14:34:25    & SP                 & 6173 IQUV (10), 8542 IQUV (11),            & 237\\
2015-04-11   & 14:14:40    & PO                 & 6173 IQUV (10x6), 8542 IQUV (11x6)         & 19\\
             & 15:01:55    & PO                 & 6173 IQUV (10x6), 8542 IQUV (11x6)         & 18\\
2015-04-30   & 13:52:44    & QS                 & 6173 IQUV (21), 8542 I (21)                  & 290\\
2015-05-01   & 14:17:52    & QS                 & 6173 IQUV (21), 8542 IQUV (21)             & 58\\
2015-05-02   & 14:25:34    & QS                 & 6173 IQUV (21), 8542 IQUV (21)             & 50\\
             & 15:08:33    & QS                 & 6173 IQUV (21), 8542 IQUV (21)             & 20\\
             & 15:27:31    & QS                 & 6173 IQUV (21), 8542 IQUV (21)             & 79\\
2015-05-03   & 14:28:42    & SP                 & 6173 IQUV (21), 8542 IQUV (21)             & 85\\
             & 17:15:05    & SP                 & 6173 IQUV (21), 8542 IQUV (21)             & 25\\
2015-05-18   & 16:38:19    & SP                 & 6173 IQUV (24), 6302 IQUV (30), 8542 I (25), 6563 I (17)             & 20\\
2016-05-13   & 13:38:48    & SP                 & 6173 IQUV (21), 8542 IQUV (21)             & 50\\
             & 14:15:09    & SP                 & 6173 IQUV (21), 8542 IQUV (21)             & 58\\
             & 15:00:27    & PO                 & 6173 IQUV (21), 8542 IQUV (21)             & 163\\
2016-05-19   & 13:37:43    & SP                 & 6173 IQUV (21), 8542 IQUV (21)             & 100\\
2016-05-20   & 13:53:06    & SP                 & 6173 IQUV (21), 8542 IQUV (21)             & 319\\
2016-10-05   & 15:06:58    & PO                 & 8542 IQUV (11), 6563 I (14)             & 200\\
2016-10-06   & 15:58:20    & QS                 & 8542 IQUV (11), 6563 I (14)             & 200\\
             & 16:55:16    & QS                 & 8542 IQUV (11), 6563 I (14)             & 176\\
2016-10-10   & 17:12:08    & PO                 & 8542 IQUV (11), 6563 I (14)             & 200\\
             & 18:07:47    & PO                 & 8542 IQUV (11), 6563 I (14)             & 100\\
2016-10-11   & 16:04:17    & QS                 & 6173 IQUV (20), 8542 IQUV (25), 6563 I (14) & 29\\
             & 16:33:07    & QS                 & 6173 IQUV (20), 8542 IQUV (25), 6563 I (14) & 31\\
2016-10-12   & 14:53:30    & QS                 & 6173 IQUV (20), 8542 IQUV (25), 6563 I (14) & 100\\
             & 16:31:47    & SP                 & 8542 IQUV (11), 6563 I (14)            & 5\\
             & 16:38:46    & SP                 & 8542 IQUV (11), 6563 I (14)            & 200\\
             & 17:32:08    & SP                 & 8542 IQUV (11), 6563 I (14)            & 85\\
2016-10-14   & 14:53:57    & QS                 & 6173 IQUV (20), 5896 IQUV (10), 6563 I (14) & 150\\
             & 16:25:24    & SP                 & 8542 IQUV (11), 6563 I (14)            & 228\\
2016-10-17   & 15:24:24    & QS                 & 6173 IQUV (20), 8542 IQUV (25), 6563 I (14) & 74\\
2016-10-18   & 15:12:40    & QS                 & 6173 IQUV (20), 8542 IQUV (25), 6563 I (14) & 58\\
2016-10-21   & 16:00:14    & QS                 & 6173 IQUV (20), 8542 IQUV (25), 6563 I (14) & 60\\
             & 17:00:34    & PO                 & 8542 IQUV (11), 6563 I (14)            & 232\\
2017-05-05   & 14:06:00    & PO                 & 6173 IQUV (21), 8542 IQUV (21)            & 85\\
             & 15:37:49    & QS                 & 6173 IQUV (21), 8542 IQUV (21)             & 107\\
2017-05-12   & 13:34:16    & QS                 & 6173 IQUV (21), 8542 IQUV (21)             & 40\\
2017-11-10   & 15:12:46    & QS                 & 8542 I (30), 5896 I (19)                   & 347\\
2018-02-01   & 17:06:03    & QS                 & 8542 I (30), 5896 I (19)                   & 149\\
             & 17:44:02    & QS                 & 8542 I (30), 5896 I (19)                   & 147\\
2018-04-11   & 16:05:05    & QS                 & 8542 I (30), 5896 I (19)                   & 293\\
2018-04-26   & 15:48:38    & PO                 & 8542 I (27), 5896 I (24)                   & 199\\
2018-05-18   & 14:01:10    & PO                 & 8542 I (30), 6563 I (17), 5896 I (19)      & 365\\
2018-05-19   & 14:01:17    & PO                 & 8542 I (29), 6563 I (27), 5896 I (19)      & 215\\
2018-06-20   & 13:53:05    & SP                 & 8542 I (29), 6563 I (27), 5896 I (19)      & 185\\
2018-06-21   & 13:57:00    & QS                 & 8542 I (29), 6563 I (27)                   & 315\\
             & 15:22:32    & QS                 & 8542 I (29), 6563 I (27)                 & 10\\
             & 15:53:12    & QS                 & 8542 I (29), 6563 I (27)                       & 10\\
             & 16:21:09    & QS                 & 8542 I (29), 6563 I (27)                       & 10\\
             & 16:32:05    & QS                 & 8542 I (29), 6563 I (27)                       & 10\\
2018-06-21   & 14:29:32    & QS                 & 8542 I (29), 6563 I (27)                       & 182\\
             & 15:18:55    & QS                 & 8542 I (29), 6563 I (27)                       & 253\\
2018-06-25   & 14:31:04    & QS                 & 8542 I (30), 6563 I (27)                       & 190\\
             & 15:22:10    & QS                 & 8542 I (30), 6563 I (27)                       & 182\\
             & 16:13:17    & QS                 & 8542 I (30), 6563 I (27)                       & 127\\
2018-06-26   & 13:35:58    & QS                 & 8542 I (30), 6563 I (27)                       & 179\\
             & 14:25:00    & QS                 & 8542 I (30), 6563 I (27)                       & 180\\
             & 15:13:44    & QS                 & 8542 I (30), 6563 I (27)                       & 177\\
             & 16:02:50    & QS                 & 8542 I (30), 6563 I (27)                       & 130\\
2018-06-27   & 13:40:50    & QS                 & 8542 I (30), 6563 I (27)                       & 213\\
             & 14:41:01    & QS                 & 8542 I (30), 6563 I (27)                       & 190\\
             & 15:36:14    & QS                 & 8542 I (30), 6563 I (27)                       & 213\\
             & 16:51:11    & QS                 & 8542 I (30), 6563 I (27)                       & 5\\
             & 16:54:46    & QS                 & 8542 I (30), 6563 I (27)                       & 5\\
             & 17:10:40    & QS                 & 8542 I (30), 6563 I (27)                       & 5\\
             & 17:14:48    & QS                 & 8542 I (30), 6563 I (27)                       & 5\\
2018-07-05   & 14:11:11    & QS                 & 8542 I (30), 6563 I (27)                       & 68\\
             & 14:35:16    & QS                 & 8542 I (30), 6563 I (27)                       & 5\\
             & 14:41:09    & QS                 & 8542 I (30), 6563 I (27)                       & 5\\
             & 14:46:33    & QS                 & 8542 I (30), 6563 I (27)                       & 5\\
             & 14:54:33    & QS                 & 8542 I (30), 6563 I (27)                       & 5\\
2018-06-16   & 13:53:21    & QS                 & 8542 I (30), 6563 I (27)                       & 106\\
             & 14:50:04    & QS                 & 8542 I (30), 6563 I (27)                       & 106\\
2018-07-17   & 13:33:46    & QS                 & 8542 I (30), 6563 I (27)                       & 122\\
             & 14:28:21    & QS                 & 8542 I (30), 6563 I (27)                       & 116\\
2018-07-23   & 14:02:33    & QS                 & 8542 I (30), 6563 I (27)                       & 3\\
             & 14:03:41    & QS                 & 8542 I (30), 6563 I (27)                       & 139\\
             & 14:41:02    & QS                 & 8542 I (30), 6563 I (27)                       & 151\\
2018-08-16   & 14:15:34    & QS                 & 8542 I (30)                                    & 573\\
2018-08-17   & 14:01:21    & QS                 & 8542 I (30)                                    & 582\\
2018-09-24   & 14:34:27    & QS                 & 8542 I (30), 6563 I (27)                       & 201\\
             & 15:33:56    & QS                 & 8542 I (30), 6563 I (27)                       & 141\\
2018-09-25   & 14:09:52    & QS                 & 8542 I (30), 6563 I (27)                       & 196\\
             & 15:32:27    & QS                 & 8542 I (30), 6563 I (27)                       & 145\\
2018-10-04   & 20:29:11    & QS                 & 8542 I (30), 6563 I (27)                       & 22\\
2018-10-10   & 15:06:09    & QS                 & 8542 I (30), 6563 I (27)                       & 146\\
2018-11-02   & 15:11:37    & QS L               & 8542 I (25)                                    & 900\\
             & 16:28:55    & QS L               & 8542 I (25)                                    & 509\\
2018-11-03   & 15:35:59    & QS L               & 8542 I (25)                                    & 900\\
             & 16:45:31    & QS L               & 8542 I (25)                                    & 900\\
             & 17:54:54    & QS L               & 8542 I (25)                                    & 900\\
2018-11-04   & 15:00:54    & QS L               & 8542 I (25)                                    & 619\\
             & 15:54:45    & QS L               & 8542 I (25)                                    & 900\\
             & 17:04:22    & QS L               & 8542 I (25)                                    & 900\\
             & 18:13:46    & QS L               & 8542 I (25)                                    & 175\\
             & 18:38:28    & QS L               & 8542 I (25)                                    & 99\\
2018-11-15   & 15:16:49    & PO                 & 8542 I (30), 6563 I (27)                       & 316\\
             & 17:20:28    & QS                 & 8542 I (30), 6563 I (27)                       & 190\\
             & 15:05:40    & PO                 & 8542 I (30), 6563 I (27)                       & 31\\
2018-11-16   & 15:13:26    & PO                 & 8542 I (30), 6563 I (27)                       & 189\\
             & 16:10:14    & QS                 & 8542 I (30), 6563 I (27)                       & 188\\
2018-11-19   & 15:33:04    & PO                 & 8542 I (30), 6563 I (27)                       & 163\\
             & 16:09:27    & PO                 & 8542 I (30), 6563 I (27)                       & 266\\
2018-11-27   & 16:58:08    & QS                 & 8542 I (30), 6563 I (27)                       & 244\\
2018-11-29   & 15:35:31    & QS                 & 8542 I (30), 6563 I (27)                       & 126\\
2018-12-04   & 16:28:59    & QS                 & 8542 I (30), 6563 I (27)                       & 279\\
2018-12-20   & 15:55:00    & QS L               & 8542 I (21), 6563 I (23) & 15\\
             & 16:04:54    & QS                 & 8542 I (21), 6563 I (23) & 10\\
             & 17:25:17    & QS                 & 8542 I (21), 6563 I (23) & 100\\
2018-12-23   & 15:15:32    & QS                 & 8542 I (21), 6563 I (23) & 120\\
             & 17:19:24    & QS                 & 8542 I (21), 6563 I (23) & 120\\
             & 17:53:06    & QS                 & 8542 I (21), 6563 I (23) & 120\\
             & 18:25:34    & QS                 & 8542 I (21), 6563 I (23) & 120\\
2019-01-25   & 15:40:56    & PO                 & 8542 I (30), 6563 I (27), 5896 I (19)          & 25\\
             & 15:52:35    & PO                 & 8542 I (30), 6563 I (27), 5896 I (19)          & 160\\
             & 17:27:47    & PO                 & 8542 I (30), 6563 I (27), 5896 I (19)          & 48\\
             & 17:46:32    & PO                 & 8542 I (30), 6563 I (27), 5896 I (19)          & 55\\
2019-02-18   & 15:39:03    & QS                 & 8542 I (30), 6563 I (27)                       & 127\\
2019-02-25   & 14:52:08    & QS L               & 8542 I (30), 6563 I (27)                       & 130\\
             & 15:25:23    & QS L               & 8542 I (30), 6563 I (27)                       & 146 \\
\end{longtable} 
\begin{quotation}
\textbf{Notes.} 
$\rm ^{(a)}$ Target: PO: pore, SP: sunspot, QS: quiet sun, QS L: Quiet Sun Limb. $\rm ^{(b)}$ Lambda: Spectral lines observed with IBIS. $\rm ^{(c)}$ Points describes the number of spectral positions sampled in the line. Observing modes with multiple exposures at each point are reported as $n \times m$, where $n$ and $m$ are the number of spectral points and the number of exposures at each point, respectively.  
\end{quotation}

\begin{longtable}{lclllc}
\caption{\label{tablev2a} Overview of the Level 2 data available in IBIS-A.}\\
\hline\hline
Date            & Time       & Target$\rm ^{(a)}$ & Data$\rm ^{(b)}$ & Other data$\rm ^{(c)}$&Scans \\
YY-MM-DD        &    [UT]    &                    &   from 6173 \AA~ IUQV & available &    \\
\hline
\endfirsthead
\caption{continued.}\\
\hline\hline
Date            & Time       & Target$\rm ^{(a)}$ & Data$\rm ^{(b)}$ & Other data$\rm ^{(c)}$&Scans \\
YY-MM-DD        &    [UT]    &                    &  from 6173 \AA~IUQV   & available &    \\
\hline
\endhead
\hline 
\endfoot
2012-04-17 & 13:58:43 & PO & $\mathrm{B, B_{inc}, B_{azi},  F_{fact},  V_{los}}$& 6302 IUQV, 8542 I & 56 \\
2012-04-17 & 15:09:06 & PO & $\mathrm{B, B_{inc}, B_{azi},  F_{fact},  V_{los}}$& 6302 IUQV, 8542 I & 68 \\
2012-04-17 & 18:43:40 & PO & $\mathrm{B, B_{inc}, B_{azi},  F_{fact},  V_{los}}$& 6302 IUQV, 8542 I & 100 \\
2013-10-07 & 17:39:08 & SP & $\mathrm{B, B_{inc}, B_{azi},  F_{fact},  V_{los}}$& 5896 IUQV& 130 \\
2014-04-21 & 14:15:33 & PO & $\mathrm{B, B_{inc}, B_{azi},  F_{fact},  V_{los}}$& 8542 IUQV& 15\\
2014-10-14 & 15:15:57 & PO & $\mathrm{B, B_{inc}, B_{azi},  F_{fact},  V_{los}}$& 8542 IUQV, 6563 I& 238 \\
2015-04-10 & 14:34:25 & SP & $\mathrm{B, B_{inc}, B_{azi},  F_{fact},  V_{los}}$& 8542 IUQV & 237\\
2015-04-11 & 14:14:40 & PO & $\mathrm{B, B_{inc}, B_{azi},  F_{fact},  V_{los}}$& 8542 IUQV & 19\\
2015-04-11 & 15:01:55 & PO & $\mathrm{B, B_{inc}, B_{azi},  F_{fact},  V_{los}}$& 8542 IUQV& 18\\
2015-04-30 & 13:52:44 & QS & $\mathrm{B, B_{inc}, B_{azi},  F_{fact},  V_{los}}$& 8542 I & 290\\
2015-05-01 & 14:17:52 & QS & $\mathrm{B, B_{inc}, B_{azi},  F_{fact},  V_{los}}$ & 8542 IUQV & 58              \\
2015-05-02 & 14:25:34 & QS & $\mathrm{B, B_{inc}, B_{azi},  F_{fact},  V_{los}}$  & 8542 IUQV & 50    \\
2015-05-02 & 15:08:33 & QS & $\mathrm{B, B_{inc}, B_{azi},  F_{fact},  V_{los}}$ & 8542 IUQV& 20 \\
2015-05-02 & 15:27:31 & QS & $\mathrm{B, B_{inc}, B_{azi},  F_{fact},  V_{los}}$  &  8542 IUQV & 79 \\ 
2015-05-03 & 14:28:42 & SP &  $\mathrm{B, B_{inc}, B_{azi},  F_{fact},  V_{los}}$ & 8542 IUQV & 85  \\
2015-05-03 & 17:15:05 & SP & $\mathrm{B, B_{inc}, B_{azi},  F_{fact},  V_{los}}$  &  8542 IUQV & 25\\ 
2016-05-13 & 13:38:48 & SP & $\mathrm{B, B_{inc}, B_{azi},  F_{fact},  V_{los}}$  & 8542 IUQV & 50        \\
2016-05-13 & 14:15:05 & SP & $\mathrm{B, B_{inc}, B_{azi},  F_{fact},  V_{los}}$  &  8542 IUQV & 58\\
2016-05-13 & 15:00:27 & PO & $\mathrm{B, B_{inc}, B_{azi},  F_{fact},  V_{los}}$ & 8542 IUQV & 163 \\
2016-05-19 & 13:37:43 & SP & $\mathrm{B, B_{inc}, B_{azi},  F_{fact},  V_{los}}$ & 8542 IUQV & 100\\
2016-05-20 & 13:53:06 & SP & $\mathrm{B, B_{inc}, B_{azi},  F_{fact},  V_{los}}$ & 8542 IUQV & 319\\
2016-10-11 & 16:04:17 & QS & $\mathrm{B, B_{inc}, B_{azi},  F_{fact},  V_{los}}$ & 8542 IUQV, 6563 I & 29\\
2016-10-11 & 16:33:07 & QS & $\mathrm{B, B_{inc}, B_{azi},  F_{fact},  V_{los}}$ &  8542 IUQV, 6563 I & 31 \\
2016-10-12 & 14:53:30 & QS & $\mathrm{B, B_{inc}, B_{azi},  F_{fact},  V_{los}}$& 8542 IUQV, 6563 I & 100\\
2017-05-05 & 14:06:00 &  PO & $\mathrm{B, B_{inc}, B_{azi},  F_{fact},  V_{los}}$ & 8542 IUQV & 85\\
2017-05-05 & 15:37:49 & QS & $\mathrm{B, B_{inc}, B_{azi},  F_{fact},  V_{los}}$ & 8542 IUQV &  107\\
2017-05-12 & 13:34:16 & QS & $\mathrm{B, B_{inc}, B_{azi},  F_{fact},  V_{los}}$ & 8542 IUQV & 40\\
\end{longtable} 
\begin{quotation}
\textbf{Notes.} 
$\rm ^{(a)}$ Target: PO: pore, SP: sunspot, QS: quiet sun, QS L: Quiet Sun Limb. $\rm ^{(b)}$ Data: Level 2 data from Fe I 6173 \AA~  IUQV observations. $\mathrm{B, B_{inc}, B_{azi}}$, and $\mathrm{F_{fact}}$ refer to the 
magnetic field strength, inclination, azimuth, and filling factor,  respectively. $\mathrm{V_{los}}$ is the longitudinal component of the velocity field. 
$\rm ^{(c)}$ Other data available: Other spectral lines observed with IBIS. Line core is given in \AA.  
\end{quotation}

\twocolumn
\end{document}